\documentclass[12pt,a4paper,authoryear]{article}

\RequirePackage[OT1]{fontenc}
\RequirePackage{graphicx}
\RequirePackage{amsthm}
\RequirePackage[cmex10]{amsmath}
\RequirePackage[round]{natbib}

\usepackage[table]{xcolor}
\usepackage{hhline}

\usepackage{indentfirst}
\usepackage{multirow}
\usepackage{tikz}
\usetikzlibrary{intersections}

\newcolumntype{Y}{>{\centering\arraybackslash}p{5em}}

\theoremstyle{definition}
\newtheorem{defi}{Definition}
\newtheorem{rem}{Remark}
\newtheorem{ex}{Example}



\theoremstyle{plain}
\newtheorem{thm}{Theorem}
\newtheorem{lem}{Lemma}
\newtheorem{prop}{Proposition}
\newtheorem{cor}{Corollary}

\usepackage{amssymb}

\usepackage{subfig} 

\usepackage{mathtools}

\usepackage{setspace}
\usepackage{geometry}
\usepackage{authblk}

\usepackage[pdftex]{hyperref}
\hypersetup{colorlinks,citecolor=blue,urlcolor=blue,linkcolor=blue}
\usepackage{hypernat}

\begin{document}

\title{Informed Principal Problems in Bilateral Trading}
\author{Takeshi Nishimura\thanks{1-23-1 Komazawa, Setagaya-ku, Tokyo, 154-8525, Japan; tnishi@komazawa-u.ac.jp.}}
\affil{Faculty of Economics, Komazawa University}
\date{February 20, 2022}
\maketitle

\begin{abstract}
	We study bilateral trade with interdependent values as an informed-principal problem. 
	The mechanism-selection game has multiple equilibria that
	differ with respect to principal's 
	payoff and trading surplus.
	We characterize 
	the equilibrium that is worst for every type of principal, 
	and 
	characterize the conditions
	under which there are no equilibria with different payoffs for the principal.
	We also show that this 
	is
	the unique 
	equilibrium 
	that survives the intuitive criterion.
\end{abstract}

\noindent
{\it Keywords: }Informed principal; 
Bilateral trade; 
Interdependent values; 
Rothschild--Stiglitz--Wilson allocation; 
Intuitive criterion

\noindent
{\bf JEL Classification} C72 $\cdot$ D82 $\cdot$ D86

\newpage

\section{Introduction}\label{Section1}

In this study, we examine bilateral trade with interdependent values as an informed-principal problem.
Examples of 
this situation
are found in decentralized markets, 
including used-cars, housing, and labor markets.
This bilateral trade
can also occur as aftermarket transactions \citep{E2020Dworczak}.
For example, 
after telecommunications companies won a spectrum auction held by the United Kingdom,
they resold 
their spectrum licenses 
through the sale of companies 
\citep{AER2008Hafalir_Krishna}. 
Our analysis 
can be applied to understand 
how an auction winner 
trades with
a third party in ``continuation'' games.
We 
then
pose the following 
fundamental questions: 
Which trading mechanism does an informed principal select in equilibrium?
What efficiency properties do equilibrium allocations have?

These and related issues have been addressed in only a few studies.
In 
independent private values (IPV) environments \`{a} la \cite{JET1983Myerson_Satterthwaite}, 
\cite{JET1999Yilankaya} and \cite{RES2014Mylovanov_Troger}
showed that
the ``full-information'' optimal mechanism is an equilibrium of the mechanism-selection game
and is ex-ante optimal. 
Hence, the privacy of the 
principal's information is irrelevant to 
equilibrium outcomes
in these IPV environments.
\cite{JET2016Koessler_Skreta} obtained a number of important results
in an interdependent-values environment.
In 
their model, 
the seller's valuation 
is zero, 
while the buyer's valuation depends, in an arbitrary way, 
on the two parties' types.\footnote{
	Hence, their model can capture an economically interesting scenario wherein 
	the seller's type determines the horizontal characteristic of a good, 
	while the buyer's type determines how these different varieties are evaluated. 
	Our model cannot describe this scenario. 
	See \cite{2012Balestrieri_Izmalkov} for a related Hotelling model with an informed seller.
}
Using these features, 
\cite{JET2016Koessler_Skreta} characterized the set of equilibrium outcomes as 
revenue-ranked allocations
and showed that some practical selling procedures (e.g., book building in sales of companies)
yield the ex-ante optimal revenue.

An 
implication of these results is that
the interdependency of valuations gives rise to
a multiplicity of 
equilibrium allocations in the mechanism-selection game.\footnote{
	\cite{TE2019Koessler_Skreta} studied 
	an informed-seller problem
	with interdependent values and
	certifiable information. 
	They showed that 
	when 
	the certifiability structure is rich enough,
	strong (unconstrained) 
	Pareto optimal (SPO) allocations 
	\citep{E1990Maskin_Tirole}
	are 
	equilibria and are ex-ante optimal.
	Further, they 
	provided an example in which
	the SPO profit vector is the unique equilibrium profit vector.
	This is in contrast to the case without certifiable information.
}
Therefore, we need some 
criteria for 
equilibrium selection or refinement
to provide definite answers to our questions.

Several solution concepts for informed-principal problems are proposed by previous studies.
\cite{E1983Myerson} and \citet{E1990Maskin_Tirole,E1992Maskin_Tirole}
developed the general theories of mechanism selection by an informed principal, 
while 
\citet{TE2012Mylovanov_Troger,RES2014Mylovanov_Troger} advanced them
by focusing on general IPV environments.\footnote{
	Other contributions include
	\cite{GEB2008Cella},
	\cite{JET2008Severinov},
	\cite{RED2011Skreta},
	and 
	\cite{GEB2017Bedard}.
	See \cite{RES2014Mylovanov_Troger} for a more detailed 
	review of the literature. 
	See also \cite{JET2015Wagner_etal} for studies on informed-principal problems with moral hazard.
}
We follow the approach developed by 
\citet[henceforth MT]{E1992Maskin_Tirole}, 
who thoroughly 
studied
these 
problems in an environment with interdependent (or common) values.
The {\it Rothschild--Stiglitz--Wilson (RSW) allocation} 
plays a crucial role in their analysis. 
This 
is 
the best {\it safe mechanism} \citep{E1983Myerson} 
for 
every type of principal.
Their main result is 
the equilibrium characterization:
If an RSW allocation is interim efficient for some interior beliefs, 
then the equilibrium set 
coincides with
the set of feasible allocations that weakly 
dominate the RSW.\footnote{
	Recently, \cite{2022Dosis} 
	found that, in some cases, 
	the condition of interim efficiency is insufficient for this characterization result. 
	To address this problem, he introduced a stronger condition called interim optimality. 
	To obtain the same characterization as in MT,
	we use the condition of undominatedness, which is 
	also stronger than interim efficiency for RSW allocations, 	
	and apply a proof method developed by \cite{E1983Myerson}.
	See 
	Remark \ref{Rem_Myerson} in Section \ref{Section5.2} for details.\label{foot_Dosis}
}
The characterization implies that
the RSW is the 
worst equilibrium 
for 
every type of principal.
Moreover, 
MT characterized the RSW allocation itself
and proved that 
the RSW is the unique allocation that passes the {\it intuitive criterion} \citep{QJE1987Cho_Kreps}, 
under the assumption 
that only the principal has private information.

Our aim is to extend MT's results to our trading environment with bilateral asymmetric information. 
For expositional simplicity, we assume that the principal is a seller and the agent is a buyer.\footnote{
	Following MT, 
	we use feminine pronouns for the principal and masculine ones for the agent.
} 
Our first theorem provides a 
simple 
characterization of RSW.
By definition, the RSW is ``belief-free'' for the buyer. 
As a result, 
his ex-post payments are pinned down by the allocation rule.  
Hence, the seller's interim revenues are determined by the buyer's {\it virtual valuation} with the allocation rule. 
A novel feature of the theorem is that it characterizes the 
allocation rule
via posterior beliefs about the seller's type. 
This posterior 
derives from the prior by shifting probability mass 
toward
lower types. 
This posterior with the prior defines the seller's {\it virtual cost}. 
A 
key finding is that
the RSW 
maximizes
the expected {\it virtual surplus},
thereby reducing the seller's overstatement incentives.

This characterization is closely related to our second theorem. 
It provides a necessary and sufficient condition 
under which the RSW allocation is undominated for a given belief.
This 
new result
has some 
implications. 
First, 
this condition holds for the RSW with 
a belief given by 
our first theorem. 
Thus, 
the RSW 
is undominated for at least one posterior 
belief. 
This belief 
generates 
posteriors 
that support the RSW as an intuitive equilibrium. 
Further, 
the second theorem 
characterizes the set of prior beliefs 
for which 
the RSW is undominated. 
In other words, 
the 
theorem provides
a necessary and sufficient condition for the existence of a {\it strong solution} 
introduced by 
\cite{E1983Myerson}.
As noted by MT, 
its existence is equivalent to the uniqueness of the principal's equilibrium payoff vector.

If the RSW allocation is dominated for the prior,
the mechanism-selection game has a continuum of equilibria.
However, the extension of MT's refinement result 
to our environment is a nontrivial task.
This 
is related to 
one of the primary concerns in
the literature on informed-principal problems: that is, 
the principal's gains from concealing her information.
To clarify this point, 
let us assume that the RSW is dominated by 
another equilibrium. 
From the definition of RSW, 
the dominating allocation 
violates 
either 
ex-post incentive compatibility (EPIC) or
ex-post individual rationality (EPIR)
for the agent.
If some principal types 
mutually
benefit
from ``exchanging slack variables'' on 
these constraints, 
these types 
lose by making 
revealing deviations from the equilibrium.\footnote{
	This concept of slack exchange is central to the analysis of \cite{E1990Maskin_Tirole}.
	They showed that the principal obtains no gain from slack exchange
	in an IPV environment with quasi-linear preferences.
	See \cite{GEB2008Cella},
	\cite{TE2012Mylovanov_Troger},
	and \cite{TE2019Koessler_Skreta}
	for further analysis.
} 
Therefore, the benefit from privacy can 
reduce the cutting power of the intuitive criterion.

We overcome 
this difficulty by 
adapting 
the 
methods 
developed by \cite{E2013Gershkov_etal}
to our informed-principal problem with interdependent values.
We prove the result of {\it interim-payoff equivalence} 
assuming that the valuation functions are additively separable in 
type variables.
Again, let us assume that the RSW is dominated by  
another equilibrium. 
Then, the equivalence result guarantees that the dominating allocation
has an interim-payoff-equivalent feasible allocation that is EPIC for the agent. 
Hence, there is no exchange of slack variables
in the equivalent allocation.
Moreover, the equivalent allocation also dominates the RSW. 
It follows that 
the agent's EPIR constraints are slack for 
at least one type 
of principal.
We demonstrate that this type can convincingly 
deviate from the original equilibrium 
to another mechanism,
and thus, 
the equilibrium is unintuitive.

In summary, by focusing on the context of bilateral trading, 
we provide a simpler characterization of RSW, 
identify a necessary and sufficient condition 
under which the RSW allocation is undominated,
and show that the 
equilibrium allocations 
passing the intuitive criterion
are interim-payoff-equivalent to the RSW allocations.
We thus contribute to the important but small body of literature on informed-principal problems in
interdependent-values environments.

In contrast to the intuitive criterion, 
some important solution concepts select undominated equilibrium allocations.
Representative concepts are ``core mechanism'' and ``neutral optimum'' proposed by 
\cite{E1983Myerson}. 
The former is the allocation with no blocking coalition of types 
and the latter is axiomatically defined as the smallest possible set of unblocked mechanisms.
\cite{2015Balkenborg_Makrisz} introduced a 
concept called ``assured allocation'' 
that
provides an easily interpreted algorithm to find a neutral optimum.
\cite{2015Balkenborg_Makrisz} as well as \cite{E1983Myerson} argued that, 
if the principal can effectively communicate with the agents, then she should be expected to offer only undominated mechanisms.
In our study, we do not oppose this argument.
This is different from 
an argument in favor of the intuitive criterion, 
and each argument is based on an implicit ``speech'' made by the principal.
It is challenging
to examine which argument is more convincing, 
but such an analysis is beyond our objective.

The 
rest of the 
paper is organized as follows: 
Section \ref{Section2} describes the model. 
Section \ref{Section3} characterizes the RSW allocations.
Section \ref{Section4} illustrates how the RSW allocation is derived
and how non-RSW allocations are eliminated as unintuitive.
Section \ref{Section5} proves that 
the intuitive equilibrium allocations are interim-payoff-equivalent to the RSW allocations
for both parties.
Section \ref{Section6} concludes the paper. 
Several technical lemmas
and 
proofs
are presented either in Appendix \hyperref[Section7]{A} or in the Supplementary material.

\section{The model}\label{Section2}

{\bf Environment.}
Consider a seller ($i=1$) who 
owns a good 
that 
a buyer ($i=2$)
wants to buy. 
Each party $i$ has a type 
$x_i \in X_i \equiv \{1,2,...,\bar{x}_i \}$.
We assume that $x \equiv (x_1, x_2)$ are independently distributed 
according to interior (i.e., full-support) probability 
distributions 
$(p_1, p_2)$
on $X \equiv X_1 \times X_2$.
The cumulative distribution function (cdf) of $x_i$ is denoted by $P_i$.
Their types affect 
each party's valuation $v_i(x) \in \mathbb{R}$ of the good. 
Let $(q, t) \in A \equiv [0,1] \times \mathbb{R}$ denote an outcome.
Here, $q$ is the probability that the buyer obtains the good, 
and $t$ is the payment from the buyer to the seller. 
Both parties are risk-neutral.
The seller's ex-post payoff is
$u_1((q,t), x) \equiv t - v_1(x)q$, 
while the buyer's is $u_2((q,t), x) \equiv v_2(x) q - t$.\footnote{
	If $v_1 \leq 0$ and $v_2 \geq 0$, 
	we can regard parties 1 and 2 as a buyer and a seller, respectively.
}

To apply the technique of \cite{E2013Gershkov_etal},
we assume that each 
$v_i$ is additively separable in $x$
(i.e., $v_i(x) \equiv v_i^1(x_1) + v_i^2(x_2)$).
We then make the generic 
assumption that
each $v_i^i$ is strictly increasing in $x_i$ 
and $v_1(x) \not= v_2(x)$ for each $x$.\footnote{
We use the assumption of 
non-zero social surplus to prove Lemma \ref{Lem_AM}. 
See Section \ref{Section5.3} for the lemma.
}
We also assume that the seller's type positively 
affects the buyer's value (i.e., $v_2^1$ is increasing in $x_1$).
For future reference, 
let $dv_1(x_1) \equiv v_1^1(x_1) - v_1^1(x_1^-)$ and $dv_2(x_2) \equiv v_2^2(x_2^+) - v_2^2(x_2)$ 
be the differences between the valuations of two adjacent types, 
with the convention that $x_i^- \equiv x_i - 1$, $x_i^+ \equiv x_i + 1$, and $dv_1(1) = dv_2(\bar{x}_2) = 1$.

{\bf Allocation and mechanism.}
An {\it allocation} $f \in A^X$ 
describes type-dependent outcomes that result from the parties' interaction.
Further, the
allocation $f=(q,t)$ can be interpreted as a direct mechanism
with allocation rule $q \in [0,1]^X$ and payment rule $t \in \mathbb{R}^X$.
Each party's interim payoff from reporting $\hat{x}_i$ in $f$ 
(when
the other party tells the truth) 
is denoted by 
\begin{align*}
	U_1^{f}(\hat{x}_1 \mid x_1) &\equiv E_{x_2}\left[ u_1(f(\hat{x}_1, x_2), x) \right],
	\\
	U_2^{f,\pi}(\hat{x}_2 \mid x_2) &\equiv E_{x_1}^\pi \left[ u_2(f(x_1, \hat{x}_2), x) \right],
\end{align*}
where $\pi \in \Delta(X_1)$ is the buyer's {\it posterior belief} about $x_1$.
It is important to note that each expectation $E_{x_i}$ is based on the prior $p_i$,
while $E_{x_1}^{\pi}$ 
is formulated
according to the posterior $\pi$ because
the buyer may update his belief through the interaction.
Following \cite{RES2014Mylovanov_Troger}, 
we call an allocation $f$ {\it $\pi$-feasible} 
if 
it satisfies the IC and IR constraints 
for the seller and the buyer with the belief $\pi$, as follows:
For each $x_1, \hat{x}_1 \in X_1$ and $x_2, \hat{x}_2 \in X_2$,
\begin{align}
	U_1^f(x_1) &\geq U_1^{f}(\hat{x}_1 \mid x_1),
	\tag{S-IC}
	\label{S-IC}
	\\
	U_1^f(x_1) &\geq 0,
	\tag{S-IR}
	\label{S-IR}
	\\
	U_2^{f, \pi}(x_2) &\geq U_2^{f,\pi}(\hat{x}_2 \mid x_2),
	\tag{B-$\pi$-IC}
	\label{B-IC}
	\\
	U_2^{f,\pi}(x_2) &\geq 0,
	\tag{B-$\pi$-IR}
	\label{B-IR}
\end{align}
where $U_1^f(x_1) \equiv U_1^f(x_1 \mid x_1)$ and $U_2^{f, \pi}(x_2) \equiv U_2^{f, \pi}(x_2 \mid x_2)$.
Further, 
$f$ is called
{\it feasible} if it is $p_1$-feasible.
Let $U_2^{f}(\cdot) \equiv U_2^{f, p_1}(\cdot)$. 
With some abuse of notation, 
we denote by $U_i^{f} \equiv (U_i^f(1),...,U_i^f(\bar{x}_i))$ party $i$'s interim payoff vector in $f$.
We observe that the IC constraints require the allocation rule $q$ to be {\it interim monotone}. 
That is, \eqref{S-IC} requires 
$Q_1(\cdot) \equiv E_{x_2}[q(\cdot, x_2)]$ to be decreasing in $x_1$, 
and \eqref{B-IC} requires 
$Q_2^\pi(\cdot) \equiv E_{x_1}^\pi[q(x_1, \cdot)]$ to be increasing in $x_2$.

We define a {\it mechanism} $G = (M, g)$ as a finite strategic game form. 
Here, $M$
is the product of the parties' finite message spaces
$M_1$ and $M_2 \cup \{ 0 \}$, 
and $g \in A^M$ is an outcome function. 
A typical message profile is denoted by $m=(m_1,m_2) \in M$.
The buyer's message $m_2 = 0$ is his opt-out option such that $g(\cdot, 0)$ always specifies the no-trade outcome $(0, 0)$.
Let $\mathcal{G}$ be the set of all mechanisms. 
The set 
includes every direct mechanism $f \in A^{X_1 \times (X_2 \cup \{ 0 \})}$ with 
$f(\cdot, 0) = (0,0)$.\footnote{
	To simplify the notation, we identify a direct mechanism with its outcome function $f$.
}

A mechanism $G$ with a belief $\pi \in \Delta(X_1)$ 
induces the static Bayesian game 
$(G, (X_1, X_2), (u_1, u_2), (\pi, p_2))$. 
We denote it by $(G, \pi)$.
Let $BN(G, \pi) \not= \varnothing$ 
be the set of 
allocations 
generated by 
Bayesian Nash equilibria (BNE) 
in the game.

{\bf Mechanism-selection game and equilibrium.}
The {\it mechanism-selection game} 
proceeds as follows:
First, $(x_1, x_2)$ are realized according to $(p_1, p_2)$, and 
each party $i$ privately observes $x_i$.
Second, the seller offers the buyer a mechanism $(M, g) \in \mathcal{G}$. 
Third, the seller and buyer simultaneously choose $m_1 \in M_1$ and $m_2 \in M_2 \cup \{ 0 \}$, respectively.
Finally, each party $i$ obtains $u_i(g(m),x)$.

For the mechanism-selection game,
we consider 
strong perfect Bayesian equilibria 
\citep{E1992Maskin_Tirole},
which would be equivalent to 
sequential equilibria 
if the set $\mathcal{G}$ was finite. 
The inscrutability principle \citep{E1983Myerson} allows us to focus on a pooling equilibrium wherein
all seller types offer the same 
feasible 
allocation.
Formally, a feasible allocation $f$ is called an {\it equilibrium} (or {\it equilibrium allocation})
if for each mechanism $G \in \mathcal{G}$, there exists a posterior $\pi \in \Delta(X_1)$ with 
a BNE allocation $f' \in BN(G, \pi)$
such that 
$U_1^f(x_1) \geq U_1^{f'}(x_1)$ for each $x_1$.
This equilibrium concept is equivalent to expectational equilibrium of \cite{E1983Myerson}.

{\bf Intuitive equilibrium.}
To eliminate equilibria supported by ``unreasonable'' posterior beliefs, 
we focus on equilibria 
that pass the intuitive criterion. 
Following MT, 
we establish the criterion in our mechanism-selection game as follows:
For each equilibrium $f$ and mechanism $G$, 
let $X_1^f(G)$ be the set of 
types $x_1$ 
such that 
\begin{align}
	U_1^f(x_1) > U_1^{f'}(x_1) \ \textnormal{ for each } \pi \in \Delta(X_1) \textnormal{ and each } f' \in BN(G, \pi).
	\label{2_IntuitiveCriterion1}
\end{align}
An equilibrium $f$ is called 
{\it intuitive} if 
there exists no mechanism $G$ 
such that, for some $x_1 \in X_1 \setminus X_1^f(G)$, 
\begin{align}
	U_1^f(x_1) < U_1^{f'}(x_1) \ \textnormal{ for each } \pi \in \Delta(X_1\setminus X_1^f(G)) \textnormal{ and each } f' \in BN(G, \pi).
	\label{2_IntuitiveCriterion2}
\end{align}
Here, $\Delta(X_1\setminus X_1^f(G))$ is called the set of {\it reasonable} posteriors.
In other words, condition \eqref{2_IntuitiveCriterion1} indicates that 
the losing type $x_1 \in X_1^f(G)$
always loses by deviating 
from $f$ to $G$,
while condition \eqref{2_IntuitiveCriterion2} indicates that the non-losing type $x_1 \not\in X_1^f(G)$ gains from the same deviation,
provided it induces the buyer to have a reasonable posterior. 
In particular, condition \eqref{2_IntuitiveCriterion2} 
requires that
the deviator $x_1$ should benefit from the deviation 
if it convinces the buyer of her true type (i.e., $\pi(x_1) = 1$).
We say that type $x_1$ can {\it convincingly (and profitably) deviate} from $f$ to $G$ if \eqref{2_IntuitiveCriterion2} is satisfied.
We then discard the 
equilibrium $f$ as unintuitive.

Condition \eqref{2_IntuitiveCriterion1} is stringent because it requires that
the losing type 
should suffer from the deviation for {\it arbitrary} posterior beliefs $\pi$. 
If 
the buyer, observing the deviation, 
believes
that the seller's type is extremely high
(i.e., the good is much valuable for the buyer), 
then the seller of lower types, expecting high revenues, 
may have incentives to deviate.
Hence, 
the existence of higher types prevents {\it middle} types 
convincingly deviating
to some mechanisms. 
This 
kind of problem 
motivates \cite{QJE1987Cho_Kreps} to introduce stronger criteria 
including
D1. 
In sum, 
the stringency of condition \eqref{2_IntuitiveCriterion1}
potentially weakens the cutting power of the intuitive criterion, 
but makes a convincing deviation more ``credible.''

\section{RSW allocation}\label{Section3}

We introduce an important concept 
for informed-principal problems with interdependent values: 
the RSW allocation 
or, equivalently, the 
best safe mechanism.

\subsection{Characterization}\label{Section3.1}
In our 
single-agent 
model, 
a feasible allocation is called {\it safe} \citep{E1983Myerson} 
if it is EPIC and EPIR for the buyer:
For each $x \in X$ and $\hat{x}_2 \in X_2$,
\begin{align}
	u_2^f(x) &\geq u_2^f(\hat{x}_2 \mid x),
	\tag{B-EPIC}
	\label{B-EPIC}
	\\
	u_2^f(x) &\geq 0,
	\tag{B-EPIR}
	\label{B-EPIR}
\end{align}
where $u_2^f(\hat{x}_2 \mid x) \equiv u_2(f(x_1, \hat{x}_2), x)$ and $u_2^f(x) \equiv u_2(f(x_1, x_2), x)$.
\begin{defi}\label{Defi_RSW}
	An allocation $f^*$ is 
	{\it RSW}
	if it is a solution to the following problem for each $x_1$: 
	\begin{align}	
		\max_{f \in A^X} \ 
		U_1^f(x_1)
		\ \ \ 
		\textnormal{s.t. } 
		f \ \textnormal{satisfies \eqref{S-IC}, \eqref{B-EPIC}, and \eqref{B-EPIR}.}
		\label{3_RSW1}
	\end{align}
\end{defi}
The allocation possesses 
three 
basic 
properties.
First, at least one RSW allocation exists.
In particular, if 
$f^{x_1}$ denotes a solution to problem \eqref{3_RSW1}, 
then the combination $(f^{x_1}(x_1, \cdot))_{x_1 \in X_1}$ of their 
components
satisfies all constraints in \eqref{3_RSW1}, 
and hence, 
it is an RSW allocation.
Second, although multiple RSW allocations may exist, the 
seller's RSW payoff vector $U_1^{f^*}$ is unique among all RSW allocations.
Third, every RSW allocation satisfies \eqref{S-IR} because the no-trade allocation is safe.

We characterize the RSW allocations in three steps. 
First, we observe that
any RSW allocation is the best safe allocation for {\it every} type of seller. 
In other words, there is no conflict of interest over safe allocations 
among different seller types.
Hence, 
$f^*$ is an RSW allocation if and only if 
it solves the following problem:
\begin{align}	
	\max_{f \in A^X} \
	E_{x_1} [ U_1^f(x_1) ]
	\ \ \ 
	\textnormal{s.t. } 
	f \ \textnormal{satisfies \eqref{S-IC}, \eqref{B-EPIC}, and \eqref{B-EPIR}.}
	\label{3_RSW2}
\end{align}
Indeed, the same equivalence holds even if the objective function in \eqref{3_RSW2} is replaced by $\sum_{x_1} w(x_1) U_1^f(x_1)$, 
provided the weights are all positive.
We 
focus on the {\it single} problem \eqref{3_RSW2}
instead of the set of problems \eqref{3_RSW1}.

Second, we formulate a relaxed problem following the suggestion of MT (footnote 33). 
Note that \eqref{B-EPIC} requires the allocation rule 
$q$ to be {\it ex-post monotone} for the buyer, 
that is, 
$q \in \mathcal{Q} \equiv \{ q' \in [0,1]^{X} \mid q' \ \textnormal{is increasing in } x_2 \}$. 
We then define the relaxed problem as follows:
\begin{align}	
	\max_{f = (q,t) \in \mathcal{Q} \times \mathbb{R}^{X}} &\ 
	E_{x_1} [ U_1^f(x_1) ]
	\label{3_RSW3}
	\\
	\nonumber
	\textnormal{s.t. } 
	\nonumber
	&\ 
	U_1^{f}(x_1) \geq U_1^{f}(x_1^+ \mid x_1) \ \ \ \forall x_1 < \bar{x}_1.
	\tag{S-IC-U}
	\label{3_S-IC-U}
	\\
	&\ 
	u_2^f(x) \geq u_2^f(x_2^- \mid x) \ \ \ \forall x \textnormal{ with } x_2 > 1.
	\tag{B-EPIC-D}
	\label{3_B-EPIC-D}
	\\
	\nonumber
	&\ 
	u_2^f(x_1,1) \geq 0 \ \ \ \forall x_1.
	\tag{B-EPIR-B}
	\label{3_B-EPIR-B}
\end{align}
This relaxed-problem approach seems quite natural. 
The buyer's local downward EPIC constraints, 
together with his EPIR at the bottom, 
immediately imply \eqref{B-EPIR}.
Further, recall that the buyer's valuation $v_2$ is 
increasing in $x_1$ and strictly increasing in $x_2$.
The sorting assumptions imply that, as in standard signaling and screening models, 
the seller wishes to overstate her type to increase revenues and
the buyer wishes to understate his type to decrease payments.

Finally, 
we characterize the solutions to problem \eqref{3_RSW3} 
by taking a Lagrangian approach. 
We then check that they satisfy the omitted constraints 
and thus are equivalent to the RSW allocations.
A simple observation is that the constraints \eqref{3_B-EPIC-D} and \eqref{3_B-EPIR-B}
are binding. 
Intuitively, 
each type $x_1$ can lower the buyer's information rents as much as possible 
by reducing allocation
probabilities in such a way that 
the adjacent type $x_1^-$ no longer mimics $x_1$.
The ex-post payments are pinned down by the allocation rule $q$ as follows: 
For each $x$, 
\begin{align}
	t(x) = v_2(x)q(x) - \sum_{\hat{x}_2 < x_2} dv_2(\hat{x}_2) q(x_1,\hat{x}_2).
	\label{3_PE}
\end{align}
Then, a standard argument implies that the seller's interim revenue 
is given by 
\begin{align}
	E_{x_2}[t(x)] = E_{x_2} \left[ \left( v_2(x) - \frac{1-P_2(x_2)}{p_2(x_2)}dv_2(x_2) \right) q(x) \right]
	= E_{x_2}[\psi_2(x)q(x)]
	\label{3_RE}
\end{align}
for each $x_1$,
where $\psi_2 \equiv v_2 - \frac{1-P_2}{p_2}dv_2$ is the buyer's {\it virtual valuation}.
If the seller's type $x_1$ was common knowledge, 
it would be optimal for her 
to set a full-information monopoly price 
to maximize her profit $E_{x_2}[(\psi_2(x)-v_1(x))q(x)]$.\footnote{
	See, for example, \cite{2015Borgers} for this benchmark result. 
}
In the RSW, the lowest-type seller obtains her full-information profit, 
while a higher type $x_1^+$ 
maximizes her interim profit subject to the 
local upward IC constraint
\begin{align}
	E_{x_2} \left[ \left( \psi_2(x) - v_1(x) \right)
	q(x) \right]
	\geq 
	E_{x_2} \left[ \left( \psi_2(x_1^+,x_2) - v_1(x) \right)
	q(x_1^+,x_2) \right].
	\label{3_S-IC-U-2}
\end{align}

Through these steps, 
we characterize the RSW allocations as follows:
\begin{thm}\label{Thm_RSW}
	An allocation $f = (q,t)$ 
	is RSW
	if and only if 
	the payment rule $t$ is determined by the formula \eqref{3_PE} given $q$ 
	and there exists a belief $\pi \in \Delta(X_1)$ with the cdf $\Pi$
	such that 
	(i) 
	the prior $P_1$ weakly first-order stochastically dominates the posterior $\Pi$;
	(ii) the allocation rule $q$ maximizes
	\begin{align}
		E_{x}^\pi \left[ \left( 
		\psi_2(x)
		- 
		v_1(x) - \frac{\Pi(x_1^-) - P_1(x_1^-)}{\pi(x_1)}dv_1(x_1) 
		\right)
		q'(x) \right]
		\label{Thm_RSW_1}
	\end{align}
	among all $q' \in \mathcal{Q}$; 
	and 
	(iii) for each $x_1 < \bar{x}_1$, $q$ satisfies the seller's local upward IC constraint \eqref{3_S-IC-U-2},
	and either the equality of \eqref{3_S-IC-U-2} or $\Pi(x_1) = P_1(x_1)$ (or both) holds.
\end{thm}

Theorem \ref{Thm_RSW} extends the ``inductive'' 
characterization 
given by
MT (Proposition 2) 
to our trading model with bilateral asymmetric information.
The essential ingredient is the posterior belief $\pi$ 
that
characterizes the RSW allocation rule $q$. 
The prior weakly first-order stochastically dominates the posterior (i.e., $P_1(x_1) \leq \Pi(x_1)$ for each $x_1$).
In fact, the proof of Theorem \ref{Thm_RSW} shows that 
the cdf is defined by 
$\Pi \equiv P_1 + \kappa$ 
given
a vector of nonnegative Lagrange multipliers $\kappa$ 
for the seller's IC constraints \eqref{3_S-IC-U-2}.
Hence, the latter condition in (iii) is a complementary slackness condition.
We can interpret the terms in parentheses in \eqref{Thm_RSW_1}
as the {\it virtual surplus}, that is, the buyer's virtual valuation minus 
the seller's {\it virtual cost}.
The virtual cost
is greater than her opportunity cost $v_1(x)$
unless she has the lowest type or the adjacent type 
has no overstatement incentive
(i.e., the multiplier $\kappa(x_1^-) = \Pi(x_1^-) - P_1(x_1^-)$ is zero).

To glean some intuition behind Theorem \ref{Thm_RSW},
we further examine the 
implications
of the virtual surplus.
As a benchmark, let us consider 
the case of private values (i.e., $v_i(x) = x_i$ for each $i$).
Then, the buyer's virtual valuation $\psi_2$ no longer depends on $x_1$,
so that 
the seller's IC constraints \eqref{3_S-IC-U-2}
are nonbinding 
as in \cite{E1990Maskin_Tirole}.
The virtual surplus reduces to
\begin{align}
	\left( x_2- \frac{1-P_2(x_2)}{p_2(x_2)} \right) 
	- \left( x_1 + \frac{P_1(x_1^-)}{p_1(x_1)} \right) + \frac{P_1(x_1^-)}{p_1(x_1)}
	\label{3_VS}
\end{align}
because $\kappa = 0$, $\Pi = P_1$, and $dv_i = 1$. 
The two bracketed terms in \eqref{3_VS} appear in 
an outsider's problem 
of 
\cite{JET1983Myerson_Satterthwaite}. 
Here, $1-P_2(x_2)$ and $P_1(x_1^-)$ are the expected information rents paid by this outsider (e.g., social planner or broker) 
to the buyer and the seller, respectively.\footnote{
	In the case of continuum type spaces, $P_1(x_1^-)$ is replaced by $P_1(x_1)$.
}
However, 
unlike in their model,
the latter information rents 
are ``repaid'' to each type of principal 
in our model.
This 
repayment corresponds to the last term in \eqref{3_VS}.

Next, let us consider the case of interdependent values.
In general, 
the buyer's virtual valuation $\psi_2$ 
is strictly increasing in $x_1$, 
so that the seller's local 
upward IC constraints \eqref{3_S-IC-U-2} bind. 
If this is the case, 
the adjustment term
\begin{align}
	\frac{\Pi(x_1^-) - P_1(x_1^-)}{\pi(x_1)}
	dv_1(x_1)
	\label{3_adjustment}
\end{align}
in \eqref{Thm_RSW_1} 
is positive for some $x_1$ with $\kappa(x_1^-) > 0$.
This adjustment is ``due to possible {\it desirable} distortions arising from redistribution of income''
in the phrase of \cite{JET2007Ledyard_Palfrey}, 
who characterized interim (incentive) efficient allocations 
in linear-IPV environments. 
In particular, 
the prior $P_1$ puts weights on high types more heavily than 
the 
posterior $\Pi$ does
as if
it is desirable for the seller from the ex-ante point of view 
to ``redistribute'' 
rents from low to high types. 
However, 
this difference in weight 
endogenously 
arises from multipliers 
for the seller's 
IC constraints \eqref{3_S-IC-U-2}, 
unlike in \cite{JET2007Ledyard_Palfrey}.

An important implication of Theorem \ref{Thm_RSW} is that 
this redistribution is achieved through
undersupply of the good. 
Specifically,
each type $x_1$ chooses 
probabilities 
$q(x_1, \cdot)$ 
to maximize 
the expected virtual surplus \eqref{Thm_RSW_1} given $x_1$.
The seller distorts 
allocation probabilities away from 
efficient (i.e., social-surplus-maximizing) 
ones 
to reduce not only the buyer's rents, but also her own overstatement incentives.
This is an important consequence of 
interdependent values.

\begin{rem}\label{Rem_regular}
	Let us consider the {\it regular case} wherein the virtual surplus is strictly increasing in the buyer's type 
	(this is equivalent to the primitive condition that 
	$\psi_2 - v_1$ is strictly increasing in $x_2$).
	In this case, the RSW allocation $(q^*, t^*)$ takes a simple form because 
	the 
	ex-post monotonicity constraints 
	$q^* \in \mathcal{Q}$
	are nonbinding.
	Specifically, the allocation rule $q^*$ maximizes the virtual surplus pointwise.
	Hence, 
	each $x_1$ 
	has a threshold type $x_2^*(x_1)$ 
	such that $q^*(x) = 0$ if $x_2 < x_2^*(x_1)$
	and 
	$q^*(x) = 1$ if $x_2 > x_2^*(x_1)$. 
	If the buyer had a continuum type space with a continuous valuation function, 
	the payment rule would determine prices for the good
	as $t^*(x) = v_2(x_1, x_2^*(x_1))$ for each $x$ with $q^*(x)=1$. 
	For each $x_1$, this price is (weakly) higher than her full-information monopoly price, 
	worsening undersupply.
\end{rem}

\begin{rem}
	\cite{1985Myerson} 
	derived a neutral optimum 
	for an informed seller
	in a 
	bilateral-trade model
	with 
	an uninformed buyer. 
	This derivation is based on 
	the seller's ``virtual cost'' (or ``virtual valuation'')
	for some 
	welfare weights on seller types. 
	\cite{2015Balkenborg_Makrisz} 
	also 
	characterized the assured allocation 
	using the ``virtual surplus''
	in a nonlinear-common-values environment 
	with one-sided asymmetric information.
	The assured allocation is inductively defined via a sequence of optimization problems that include 
	the principal's ``assured claim'' constraints
	and the agent's non-ex-post participation constraints.
	Some essential differences between 
	these allocations and the RSW allocation 
	are
	discussed in Section \ref{Section5.4}.
\end{rem}

\subsection{Undominatedness}\label{Section3.2}
Now, 
following \cite{E1983Myerson} and MT, 
we introduce the Pareto dominance relation on allocations among all seller types:
\begin{defi}\label{Defi_DOM}
	Let $f$ and $f'$ be a pair of allocations. 
	We say that (i) $f'$ {\it weakly dominates} $f$ if 
	$U_1^{f'}(x_1) \geq U_1^{f}(x_1)$ for each $x_1$;
	(ii) $f'$ {\it dominates} $f$ 
	if 
	$U_1^{f'}(x_1) \geq U_1^{f}(x_1)$ for each $x_1$, with strict inequality for at least one $x_1$; 
	(iii) $f$ is {\it undominated for $\pi \in \Delta(X_1)$} 
	if $f$ is $\pi$-feasible and 
	$f$ is
	not dominated by any other $\pi$-feasible allocation;
	and (iv) $f$ is {\it undominated} if it is undominated for the prior $p_1$.
\end{defi}

Given any belief $\pi$, 
the following theorem provides a necessary and sufficient condition 
under which the RSW allocation is undominated for $\pi$.
\begin{thm}\label{Thm_RSW2}
	Let $f = (q,t)$ be an RSW allocation 
	and $\pi \in \Delta(X_1)$ be a belief 
	with the cdf $\Pi$.
	Then, 
	$f$ is undominated for $\pi$
	if and only if
	there exists an interior belief $w \in \Delta(X_1)$ with the cdf $W$
	such that 
	(i) 
	$W$ weakly first-order stochastically dominates 
	$\Pi$;
	(ii) the allocation rule $q$ maximizes
	\begin{align}
		E_{x}^\pi \left[ \left( 
		\psi_2(x)
		- 
		v_1(x) - \frac{\Pi(x_1^-) - W(x_1^-)}{\pi(x_1)}dv_1(x_1) 
		\right)
		q'(x) \right]
		\label{Thm_RSW2_1}
	\end{align}
	among all $q' \in \mathcal{Q}$; 
	and (iii) for each $x_1 < \bar{x}_1$, 
	either the equality of the seller's local upward IC constraint \eqref{3_S-IC-U-2} or $W(x_1) = \Pi(x_1)$ (or both) holds.
\end{thm}

The idea behind the proof is 
as follows:
A standard argument based on 
the supporting hyperplane theorem 
establishes that an allocation $f$ is undominated for the given belief $\pi$
if and only if 
there exists an interior belief $w \in \Delta(X_1)$
such that
$f$ is a solution to the following problem:
\begin{align}	
	\max_{f \in A^{X}} \
	E_{x_1}^w [ U_1^{f}(x_1) ]
	\ \ \
	\textnormal{s.t. } 
	f \ \textnormal{is $\pi$-feasible}.
	\label{3_pi-undominated}
\end{align}
Hence, it is sufficient to show that,
given any interior 
$w$,  
the RSW allocation 
$f = (q,t)$ is a solution to 
\eqref{3_pi-undominated}
if and only if the belief $w$ with the rule $q$ satisfies conditions (i)--(iii) in Theorem \ref{Thm_RSW2}. 
The intuition behind this characterization is similar to that for Theorem \ref{Thm_RSW}. 
Importantly, 
the cdf $W$ must put weights on high types more heavily than the given cdf $\Pi$ 
by 
Lagrange multipliers 
for the seller's local upward 
IC \eqref{3_S-IC-U-2}.
Otherwise, 
the redistribution of 
rents among seller types 
could increase her weighted average payoff in 
\eqref{3_pi-undominated}.

As a technical step, we must show that relaxing the buyer's ex-post monotonicity constraints 
to the interim ones
cannot increase the expected virtual surplus \eqref{Thm_RSW2_1}. 
Note that
the virtual surplus is additively separable in $x$. 
This leads us to apply a transformation method developed by \cite{E2013Gershkov_etal}. 
This method ensures that 
each interim-monotone allocation rule 
has an ex-post-monotone 
rule 
with the same interim probabilities as the original rule. 

We now apply Theorem \ref{Thm_RSW2} to obtain two important results. 
First, we denote the prior by $w = p_1$. 
Theorem \ref{Thm_RSW} then guarantees the existence of a belief $\pi$ 
such that the interior prior $w$ with the RSW allocation rule $q$ 
satisfies conditions (i)--(iii) in Theorem \ref{Thm_RSW2} given $\pi$. 
This belief $\pi$ plays a key role in proving 
that the RSW is 
an intuitive equilibrium. 
Additionally, using Theorem \ref{Thm_RSW2} twice, we can show that 
the set of beliefs $\pi$ for which the RSW is undominated is convex.
We thus obtain the following result,
which corresponds to the corollary in Section 3.B of MT.\footnote{
	Unlike 
	MT, 
	we consider the condition of undominatedness, 
	which is stronger than interim efficiency for RSW allocations.
	See 
	footnote \ref{foot_Dosis}
	for why we need the stronger condition. 
}
\begin{cor}\label{Cor_undominated}
	For any RSW allocation, 
	the set of beliefs $\pi$ for which 
	this allocation 
	is undominated is nonempty and convex.
\end{cor}

Second, we denote the prior by $\pi = p_1$. 
Theorem \ref{Thm_RSW2} then provides a necessary and sufficient condition 
for the RSW allocation to be undominated for the prior $\pi = p_1$. 
As noted by MT, 
the undominatedness of the RSW 
is equivalent to the existence of a {\it strong solution} 
(i.e., safe and undominated mechanism) introduced by \cite{E1983Myerson}. 
Moreover, we show in Section \ref{Section5.2} that 
the equilibrium characterization 
of MT 
holds in our model.
That is, 
a feasible allocation is an equilibrium if and only if it weakly dominates the RSW. 
Together with this result,  
Theorem \ref{Thm_RSW2} allows us 
to characterize the set of prior beliefs for which 
the seller's equilibrium payoff vector is uniquely determined by her RSW payoffs. 
Roughly,
if the interior prior $p_1$ 
assigns a higher probability to the lowest type, 
then it is easier to find an interior belief $w$ with which the RSW allocation rule $q$ satisfies conditions (i)--(iii)
in Theorem \ref{Thm_RSW2}.
The next section illustrates this characterization.

\section{Illustrative example}\label{Section4}

We now consider a regular-case example to illustrate how the RSW allocation is derived 
and how non-RSW equilibrium allocations 
are eliminated as unintuitive.

Assume that 
$X_i = \{1,2\}$ for each $i$ 
and $p_2$ is uniform, while $p_1$ is an arbitrary interior belief.
The valuations are 
$v_1(x) = 100(x_1-1)$ and $v_2(x) = 300 x_1 + 100 x_2$.
It is 
efficient to always allocate the good to the buyer.
For each type $x_1$, 
her full-information monopoly price is 
$300x_1+100$.
However, these prices 
are infeasible 
for the informed seller 
because the low-type seller mimics the high type.

From Theorem \ref{Thm_RSW}, 
the RSW $f^* = (q^*, t^*)$ is uniquely determined by Table \ref{Tbl1}, 
showing both 
$q^*(x)$ 
and 
$t^*(x)$ in each cell.
The low-type seller selects her full-information 
price, 
while the high-type seller 
sells the good less often by raising the price 
to make 
the low type indifferent 
between truth-telling and lying.

From Theorem \ref{Thm_RSW2}, 
$f^*$ is undominated (for the prior $p_1$) if and only if 
there exists an interior belief $w \in \Delta(X_1)$ 
such that $W \leq P_1$ and 
$q^*$ maximizes the virtual surplus in \eqref{Thm_RSW2_1} pointwise, given $\pi = p_1$.
Note that condition (iii) (i.e., complementary slackness condition) in Theorem \ref{Thm_RSW2} 
is trivially satisfied because the low-type seller's 
IC
binds in the RSW $f^*$.
The virtual-surplus 
maximization is achieved
if and only if the virtual surplus is nonpositive at $x=(2,1)$ and nonnegative at $x=(2,2)$. 
This implies that
the undominatedness of $f^*$ is equivalent to $p_1(1) \in (5/6, 1)$.\footnote{
If $p_1(1)=5/6$, 
there exists no {\it interior} $w$ such that the virtual surplus is nonpositive at $x=(2,1)$. 
Then, $f^*$ is dominated by the constant feasible allocation $(1,450)$.
Further, if $\pi(1) = 1$, 
the virtual surplus given any interior $w$ is negative infinity at $x=(2,2)$. 
Then, $f^*$ is dominated by
the $\pi$-feasible allocation $f$ such that 
$f(1, \cdot) = (1, 400)$ and $f(2, \cdot) = (0, 400)$.
}
\begin{table}[htb]
\begin{minipage}{0.5\hsize}
\begin{center}
	\begin{tikzpicture}[x=1,y=1]			
		\path[draw, black, thick] (0,36) -- (140,36);
		\path[draw, black, thick] (0,18) -- (140,18);
		\path[draw, black, thick] (0,0) -- (140,0);
		\path[draw, black, thick] (0,0) -- (0,36);
		\path[draw, black, thick] (70,0) -- (70,18);
		\path[draw, black, thick] (140,0) -- (140,36);
	
		\path (0,26) node[left] {$x_1 = 1$};
		\path (0,8) node[left] {$x_1 = 2$};		
		\path (35,36) node[above] {$x_2 = 1$};
		\path (105,36) node[above] {$x_2 = 2$};
	
		\path (70,26) node {$1, \ \ \ 400$};
		\path (35,8) node {$0, \ \ \ 0$};
		\path (105,8) node {$1, \ \ \ 800$};
	\end{tikzpicture} 
\end{center}
	\caption{RSW allocation $f^*$}
	\label{Tbl1}
\end{minipage}
\begin{minipage}{0.5\hsize}
\begin{center}
	\begin{tikzpicture}[x=1,y=1]			
		\path[draw, black, thick] (0,36) -- (140,36);
		\path[draw, black, thick] (0,18) -- (140,18);
		\path[draw, black, thick] (0,0) -- (140,0);
		\path[draw, black, thick] (0,0) -- (0,36);
		\path[draw, black, thick] (70,0) -- (70,36);
		\path[draw, black, thick] (140,0) -- (140,36);
	
		\path (0,26) node[left] {$x_1 = 1$};
		\path (0,8) node[left] {$x_1 = 2$};		
		\path (35,36) node[above] {$x_2 = 1$};
		\path (105,36) node[above] {$x_2 = 2$};
	
		\path (35,26) node {$4/5, \ \ \ 30$};
		\path (105,26) node {$1, \ \ \ 960$};
		\path (35,8) node {$1, \ \ \ 990$};
		\path (105,8) node {$4/5, \ \ \ 0$};
	\end{tikzpicture} 
\end{center}
	\caption{Equilibrium allocation $f'$}
	\label{Tbl2}
\end{minipage}
\end{table}
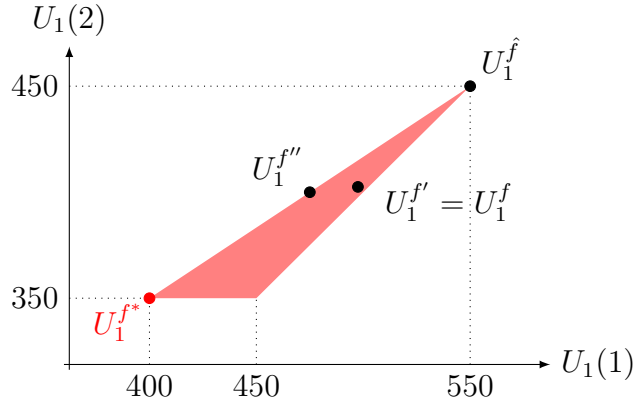
\begin{figure}[htb]
\begin{center}
\begin{tikzpicture}[x=1,y=1]
	\path[draw,->,>=latex] (0,-2) -- (0,120);
	\path[draw,->,>=latex] (-2,0) -- (180,0);
	\path (180,0) node[right] {$U_1(1)$};
	\path (0,120) node[above] {$U_1(2)$};

		\coordinate (A) at (30,25);
	\coordinate (B) at (70,25);
	\coordinate (C) at (150,105);
	\fill[red, opacity=0.5] (A)--(B)--(C)--cycle;

		\path[draw, dotted] (0,25) -- (30,25);
	\path (0,25) node[left] {$350$};
	\path[draw, dotted] (30,0) -- (30,25);
	\path (30,0) node[below] {$400$};
	\path[draw, dotted] (0,105) -- (150,105);
	\path (0,105) node[left] {$450$};
	\path[draw, dotted] (70,0) -- (70,25);
	\path (70,0) node[below] {$450$};
	\path[draw, dotted] (150,0) -- (150,105);
	\path (150,0) node[below] {$550$};

		\fill[draw, red](30,25) circle[radius=2];
	\path (32,27) node[below left] {\color{red}$U_1^{f^*}$};

		\fill[draw](108,67) circle[radius=2];
	\path (114,73) node[below right] {$U_1^{f'} = U_1^f$};

		\fill[draw](150,105) circle[radius=2];
	\path (150,103) node[above right] {$U_1^{f^{**}}$};
\end{tikzpicture}
\caption{Seller's equilibrium payoff vectors}\label{Fig2}
\end{center}
\end{figure}

The equilibrium characterization 
in Section \ref{Section5.2} 
implies that
the seller's equilibrium payoff vector 
is unique if and only if the RSW is undominated (i.e., $p_1(1) \in (5/6, 1)$). 
Let us assume that $p_1$ is uniform.
Then, the set of equilibrium payoff vectors for the seller is 
characterized as 
the red triangle in Figure \ref{Fig2}.\footnote{
	See Appendix B in the Supplementary material for this characterization. 
}
For example, consider the allocation $f' = (q',t')$ in Table \ref{Tbl2}.
Simple computations show that $f'$ is feasible and $U_1^{f'} = (495, 405) > (400, 350) = U_1^{f^*}$. 
Hence, 
$f'$ is an equilibrium.
We show that the high-type seller can convincingly deviate from the equilibrium $f'$ to another mechanism,
so that
$f'$ is unintuitive.

We prove this in two steps. 
First, we transform $f'$ into a well-behaved allocation.
Note that $f'$ is {\it neither EPIR nor EPIC for the buyer}.
In particular, the decreasing function $q'(2,\cdot)$ 
violates the ex-post 
monotonicity. 
In the equilibrium $f'$,
if the high-type seller gains from exchanging slack variables on 
the 
EPIC and EPIR constraints
with the low-type seller, 
the high type is reluctant to make a revealing deviation.
However, she has no such benefit.
To see why, consider the constant feasible allocation $f$ in Table \ref{Tbl3}.
It still violates \eqref{B-EPIR}, but does satisfy \eqref{B-EPIC}.
Moreover, $f$ is {\it interim-payoff-equivalent} to the original allocation $f'$ 
(i.e., $U_i^f = U_i^{f'}$ for each $i$).

\begin{table}[htb]
\begin{minipage}{0.5\hsize}
\begin{center}
	\begin{tikzpicture}[x=1,y=1]			
		\path[draw, black, thick] (0,36) -- (140,36);
		\path[draw, black, thick] (0,0) -- (140,0);
		\path[draw, black, thick] (0,0) -- (0,36);
		\path[draw, black, thick] (140,0) -- (140,36);
	
		\path (0,26) node[left] {$x_1 = 1$};
		\path (0,8) node[left] {$x_1 = 2$};		
		\path (35,36) node[above] {$x_2 = 1$};
		\path (105,36) node[above] {$x_2 = 2$};
	
		\path (70,18) node {$9/10, \ \ \ 495$};
	\end{tikzpicture} 
\end{center}
	\caption{Payoff-equivalent $f$}
	\label{Tbl3}
\end{minipage}
\begin{minipage}{0.5\hsize}
\begin{center}
	\begin{tikzpicture}[x=1,y=1]			
		\path[draw, black, thick] (0,36) -- (140,36);
		\path[draw, black, thick] (0,0) -- (140,0);
		\path[draw, black, thick] (0,0) -- (0,36);
		\path[draw, black, thick] (140,0) -- (140,36);
	
		\path (0,26) node[left] {$x_1 = 1$};
		\path (0,8) node[left] {$x_1 = 2$};		
		\path (35,36) node[above] {$x_2 = 1$};
		\path (105,36) node[above] {$x_2 = 2$};
	
		\path (70,18) node {$1, \ \ \ 550$};
	\end{tikzpicture} 
\end{center}
	\caption{Undominated equilibrium $f^{**}$}
	\label{Tbl4}
\end{minipage}
\end{table}

Next, 
we construct a ``less-trading'' mechanism from 
$f$. 
Because the buyer's EPIR constraints are slack for the high-type seller in $f$, 
she 
can deliver the good less often.\footnote{
	Indeed, the high-type seller must lower interim allocation probabilities to make a convincing and profitable deviation. 
	This is because 
	the low-type seller's IC binds in the allocation $f$.
}
However, this requires 
her to 
lower the buyer's payments 
to prevent the low-type seller's deviation.
Specifically, we define a convex combination 
\begin{align}
	\tilde{f} \equiv (1-\delta)f + \delta(0, \tau)
	\label{4_convex}
\end{align}
of the two (constant) allocations, 
where $\tau \in (405, 495) = (U_1^f(2),U_1^f(1))$ is a fixed fee 
and $\delta \in (0,1)$ is a small weight with $u_2^{\tilde{f}}(2,1) > 0$.
The interim-payoff equivalence implies that 
the low-type seller prefers the equilibrium $f'$ 
to the new allocation $\tilde{f}$,
while the high-type seller has the opposite preference:
\begin{align}
	U_1^{f'}(1) = U_1^{f}(1) &> (1-\delta) U_1^{f}(1) + \delta \tau = U_1^{\tilde{f}}(1), 
	\label{4_opposite_preference1}
	\\
	U_1^{f'}(2) = U_1^{f}(2) &< (1-\delta) U_1^{f}(2) + \delta \tau = U_1^{\tilde{f}}(2).
	\label{4_opposite_preference2}
\end{align}
In fact, the low type 
loses by 
deviating to the less-trading mechanism $\tilde{f}$ regardless of the buyer's posterior and response.
The buyer, believing that $\tilde{f}$ is offered by 
the non-losing high-type seller, 
is willing to accept it.
Thus, the high-type seller 
can convincingly 
deviate from the equilibrium $f'$ to 	
the direct mechanism $\tilde{f}$.

After all, the equilibrium 
$f'$ is supported by unreasonable posteriors.
To support it, 
the buyer must believe that 
$\tilde{f}$ is sometimes offered by the low-type seller, who always suffers from the deviation.
Similarly, the undominated 
equilibrium 
$f^{**}$ in Table \ref{Tbl4}
is also eliminated as unintuitive.
It should be noted that the unintuitive equilibrium $f^{**}$ is efficient,
while the RSW allocation $f^*$ is not.

\section{Intuitive equilibrium allocation}\label{Section5}

In this section, we 
provide a justification for the RSW allocation 
using the intuitive criterion. 
We show that 
every RSW allocation is an intuitive equilibrium and
every intuitive equilibrium is interim-payoff-equivalent to an RSW allocation. 

\subsection{Existence}\label{Section5.1}

The following theorem 
ensures that 
every RSW allocation is an intuitive equilibrium.
Therefore, 
our mechanism-selection game has at least one intuitive 
equilibrium.
\begin{thm}\label{Thm_EofIA}
	Every RSW allocation
	is an intuitive equilibrium.
\end{thm}
To obtain the intuition behind the proof, let us assume that the RSW allocation is an equilibrium 
supported by {\it reasonable} posteriors. 
We then observe that
no type of seller can convincingly deviate from the RSW to any other mechanism, 
and hence, the RSW is intuitive. 
The reason is simple: 
If some type 
had a convincing deviation, 
the reasonable posterior could no longer support the RSW 
as an equilibrium.\footnote{
	This observation immediately follows from the definitions of reasonable posterior and convincing deviation
	at the end of Section \ref{Section2}.
	See \cite{E1987Cho} for a more detailed 
	analysis.
} 
Thus, to prove Theorem \ref{Thm_EofIA}, 
it is sufficient to find reasonable posteriors
that support the RSW 
as an equilibrium. 
Here, 
a 
belief $\pi$ for which the RSW is undominated plays a crucial role. 
Note that the existence of such a belief is guaranteed by Corollary \ref{Cor_undominated}.
The idea is to derive, for each off-path mechanism $G$, 
a reasonable 
posterior $\pi^G$
from 
this 
belief $\pi$.

We outline 
the arguments
as follows: 
In the first step, given any off-path mechanism $G$, 
we formulate an auxiliary mechanism-selection game wherein 
$(x_1, x_2)$ are realized according to {\it not $(p_1, p_2)$}, {\it but $(\pi, p_2)$},
and subsequently the seller selects either offering $G$ or ending up with the status quo 
$f^*$.
We find a trembling-hand perfect equilibrium 
of the auxiliary game.
The trick is to choose a sequence of perturbed games 
wherein the losing types $x_1 \in X_1^{f^*}(G)$ 
``mistakenly'' select $G$ less often than 
the non-losing types $x'_1 \not\in X_1^{f^*}(G)$ do.
Naturally, this difference in the mistake likelihood 
ensures 
that 
any limit 
$\pi^G$ of equilibrium posteriors 
is reasonable, 
that is, the buyer believes that 
$G$ is never offered by losing types.

In the second step, we show that this reasonable posterior $\pi^G$ 
supports the RSW allocation as an equilibrium of the original mechanism-selection game. 
Here, we follow the argument of \citet[Theorem 2]{E1983Myerson}.
To clarify this argument, we denote the perfect equilibrium allocation 
by
\begin{align}
	f'(x)
	&\equiv \gamma^G(x_1) f(x)
	+ (1-\gamma^G(x_1)) f^*(x)
	\label{5_PEA}
\end{align}
for each $x$, 
where $\gamma^G(x_1)$ is the probability with which type $x_1$ selects $G$
and $f \in BN(G, \pi^G)$ is the continuation BNE in the perfect equilibrium. 
Because each type of seller selects her favorable allocation in equilibrium, 
the allocation $f'$ weakly dominates $f^*$. 
Moreover, $f'$ is $\pi$-feasible because 
the parties use equilibrium strategies in the auxiliary game 
given the belief $\pi$
and the RSW is safe. 
At this point, we use the 
undominatedness property:
The RSW is not dominated by any $\pi$-feasible allocation. 
Hence, $U_1^{f^*} = U_1^{f'}$. 
This implies that 
$f^*$ weakly dominates $f$. 
Thus, the reasonable posterior $\pi^G$ with the continuation BNE $f$ prevents the seller's deviation from the RSW
to the off-path mechanism.

\begin{rem}
We cannot apply these arguments to non-RSW allocations.  
To clarify this point, let us 
consider the equilibrium $f$ in Table \ref{Tbl3}.
We have already shown that $f$ is unintuitive. 
We take 
$\tilde{f}$ defined by \eqref{4_convex} as an off-path mechanism, 
and consider an auxiliary game wherein the seller selects either $\tilde{f}$ or $f$.
If a perfect equilibrium in this auxiliary game 
generates the reasonable posterior (i.e., $\pi^{\tilde{f}}(2)=1$) as in the first step, 
it dominates $f$ because 
the high-type seller selects 
the new 
$\tilde{f}$ rather than the status quo $f$.
Hence, 
the argument in the second step fails for this {\it particular} perfect equilibrium.
However, another perfect equilibrium generates an unreasonable posterior 
that supports $f$ as an equilibrium.\footnote{
	By contrast, in the case of the RSW, the argument in the second step is valid for {\it every} perfect equilibrium.
	This property is closely 
	related to strategic stability
	\citep{E1986Kohlberg_Mertens}.
} 
\end{rem}

\subsection{Equilibrium characterization}\label{Section5.2}

To highlight the effectiveness of the intuitive 
criterion as a refinement concept,
we characterize the set of equilibrium allocations including unintuitive ones.
Due to Corollary \ref{Cor_undominated},
the following proposition 
yields the same 
conclusion 
as that of 
Theorem $1^*$ in MT.
\begin{prop}
\label{Prop_EqmAllocations}
	Suppose that an RSW allocation is undominated for some $\pi \in \Delta(X_1)$. 
	Then, the set of equilibrium allocations of the mechanism-selection game 
	is the set of feasible 
	allocations
	that weakly 
	dominate 
	the RSW allocations. 
\end{prop}
As a corollary of Theorem \ref{Thm_EofIA}, 
every feasible allocation that weakly dominates the RSW allocations
is an equilibrium. 
Thus, to prove Proposition \ref{Prop_EqmAllocations}, 
we need only show that each type of seller obtains at least her RSW payoff in equilibrium.
Although every RSW allocation is by definition safe,
simply selecting it may not guarantee the RSW payoff. 
This is due to the multiplicity of continuation equilibria.
For example, the direct mechanism $f^*$ in 
Table \ref{Tbl1}
has an untruthful BNE 
wherein each party always reports the low type.
This is an ex-post equilibrium, 
and thus remains a BNE regardless of the buyer's posterior. 
In the untruthful BNE, 
the high-type seller obtains $400 - 100 = 300$. 
This is less than her RSW payoff
$U_1^{f^*}(2) = 350$.
We address the problem of multiple continuation equilibria 
by constructing an Abreu--Matsushima (AM) 
mechanism
\citep{E1992Abreu_Matsushima,1992Abreu_Matsushima}.
This methodology is described in Section \ref{Section5.3}.

\begin{rem}\label{Rem_Myerson}
	The hypothesis of Theorem $1^*$ of 
	MT 
	is that the RSW allocation is 
	{\it interim efficient}
	for
	some {\it interior} 
	beliefs.
	The hypothesis corresponds to that of Proposition \ref{Prop_EqmAllocations}, 
	but there are two differences.
	First, 
	we strengthen the condition of interim efficiency to 
	undominatedness (for some beliefs).
	As argued by 
	Dosis (2022), 
	the 
	interim efficiency 
	is insufficient 
	for the 
	characterization result of MT
	in some cases.	
	Second, we prove Proposition \ref{Prop_EqmAllocations} without the interior-belief requirement. 
	In effect, 
	we prove in the second step of Theorem \ref{Thm_EofIA} that, 
	if a safe mechanism is undominated for some 
	posteriors $\pi$, 
	the mechanism (i.e., RSW) is an equilibrium.
	This proof is based on that of \citet[Theorem 2]{E1983Myerson}. 
	He shows that, 
	if a safe mechanism is undominated for the prior, 
	the mechanism (i.e., strong solution) is an equilibrium. 
	The essential difference 
	from Myerson's original proof
	is that we formulate an auxiliary mechanism-selection game 
	by using the posterior $\pi$, not the prior $p_1$.
\end{rem}

As emphasized by MT, 
the equilibrium characterization 
implies that 
the principal's equilibrium payoff vector 
is unique if and only if the RSW allocation is undominated.
This condition is characterized by Theorem \ref{Thm_RSW2}.
In the example of Section \ref{Section4}, 
the RSW is undominated if and only if the prior assigns a sufficiently high probability to the low type (i.e., $p_1(1) \in (5/6, 1)$). 
If the RSW is dominated, the mechanism-selection game has infinitely many equilibrium allocations.
Therefore, we need to refine equilibria to obtain a stronger prediction.

\subsection{Uniqueness}\label{Section5.3}

The intuitive criterion is 
weaker than well-known concepts including
(universal) divinity \citep{E1987Banks_Sobel}, 
perfect sequential equilibrium \citep{JET1986Grossman_Perry}, 
neologism-proofness \citep{GEB1993Farrell}, 
and strong neologism-proofness \citep{TE2012Mylovanov_Troger, RES2014Mylovanov_Troger}.
The following theorem, together with Theorem \ref{Thm_EofIA},
shows that 
this reasonable criterion selects only
a feasible allocation 
that yields the seller's RSW payoffs.

\begin{thm}\label{Thm_IC1}
	For every intuitive equilibrium $f$, 
	$U_1^{f}$ is equal to the RSW payoff vector $U_1^{f^*}$. 
\end{thm}
The idea of the proof 
is illustrated in Section \ref{Section4}
and outlined as follows:
From
the equilibrium characterization (Proposition \ref{Prop_EqmAllocations}), 
we need only show that an equilibrium $f'$ dominating the RSW is unintuitive.
First, we prove the interim-payoff equivalence. 
This result provides 
a new feasible allocation $f$ that 
satisfies \eqref{B-EPIC}
and is interim-payoff-equivalent to $f'$. 
As $f'$ dominates the RSW, so does the equivalent $f$.
Hence, $f$ must violate \eqref{B-EPIR}. 
This implies that 
the lowest-type buyer loses in $f$
for some seller types.
However, $f$ satisfies 
the constraints (\hyperref[B-IR]{B-$p_1$-IR}).
This in turn implies that 
the lowest-type buyer 
benefits 
from $f$
for another type $\tilde{x}_1$. 
We call this type $\tilde{x}_1$ 
a {\it candidate deviator}.
Second, 
we construct a less-trading allocation $\tilde{f}$ nearby $f$,
as in \eqref{4_convex}. 
Further, we design an indirect mechanism $G$ 
that 
``virtually''
implements 
$\tilde{f}$.
This implementation is necessary only 
if
the candidate should 
offer a mechanism 
wherein both parties make reports.
Finally, we show that the candidate $\tilde{x}_1$ can convincingly deviate from 
the original equilibrium 
$f'$ 
to the mechanism $G$.

\begin{figure}[htb]
\begin{center}
\begin{tikzpicture}
    \node (sequence) at (0,0) {
    $
	f' \xmapsto[\textnormal{\small \parbox[c][3em]{8em}{
		\centering
		payoff-equivalent
		allocation
	}} ]{}
	f \xmapsto[\textnormal{\small \parbox[c][3em]{8em}{
		\centering 
		less-trading allocation
		}} ]{}
	\tilde{f} \xmapsto[\textnormal{\small \parbox[c][3em]{8em}{\centering 
		virtual
		implementation
		}} ]{}
	G$
    };
\end{tikzpicture}
\caption{Construction of the deviation mechanism $G$}
\label{Fig3}
\end{center}
\end{figure}
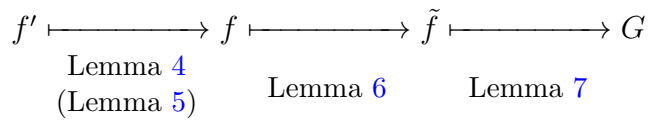
These construction procedures are summarized in Figure \ref{Fig3}.
Each step is described in detail below,
with reasons as to why these 
approaches are needed.

{\bf Payoff-equivalent allocation.}
First, we prove the following result:
\begin{prop}
\label{Prop_PayoffEq}
	Suppose that an allocation $f'$ satisfies \eqref{S-IC} and 
	\textnormal{(\hyperref[B-IC]{B-$p_1$-IC})}.
	Then, there exists an allocation $f$ that satisfies \eqref{S-IC}, \eqref{B-EPIC},
	and $U_i^{f} = U_i^{f'}$ for each $i = 1,2$.
\end{prop}
Such results are essential to
eliminate 
all equilibria dominating the RSW 
as unintuitive. 
As discussed in Sections \ref{Section1} and \ref{Section4}, 
if some seller types 
mutually
benefit
from exchanging slack variables on the buyer's EPIC and EPIR constraints, 
these types 
cannot 
make a convincing deviation.
This is because 
the separation of types
hurts them. 
Proposition \ref{Prop_PayoffEq} ensures that 
seller types have no benefit from slack exchange in any equilibrium.

The intuition behind Proposition \ref{Prop_PayoffEq} is as follows:
As in Theorem \ref{Thm_RSW2}, 
the transformation method of \cite{E2013Gershkov_etal} plays a key role.
In the original allocation $f' = (q',t')$, 
the rule $q'$ may violate the ex-post monotonicity for the buyer.
Then, their lemma 
ensures that there exists an ex-post-monotone allocation rule $q$ 
that has the same interim probabilities as the original rule $q'$. 
The equivalence of the interim 
probabilities immediately implies the equivalence of ex-ante social surpluses
(i.e., $E_{x}[(v_2(x)-v_1(x))q(x)] = E_{x}[(v_2(x)-v_1(x))q'(x)]$),
under the assumption that each valuation 
$v_i$ is additively separable in $x$.
This allows us to construct a payment rule $t$ 
such that 
the new allocation $f\equiv(q, t)$ satisfies both \eqref{S-IC} and \eqref{B-EPIC} 
and 
is 
interim-payoff-equivalent to the original $f'$. 

\begin{rem}
\cite{E2013Gershkov_etal} considered a linear-IPV environment 
wherein an uninformed principal designs a mechanism.
They showed that 
each 
IC mechanism has a dominant strategy IC mechanism 
that yields the same interim payoffs for each agent 
and the same ex-ante social surplus.
They also provided a counterexample to the equivalence result
in an interdependent-values auction environment 
by showing that
there exists no EPIC mechanism that yields the same ex-ante social surplus
as an 
IC mechanism.
Nevertheless, \citet[p. 213]{E2013Gershkov_etal} stated that
each IC mechanism 
has an EPIC mechanism that yields 
the same interim payoffs for each agent.
In our model, 
the informed principal herself participates in a mechanism 
and no third party balances the budget.
Because of this feature, 
we cannot
extend Proposition \ref{Prop_PayoffEq} 
in such a way that the interim-payoff-equivalent allocation $f$ is EPIC for {\it both parties}.
\end{rem}

{\bf Less-trading allocation.}
Next, we construct a less-trading allocation $\tilde{f}$ from 
the equivalent $f$.
This step is simple if the seller's type is binary.
In this case, 
the candidate deviator $\tilde{x}_1$ 
must be the high type.\footnote{
If the buyer's 
EPIR at the bottom
was slack for the low-type seller, 
she would obtain less than her full-information payoff
because the allocation $f$ is EPIC for the buyer.}
As in Section \ref{Section4}, 
the high-type seller 
can offer 
the buyer a 
``menu'' 
of less-trading 
outcomes, 
reporting nothing after the offer. 
If this menu is appropriately designed, 
the low-type seller 
prefers $f$
to the menu 
{\it regardless of the buyer's posterior}. 
The high-type seller can convincingly deviate from 
the equilibrium $f'$ (and the equivalent $f$) to this kind of menu.

Unfortunately, this argument fails if the seller has more than two types. 
In some cases, 
the highest-type seller is not a candidate deviator. 
Then, recall the stringency of condition \eqref{2_IntuitiveCriterion1} for the losing types. 
A deviation to a simple menu 
can be profitable for lower types $x_1 < \tilde{x}_1$ 
if the buyer, believing that the menu is offered by higher types $x_1 > \tilde{x}_1$, 
purchases the good more often.
Hence, 
to prevent the deviation of the lower types, 
the candidate $\tilde{x}_1$ should offer a {\it mechanism} 
in which 
the higher types 
choose their own outcomes
by making reports.\footnote{
	Appendix C in the Supplementary material
	provides an example 
	of an equilibrium
	wherein 
	the seller 
	cannot make a convincing deviation to any menu.
	See 
	\cite{QJE1987Cho_Kreps} 
	for a related example in the Spence signaling game
	with three worker types.
}

Then, we construct a less-trading 
mechanism $\tilde{f}$ 
in the same manner as in Section \ref{Section4}.
As in \eqref{4_convex}, 
this allocation $\tilde{f}$ is defined as a convex combination of the original $f$ and a fixed fee $\tau$.
This fee does not depend on reports. 
Hence, the allocation $\tilde{f}$ satisfies the desired property
\eqref{S-IC}. 
Further, as $f$ 
satisfies \eqref{B-EPIC}, so does $\tilde{f}$. 
If this fixed fee is appropriate (i.e., $U_1^{f}(\tilde{x}_1) < \tau < U_1^{f}(\tilde{x}_1^-)$), 
then 
all lower types $x_1 < \tilde{x}_1$ prefer the original allocation $f$ to the less-trading $\tilde{f}$, 
while all higher types $x_1 \geq \tilde{x}_1$ have the opposite preference, 
as shown by \eqref{4_opposite_preference1} and \eqref{4_opposite_preference2}.
This difference in the preference follows from the general result that higher seller types obtain lower interim payoffs 
in the IC allocation $f$. 
The candidate 
$\tilde{x}_1$ 
is better off deviating from the equilibrium $f'$ to the nearby mechanism $\tilde{f}$, 
provided that {\it the buyer participates in $\tilde{f}$}
and {\it both parties tell the truth in $\tilde{f}$}.

{\bf Remaining steps.}
We have two technical problems. 
First, in the construction of $\tilde{f}$, 
we implicitly assumed 
that 
the constraint \eqref{3_B-EPIR-B} is slack in the original $f$
for each type $x_1 > \tilde{x}_1$ 
as well as for the candidate $\tilde{x}_1$. 
Without this assumption, 
the buyer may opt out of the mechanism $\tilde{f}$ 
in the case that 
he believes that the seller has a non-losing type $x_1 > \tilde{x}_1$.
We address this problem 
by reconstructing both allocations $f$ and $\tilde{f}$. 
Roughly, we construct an allocation $f$ wherein
the candidate forms a ``coalition'' $\{ \tilde{x}_1,...,\bar{x}_1 \}$ of the types 
for which the buyer's EPIR constraints are slack. 
This 
construction 
is possible because the buyer's valuation is increasing in $x_1$.\footnote{
	\label{foot_coalition}However, 
	some high types $x_1 > \tilde{x}_1$ 
	whose valuations are so high that the no-trade outcome is favorable for them
	should be removed from this coalition.
} 
We also reconstruct the nearby $\tilde{f}$ as a convex combination of this modified $f$ and a fixed fee.
As before, $\tilde{f}$ satisfies both \eqref{S-IC} and \eqref{B-EPIC}. 
The buyer has incentives to participate in this mechanism $\tilde{f}$
as long as he believes that 
the seller's type is a coalition member. 
Although some members $x_1 > \tilde{x}_1$ 
may be worse off in this 
allocation $\tilde{f}$ than in the equilibrium $f'$, 
the candidate's interim payoff is higher in $\tilde{f}$. 
See Lemmas \ref{Lem_higher} and \ref{Lem_perturb} in Appendix \hyperref[Section7]{A}
for details.

Second, the seller's reporting opportunities 
cause the problem of multiple continuation equilibria in the direct mechanism $\tilde{f}$.\footnote{
	Appendix D in the Supplementary material 
	illustrates this problem. 
} 
To address this problem, 
we apply the methodology of \citet{E1992Abreu_Matsushima,1992Abreu_Matsushima}.
They showed that, 
in a complete (incomplete) information environment,
every allocation 
(every IC allocation satisfying a weak condition)
is virtually implementable 
in iteratively undominated strategies. 
Note that their mechanisms, which are finite strategic game forms, are admissible in our model.
Our idea is simple: Using the equilibrium hypothesis---that the buyer's off-path beliefs are common knowledge---we design an AM mechanism $G \in \mathcal{G}$ that elicits his posterior $\pi^G \in \Delta(X_1)$ after it is offered.
Then, the elicited belief is used to decide which allocation is 
virtually implemented.\footnote{
	This idea of using AM mechanisms is inspired by a discussion with Takuro Yamashita.
	This posterior-elicitation mechanism is unrealistic, 
	but it is used only to upset an unintuitive equilibrium. 
	The idea that a well-designed mechanism can elicit the agent's posterior 
	is suggested by \citet[Proposition 7]{E1990Maskin_Tirole}. 
	Using the idea, 
	they showed that 
	any equilibrium of the mechanism-selection game is ``strongly 
	unconstrained 
	Pareto optimal.''
} 
Roughly, if 
the posterior $\pi^G$ 
assigns a high probability to the coalition $\{ \tilde{x}_1,...,\bar{x}_1 \}$, then 
the allocation $\tilde{f}$ is implemented, 
and 
otherwise the no-trade outcome is implemented. 
The former implementation is possible because $\tilde{f}$ is both 
IC and IR 
for the buyer 
given coalition members. 
Thus, the non-coalition members 
are losing types,
and the candidate $\tilde{x}_1$ can convincingly deviate from 
the original equilibrium $f'$ to the AM mechanism $G$. 
Although these ideas are simple, a formal analysis is complicated.
See Lemma \ref{Lem_AM} in Appendix \hyperref[Section7]{A} for details. 

\begin{rem}
	MT (Proposition 7) 
	showed the same refinement result as Theorems \ref{Thm_EofIA} and \ref{Thm_IC1} 
	in the case of one-sided asymmetric information.
	They also proved that, under sorting assumptions, 
	(a) the RSW allocation passes the Farrell--Grossman--Perry (FGP) criterion 
	if and only if it is 
	interim efficient 
	for the prior; 
	and (b) no other allocation passes the FGP.\footnote{
		Hence, the FGP implies the intuitive criterion. 
		The relation holds in standard signaling games, 
		as shown by \cite{JET1986Grossman_Perry}.
		See \cite{1991vanDamme} for other criteria.
	}	
	Moreover, MT (Proposition 11) 
	regarded their model 
	as a renegotiation game 
	and used the 
	intuitive criterion with a renegotiation proofness to refine the equilibria of the game.
	Interestingly, they showed that the cutting power of the intuitive criterion is weak in the renegotiation game.
	An analysis of the renegotiation proofness with the intuitive criterion is beyond the scope of this study.
\end{rem}

Our last theorem shows that
every intuitive equilibrium is also equivalent to an RSW allocation
with respect to the buyer's interim payoffs. 
\begin{thm}\label{Thm_IC2}
	For every intuitive equilibrium $f$, 
	there exists an RSW allocation $f^*$ with $U_2^{f} = U_2^{f^*}$.
\end{thm}
The intuition behind the result is simple:
From the interim-payoff equivalence (Proposition \ref{Prop_PayoffEq}), 
every intuitive equilibrium has an interim-payoff-equivalent feasible allocation $f$ that satisfies \eqref{B-EPIC}.
From the uniqueness result for the seller (Theorem \ref{Thm_IC1}), the equivalent $f$ yields the RSW payoffs to her. 
If the constraint \eqref{3_B-EPIR-B} is slack for a seller type $x_1$ in the equivalent $f$, 
she must deliver the good more often in $f$ than in the RSW 
to earn her RSW payoff $U_1^{f^*}(x_1)$. 
However, the adjacent $x_1^-$ then mimics $x_1$ to obtain higher than her RSW payoff $U_1^{f^*}(x_1^-)$. 
Hence, 
the constraints \eqref{3_B-EPIR-B} in $f$ must bind,
and $f$ is an RSW allocation.

\subsection{Inefficiency}\label{Section5.4}

Finally, we 
discuss
the economic implications of Theorems \ref{Thm_RSW}--\ref{Thm_IC2}.
In particular, we investigate the efficiency properties of intuitive equilibrium allocations.

The ex-ante social surpluses in intuitive equilibria are fully characterized by those in RSW allocations
as follows:
First, Theorem \ref{Thm_EofIA} shows 
that every RSW allocation is an intuitive equilibrium.
Second, Theorems \ref{Thm_IC1} and \ref{Thm_IC2} imply that 
every intuitive equilibrium $f=(q,t)$ 
has an RSW allocation $f^*=(q^*,t^*)$ 
with the same ex-ante social surplus:
\begin{align}
	E_x[(v_2(x)-v_1(x))q(x)] 
	= E_x[(v_2(x)-v_1(x))q^*(x)].
	\label{5_eff}
\end{align}

With this equivalence result, 
the characterization of RSW (Theorem \ref{Thm_RSW}) implies that, 
in general, 
the intuitive equilibrium allocations are 
inefficient due to undersupply. 
Moreover, 
even if the mechanism-selection game has an efficient equilibrium allocation dominating the RSW, 
it is eliminated as unintuitive.\footnote{
	From the equilibrium characterization, 
	an efficient feasible allocation is an equilibrium 
	if it weakly dominates the RSW. 
	See \cite{JET2003Fieseler_etal} for a necessary and sufficient condition for 
	the existence of an efficient feasible 
	allocation 
	in an interdependent-values environment 
	with continuous type spaces. 
}
These inefficiency results are clearly illustrated by the example in Section \ref{Section4}. 
The unique RSW allocation $f^*$ 
in
Table \ref{Tbl1} is inefficient, 
and thus, 
no intuitive equilibrium is efficient.
The equilibrium $f^{**}$ 
in Table \ref{Tbl4}
is efficient, 
but it is unintuitive.

The inefficiency of intuitive equilibria 
might increase the relative attractiveness of undominated equilibria.
Then, alternative solution concepts 
should be 
the core mechanism 
and neutral optimum proposed by \cite{E1983Myerson}, 
and 
the assured allocation 
introduced by \cite{2015Balkenborg_Makrisz}.
In the example 
of 
Section \ref{Section4}, 
the efficient equilibrium $f^{**}$ is selected by the core mechanism and the neutral optimum, 
and other dominated allocations are eliminated by the two concepts.
Further, following \citet{2015Balkenborg_Makrisz}, we can naturally extend their assured allocation to our environment 
with bilateral asymmetric information.
This 
assured allocation is undominated, 
and hence, 
it also selects the efficient equilibrium $f^{**}$. 

Nevertheless, the inefficiency itself should not be a reason to discard intuitive equilibria.
Whenever an unintuitive equilibrium 
is common knowledge, 
some convincing deviations upset the equilibrium.
While \cite{QJE1987Cho_Kreps} established the intuitive criterion using introspective arguments, 
\cite{GEB2020Fudenberg_He} introduced a learning-based equilibrium selection criterion
that is stronger than the intuitive criterion and weaker than divinity.\footnote{
	By extending Theorem \ref{Thm_EofIA}, 
	we can 
	show that every RSW allocation is a D1 (and hence divine) equilibrium.
	See Theorem 1 of \cite{2019Nishimura} for this 
	extension.
} 
If the intuitive equilibrium allocations also describe long-run outcomes 
in our bilateral-trade environment, 
highly inefficient trading outcomes will emerge in the long run.

\section{Concluding remarks}\label{Section6}

We provided a simple characterization of the RSW allocation 
in our bilateral-trade model with interdependent values.
The RSW allocation is derived from the maximization of 
the expected virtual surplus 
given some posterior beliefs.
We also 
identified
a necessary and sufficient condition 
under which the RSW is undominated.
If 
this condition does not hold, 
the RSW is dominated by infinitely 
many equilibrium allocations. 
However, 
the RSW allocations
are interim-payoff-equivalent to the intuitive equilibrium allocations.
From a normative viewpoint, the inefficient 
RSW allocation should not be selected. 
Nevertheless, we can expect that the intuitive RSW allocation describes well 
actual trading outcomes.

Bilateral trade is a fundamental economic activity,
and therefore, 
our results can be applied to many problems.
For example, in the growing literature on aftermarkets, 
a combination of mechanism and information design is an important problem
\citep{E2020Dworczak}.
Then, let us 
consider a post-auction resale with a third party.
Our results demonstrate how information disclosure rules in the auction 
affect resale outcomes. 
In particular, 
if 
an auction winner has
full bargaining power in the resale
and the resale always 
results in
the belief-free RSW allocation, 
then disclosure rules are irrelevant to parties' payoffs and resale surplus.

We have obtained 
clear results by focusing on the simple bilateral-trade model. 
While the assumption of additively separable values 
excludes some situations, 
it is useful for establishing
the result of interim-payoff equivalence.
This separability with sorting assumptions
allows us to
extend many of our results to 
other single-agent environments. 
However, 
the extension 
to multiple-agents environments (e.g., auction, collusion, and subcontracting) 
involves conceptual issues. 
Specifically,
in the definition of the best safe mechanism, 
the agents' EPIC and EPIR constraints should be replaced by 
their IC and IR constraints given each type of principal.
An interesting question is how
this best safe mechanism is characterized in 
multiple-agents
environments. 
This analysis is left for future work.

%
%
%
%
%
%
%


\renewcommand{\thesection}{A}
\section{Appendix}\label{Section7}

\makeatletter
\@addtoreset{equation}{section}
\def\theequation{\thesection.\arabic{equation}}
\makeatother

\setcounter{equation}{0}

The following lemma shows that
the IC constraints are characterized by
the local upward and local downward ones. 
See, for example, Lemma 1 of \cite{2015Balkenborg_Makrisz} for the proof.
\begin{lem}\label{Lem_local}
(i) An allocation $f$ satisfies \eqref{S-IC} if and only if for each $x_1 < \bar{x}_1$,
\begin{align*}
	U_1^f(x_1) &\geq U_1^f(x_1^+ \mid x_1), 
	\tag{S-IC-U}
	\label{S-IC-U}
	\\
	U_1^f(x_1^+) &\geq U_1^f(x_1 \mid x_1^+). 
	\tag{S-IC-D}
	\label{S-IC-D}
\end{align*}
(ii) An allocation $f$ satisfies \eqref{B-EPIC} if and only if for each $x$ with $x_2 > 1$,
\begin{align*}
	u_2^f(x_1,x_2^-) &\geq u_2^f(x_2 \mid x_1,x_2^-),
	\tag{B-EPIC-U}
	\label{B-EPIC-U}
	\\
	u_2^f(x) &\geq u_2^f(x_2^- \mid x). 
	\tag{B-EPIC-D}
	\label{B-EPIC-D}
\end{align*}
\end{lem}

We 
present a transformation method developed by \citet{E2013Gershkov_etal}. 
While their method is 
general, the following result is sufficient for our analysis.
See Lemmas 1--3 of their article (or Lemma 4 of \cite{2019Nishimura}) for the proof. 
\begin{lem}\label{Lem_QMon}
	Let $q^\prime \in [0,1]^X$ be an allocation rule 
	and fix any belief $\pi \in \Delta(X_1)$.
	Suppose that $Q_2^{\prime \pi}$ is increasing in $x_2$.
	Let $q$ denote a solution to the problem: 
	\begin{align}	
		\min_{q \in [0,1]^{X}} \ E_{x}^\pi \left[ (q(x))^2 \right]
		\label{3.1_PMin}
		\ \ \ 
		\textnormal{s.t. } 
		Q_1 = Q_1^\prime, \ 
		Q_2^\pi = Q_2^{\prime \pi}.
	\end{align}
	Then, 
	$q(x_1, \cdot)$ is increasing in $x_2$ for each $x_1$ with $\pi(x_1) > 0$.\footnote{
	Under the additional hypothesis that $Q'_1$ is decreasing, 
	the solution $q$ to problem \eqref{3.1_PMin} is decreasing in $x_1$. 
	However, the ex-post monotonicity for the seller is unnecessary for our results.}
\end{lem}

\begin{proof}[Proof of Theorem \ref{Thm_RSW}]
	By taking a Lagrangian approach, 
	we show that the solutions to problem \eqref{3_RSW3} in Section \ref{Section3.1}
	are characterized by the conditions in the statement. 
	Then, we prove that an allocation is a solution to problem \eqref{3_RSW2} (i.e., RSW) if and only if 
	it is a solution to the relaxed problem 
	\eqref{3_RSW3}.

	\noindent
	{\bf Step 1.}	
	We define the Lagrangian function $L$ for problem \eqref{3_RSW3} as 
	\begin{align}
		L(f, \kappa, \lambda) &\equiv 
		E_{x_1} \left[ U_1^f(x_1) \right] 
		+ \sum_{x_1 < \bar{x}_1} \kappa(x_1) \left[ U_1^f(x_1) - U_1^f(x_1^+ \mid x_1) \right]
		\label{Thm_RSW_Lag}
		\\
		&\ \ \ + \sum_{x_1}\sum_{x_2> 1} \lambda(x) \left[ u_2^f(x) - u_2^f(x_2^- \mid x) \right]
		+ \sum_{x_1} \lambda(x_1, 1) u_2^f(x_1, 1),
		\nonumber
	\end{align}
	where $(\kappa, \lambda) \in \mathbb{R}_+^{X_1} \times \mathbb{R}_+^{X}$ 
	is a vector of Lagrange multipliers 
	(with $\kappa(\bar{x}_1) \equiv 0$).
	The domain $\mathcal{Q} \times \mathbb{R}^{X}$ of \eqref{3_RSW3} 
	is convex 
	and contains an allocation 
	that satisfies 
	the constraints in \eqref{3_RSW3} with strict inequality.\footnote{
		For example, $(q,t)$ defined by $q(x) = 0$ and $t(x) = -(x_1 + x_2)$ for each $x$ works well.
	}
	The functionals 
	$U_1^f(\cdot), U_1^f(\cdot \mid \cdot), u_2^f(\cdot \mid \cdot)$, and $u_2^f(\cdot)$ 
	are linear in $f$. 
	It then 
	follows from the ``saddle-point theorem'' (see, for example, \citet[Sections 8.3 and 8.4]{1969Luenberger})
	that 
	$(q,t)$ is a solution to \eqref{3_RSW3}
	if and only if
	there exists a nonnegative vector $(\kappa, \lambda)$ 
	such that the Lagrangian
	$L$ has a saddle point at $((q, t), (\kappa, \lambda))$:
	\begin{align}
		L(q', t', \kappa, \lambda) \leq L(q, t, \kappa, \lambda) \leq L(q, t, \kappa', \lambda')
		\label{Thm_RSW_Sad}
	\end{align}
	for each $(q',t') \in \mathcal{Q} \times \mathbb{R}^{X}$ and
	$(\kappa', \lambda') \in \mathbb{R}_+^{X_1} \times \mathbb{R}_+^{X}$.

	Because payments can be any real number, 
	the saddle-point condition \eqref{Thm_RSW_Sad}
	requires that no payment rule 
	should affect the Lagrangian $L$ 
	at $(\kappa, \lambda)$.
	This is equivalent to the condition that
	\begin{align*}
		\left(p_1(x_1)+\kappa(x_1)-\kappa(x_1^-)\right)p_2(x_2) = \lambda(x_1, x_2)-\lambda(x_1,x_2^+)
	\end{align*}
	for each $x$, where $\kappa(0) \equiv 0$ and $\lambda(x_1, \bar{x}_2^+) \equiv 0$.
	This condition pins down the multipliers $\lambda$ 
	as $\lambda(x) = (p_1(x_1)+\kappa(x_1)-\kappa(x_1^-) ) (1 - P_2(x_2^-))$
	for each $x$.
	\\

	In Steps 2--6,
	we derive necessary conditions for a solution 
	$f = (q,t)$ to problem 
	\eqref{3_RSW3}. 
	Let $(\kappa, \lambda)$ 
	be a nonnegative vector with which the allocation $f$ satisfies the saddle-point condition \eqref{Thm_RSW_Sad}.
	We define $\pi(x_1) \equiv p_1(x_1) + \kappa(x_1) - \kappa(x_1^-)$
	for each $x_1$. 

	\noindent
	{\bf Step 2.}	
	We show that the vector $\pi$ is a belief 
	satisfying condition (i). 
	By definition, $\sum_{x_1} \pi(x_1) = \sum_{x_1}p_1(x_1) = 1$.
	From Step 1, the multipliers $\lambda$ 
	are given by
	$\lambda(x) = \pi(x_1)(1-P_2(x_2^-))$ for each $x$.
	As the vector $\lambda$
	is nonnegative, 
	so is the vector $\pi$. 
	Hence, $\pi \in \Delta(X_1)$. 
	Its cdf is given by 
	$\Pi(x_1) \equiv \sum_{\hat{x}_1 \leq x_1}\pi(\hat{x}_1) = P_1(x_1) + \kappa(x_1) \geq P_1(x_1)$ for each $x_1$.

	\noindent
	{\bf Step 3.}	
	We claim that the allocation rule $q$ with the belief $\pi$ satisfies condition (ii).
	By substituting the multipliers $(\kappa, \lambda)$ into the Lagrangian and 
	interchanging summations, we observe that, 
	for each $q'$, 
	$L(q', t, \kappa, \lambda)$ is equal to 
	the expected virtual surplus \eqref{Thm_RSW_1}. 
	Then, the saddle-point condition \eqref{Thm_RSW_Sad} verifies our claim, that is, 
	$L(q, t, \kappa, \lambda) = \max_{q' \in \mathcal{Q}} L(q', t, \kappa, \lambda)$. 

	\noindent
	{\bf Step 4.}
	We claim that the payment rule $t$ is determined by the formula \eqref{3_PE} given $q$.
	Fix any $x$. 
	We have to prove that $u_2^{f}(x) = u_2^{f}(x_2^- \mid x)$ if $x_2 > 1$, and $u_2^{f}(x) = 0$ if $x_2 =1$. 
	First, suppose $\lambda(x) > 0$. 
	The complementary slackness condition then verifies our claim.
	Next, suppose $\lambda(x) = 0$. 
	It follows from 
	$\lambda(x) = \pi(x_1)(1-P_2(x_2^-))$
	that 
	$\pi(x_1) = 0$ and $\Pi(x_1^-) > P_1(x_1^-)$. 
	Because $q$ maximizes the expected virtual surplus \eqref{Thm_RSW_1}, 
	$q(x_1,\cdot) \equiv 0$.
	Then, $t(x_1, \cdot) \leq 0$ because $(q, t)$ satisfies 
	\eqref{3_B-EPIC-D} and \eqref{3_B-EPIR-B}.
	This in turn implies $t(x_1, \cdot) \equiv 0$ because $(q, t)$ 
	solves 
	\eqref{3_RSW3}. 
	Thus, 
	$(q,t)$ satisfies 
	both \eqref{3_B-EPIC-D} and \eqref{3_B-EPIR-B} with equality.

	\noindent
	{\bf Step 5.}
	We show that $q$ with $\pi$ satisfies condition (iii).
	From Step 4, the seller's interim revenue is given by \eqref{3_RE}. 
	Substituting these revenues into 
	\eqref{3_S-IC-U} in \eqref{3_RSW3}, 
	we observe that	
	$q$ satisfies the inequality \eqref{3_S-IC-U-2}.
	Moreover, 
	the latter condition in (iii)
	is equivalent to the complementary slackness condition 
	given $\kappa = \Pi - P_1$.

	\noindent
	{\bf Step 6.}
	We show that $f$ is both EPIC and EPIR for the buyer.
	The allocation $f$ satisfies both \eqref{3_B-EPIR-B} and \eqref{3_B-EPIC-D}, 
	and hence, it is EPIR for the buyer. 
	From Step 4, $f$ satisfies \eqref{3_B-EPIC-D} with equality. 
	This, together with the ex-post monotonicity $q \in \mathcal{Q}$ for the buyer, 
	implies that $f$ is EPIC for the buyer.

	We also show that $f$ is IC for the seller.
	Because $f$ satisfies \eqref{3_S-IC-U}, 
	it is sufficient to show that it satisfies \eqref{S-IC-D}. 
	Fix any $x_1 < \bar{x}_1$.
	To derive a contradiction, suppose $U_1^{f}(x_1^+) < U_1^{f}(x_1 \mid x_1^+)$. 
	Let us define a new allocation $f' = (q',t')$ as $f'(x_1^+, \cdot) \equiv f(x_1, \cdot)$ and 
	$f'(x'_1, \cdot) \equiv f(x'_1, \cdot)$ for each $x'_1 \not= x_1^+$. 
	By definition, $f'$ 
	satisfies \eqref{3_S-IC-U}. 
	It 
	also satisfies \eqref{3_B-EPIC-D} because 
	\begin{align*}
		t'(x_1^+, x_2) - t'(x_1^+, x_2^-)
		&=
		\left( q(x_1, x_2) - q(x_1, x_2^-) \right) v_2(x_1, x_2) 
		\\
		&\leq 
		\left( q'(x_1^+, x_2) - q'(x_1^+, x_2^-) \right) v_2(x_1^+, x_2)
	\end{align*}
	for each $x_2 > 1$, 
	where the equality follows from 
	the binding \eqref{3_B-EPIC-D} for $(q,t)$
	and
	the inequality from the monotonicity of $q$ in $x_2$ and that of $v_2$ in $x_1$. 
	Similarly, $f'$ satisfies \eqref{3_B-EPIR-B}. 
	Thus, $f'$ satisfies all the constraints in problem \eqref{3_RSW3}. 
	This contradicts the hypothesis that 
	$f$ is a solution to \eqref{3_RSW3}.
	\\

	From Steps 2--5, 
	if $f = (q,t)$ solves \eqref{3_RSW3}, 
	the payment rule $t$ is determined by the formula \eqref{3_PE}
	and there exists a belief $\pi \in \Delta(X_1)$ 
	with which $q$ satisfies conditions (i)--(iii).
	Conversely, if $f$ satisfies these conditions, 
	then the Lagrangian $L$ for \eqref{3_RSW3} has a saddle point at $(f,(\kappa, \lambda))$
	given 
	the nonnegative vector $(\kappa, \lambda)$ 
	defined by 
	$\kappa \equiv \Pi - P_1$ and $\lambda(x) \equiv \pi(x_1) (1-P_2(x_2^-))$ for each $x$. 
	From Step 6,
	every solution to problem \eqref{3_RSW3} 
	satisfies all the constraints in problem \eqref{3_RSW2}.
	Hence, $f$ solves \eqref{3_RSW3} 
	if and only if it solves \eqref{3_RSW2}
	(i.e., $f$ is RSW). 
	This completes the proof.
\end{proof}

\begin{proof}[Proof of Theorem \ref{Thm_RSW2}]
	\noindent
	{\bf Step 1.}
	We claim that 
	an allocation $f$ is undominated for $\pi$ 
	if and only if 
	there exists an interior belief $w \in \Delta(X_1)$ such that 
	$f$ solves problem \eqref{3_pi-undominated} in Section \ref{Section3.2}. 
	As the set of all $\pi$-feasible allocations 
	is a 
	convex polyhedron, 
	so is the set of all interim payoff vectors $U_1^f$ in $\pi$-feasible allocations $f$. 
	Further, the latter set is bounded due to 
	\eqref{B-IR} and \eqref{S-IC}. 
	With these facts, 
	the supporting hyperplane theorem 
	verifies our claim, 
	as in \cite{E1983Myerson}.

	In Steps 2 and 3, 
	fix any RSW allocation $f = (q,t)$, 
	any belief $\pi \in \Delta(X_1)$, 
	and any interior belief $w \in \Delta(X_1)$.
	We prove that the RSW $f$ solves problem \eqref{3_pi-undominated}
	if and only if the interior belief $w$ with the rule $q$ satisfies conditions (i)--(iii).

	\noindent
	{\bf Step 2.}
	Denote by
	$\mathcal{Q}^{\pi}$ the set of interim-monotone allocation rules for the buyer:
	\begin{align*}
		\mathcal{Q}^{\pi} \equiv \{ q' \in [0,1]^{X} \mid Q_2^{\prime \pi} \ \textnormal{is increasing in } x_2 \}.
	\end{align*}	
	By definition, the RSW $f$ is safe. 
	In particular, it is $\pi$-feasible. 
	Hence, $f$ solves problem \eqref{3_pi-undominated} if and only if 
	it solves the following relaxed problem: 
	\begin{align}
		\max_{f' \in \mathcal{Q}^{\pi} \times \mathbb{R}^{X}} &\ 
		E_{x_1}^w \left[ U_1^{f'}(x_1) \right]
		\label{Thm_RSW2_Undom_2}
		\\
		\nonumber
		\textnormal{s.t. } 
		\nonumber
		&\ 
		U_1^{f'}(x_1) \geq U_1^{f'}(x_1^+ \mid x_1) \ \ \ \forall x_1 < \bar{x}_1.
		\\
		\nonumber
		&\ 
		U_2^{f', \pi}(x_2) \geq U_2^{f', \pi}(x_2^- \mid x_2) \ \ \ \forall x_2 > 1.
		\\
		\nonumber
		&\ 
		U_2^{f', \pi}(1) \geq 0.
	\end{align}

	We define the Lagrangian function $L$ for \eqref{Thm_RSW2_Undom_2} as 
	\begin{align}
		L(f', \kappa', \lambda') &\equiv 
		E_{x_1}^w \left[ U_1^{f'}(x_1) \right] 
		+ \sum_{x_1 < \bar{x}_1} \kappa'(x_1) \left[ U_1^{f'}(x_1) - U_1^{f'}(x_1^+ \mid x_1) \right]
		\\
		&\ \ \ + \sum_{x_2 > 1} \lambda'(x_2) \left[ U_2^{f', \pi}(x_2) - U_2^{f', \pi}(x_2^- \mid x_2) \right]
		+ \lambda'(1) U_2^{f', \pi}(1),
		\nonumber
	\end{align}	
	where $(\kappa', \lambda') \in \mathbb{R}_+^{X_1} \times \mathbb{R}_+^{X_2}$ 
	(with $\kappa'(\bar{x}_1) \equiv 0$). 
	As in 
	Theorem \ref{Thm_RSW}, 
	the saddle-point theorem implies that 
	$f = (q,t)$ solves problem \eqref{Thm_RSW2_Undom_2}
	if and only if
	there exists a nonnegative vector $(\kappa, \lambda)$ 
	such that 
	$L$ has a saddle point at $((q, t), (\kappa, \lambda))$:
	\begin{align}
		L(q', t', \kappa, \lambda) \leq L(q, t, \kappa, \lambda) \leq L(q, t, \kappa', \lambda')
		\label{Thm_RSW2_Sad}
	\end{align}
	for each $(q',t') \in \mathcal{Q}^{\pi} \times \mathbb{R}^{X}$ and 
	$(\kappa', \lambda') \in \mathbb{R}_+^{X_1} \times \mathbb{R}_+^{X_2}$.

	As in Theorem \ref{Thm_RSW}, 
	the saddle-point condition \eqref{Thm_RSW2_Sad} requires that 
	no payment rule 
	should affect 
	$L$ 
	at $(\kappa, \lambda)$.
	This is equivalent to the condition that
	\begin{align*}
		\left( w(x_1)+\kappa(x_1)-\kappa(x_1^-) \right)p_2(x_2) = \pi(x_1) \left( \lambda(x_2)-\lambda(x_2^+) \right)
	\end{align*}
	for each $x$, where $\kappa(0) \equiv 0$ and $\lambda(\bar{x}_2^+) \equiv 0$.
	This condition pins down the multipliers 
	as $\kappa = \Pi - W$ and $\lambda(x_2) = 1 - P_2(x_2^-)$ for each $x_2$.

	\noindent
	{\bf Step 3.}
	To complete the proof, 
	we show that, given the interior belief $w$, 
	the RSW 
	$f$ satisfies the saddle-point condition \eqref{Thm_RSW2_Sad} for some nonnegative multipliers $(\kappa, \lambda)$
	if and only if $w$ with $q$ satisfies conditions (i)--(iii) in the statement.

	First, we assume that $w$ with $q$ satisfies conditions (i)--(iii).
	Define $\kappa \equiv \Pi - W$ and $\lambda(x_2) \equiv 1-P_2(x_2^-)$ for each $x_2$.
	Condition (i) implies that $\kappa$ is nonnegative. 
	By substituting $(\kappa, \lambda)$ into $L$ and 
	interchanging summations, we obtain
	\begin{align*}
		L(q', t, \kappa, \lambda) 
		&= \sum_{x_1} \pi(x_1) \left( v_2^1(x_1)-v_1^1(x_1) - \frac{
		\Pi(x_1^-)-W(x_1^-)
		}{\pi(x_1)}dv_1(x_1) \right) Q'_1(x_1) 
		\nonumber
		\\
		&\ \ 
		+ \sum_{x_2} p_2(x_2) \left( v_2^2(x_2)-v_1^2(x_2) - \frac{1-P_2(x_2)}{p_2(x_2)} dv_2(x_2) \right) Q_2^{\prime \pi}(x_2)
	\end{align*}
	for each $q' \in \mathcal{Q}^{\pi}$. 
	It then holds that
	\begin{align*}
		L(q, t, \kappa, \lambda) 
		= \max_{q' \in \mathcal{Q}} L(q', t, \kappa, \lambda)
		= \max_{q' \in \mathcal{Q}^{\pi}} L(q', t, \kappa, \lambda),
	\end{align*} 
	where the first equality follows from condition (ii)
	and the second from the transformation method of \cite{E2013Gershkov_etal} (Lemma \ref{Lem_QMon}). 
	Now, the RSW satisfies the 
	constraints for the buyer in problem \eqref{Thm_RSW2_Undom_2}
	with equality,
	and the belief $w$ with the rule $q$ satisfies condition (iii) 
	(i.e., the complementary slackness condition for the seller's local upward IC \eqref{3_S-IC-U-2}).
	Hence, the right inequality of the saddle-point condition \eqref{Thm_RSW2_Sad} is satisfied.
	Thus, $((q,t),(\kappa, \lambda))$ satisfies the 
	condition \eqref{Thm_RSW2_Sad}.

	Second, we assume that 
	the RSW $f$ satisfies \eqref{Thm_RSW2_Sad} for some nonnegative $(\kappa, \lambda)$.
	Step 2 then implies that
	$\kappa = \Pi - W$ and $\lambda(x_2) = 1 - P_2(x_2^-)$ for each $x_2$.
	We thus obtain condition (i). 
	The optimality condition $L(q, t, \kappa, \lambda) = \max_{q' \in \mathcal{Q}^{\pi}} L(q', t, \kappa, \lambda)$
	with the 
	ex-post monotonicity $q \in \mathcal{Q}$ 
	implies condition (ii). 
	Finally, 
	condition (iii) follows from 
	the right inequality of the saddle-point condition \eqref{Thm_RSW2_Sad}
\end{proof}

\begin{proof}[Proof of Corollary \ref{Cor_undominated}]
	Fix any RSW allocation $f=(q,t)$. 
	Given the belief $\pi$ in Theorem \ref{Thm_RSW}, 
	the interior prior $w = p_1$ with the rule $q$ satisfies conditions (i)--(iii) in Theorem \ref{Thm_RSW2}. 
	Thus, the set of beliefs for which $f$ is undominated is nonempty.

	We show that this set is convex. 
	Suppose that the RSW is undominated for beliefs $\pi^0$ and $\pi^1$. 
	From Theorem \ref{Thm_RSW2}, for each $k\in \{0,1\}$, 
	there exists an interior belief $w^k$ with 
	which the allocation rule $q$ satisfies conditions (i)--(iii) given $\pi^k$. 
	For any $\alpha \in (0,1)$, the convex combination $\alpha w^0 + (1-\alpha)w^1$ is an interior belief 
	with which $q$ satisfies conditions (i)--(iii) given $\alpha \pi^0 + (1-\alpha) \pi^1$. 
	This, together with Theorem \ref{Thm_RSW2}, implies that the RSW is undominated for $\alpha \pi^0 + (1-\alpha) \pi^1$.
\end{proof}

\begin{proof}[Proof of Theorem \ref{Thm_EofIA}]
	By using Corollary \ref{Cor_undominated},
	we assume that the RSW allocation $f^*$ is undominated for a belief $\pi$.
	Suppose that all seller types select the same mechanism $f^*$.  
	Fix any off-path mechanism $G = ( M, g ) \in \mathcal{G}$.
	The proof consists of two steps. 
	In Step 1, we formulate an auxiliary game given $G$ with $\pi$ 
	and find a perfect equilibrium 
	that generates a reasonable posterior.
	In Step 2, 
	we show that this reasonable posterior supports the RSW as an equilibrium of the mechanism-selection game.

	\noindent
	{\bf Step 1.}	
	We define the {\it auxiliary game} as follows: 
	First, $x=(x_1, x_2)$ are realized according to $(\pi, p_2)$, and 
	each party $i$ privately observes $x_i$.
	Second, the seller selects either $f^*$ or $G$. 
	If the seller selects the status quo $f^*$, each party $i$ obtains $u_i(f^*(x), x)$.
	If the seller selects the new mechanism $G$, then the seller and buyer simultaneously choose 
	$m_1 \in M_1$ and $m_2 \in M_2 \cup \{ 0 \}$, respectively, and
	each party $i$ obtains $u_i(g(m), x)$.
	Note that 
	$x_1$ is realized not according to $p_1$, but $\pi$.

	For each $k \in \mathbb{N}$, 
	we define {\it perturbed game $k$} of the ``agent'' strategic form of the auxiliary game as follows:
	Each type of seller has two agents, and each type of buyer has one agent. 
	An agent of $x_1$ chooses a probability 
	$\gamma^k(x_1) \in [\delta^k(x_1), 1- \delta^k(x_1)]$ of selecting the mechanism $G$, 
	where 
	$\delta^k(x_1) \equiv 1/(2k^3)$ if $x_1 \in X_1^{f^*}(G)$,
	and $\delta^k(x_1) \equiv 1/(2k)$ otherwise. 
	Note that the losing types $x_1 \in X_1^{f^*}(G)$ mistakenly select $G$ less often than the non-losing types.
	The other agent of $x_1$ chooses a distribution 
	$\sigma_1^k(\cdot \mid x_1) \in \Delta(M_1)$ such that $\sigma_1^k(\cdot \mid x_1) \geq 1/(|M_1|k)$. 
	The agent of $x_2$ chooses a distribution 
	$\sigma_2^k(\cdot \mid x_2) \in \Delta(M_2 \cup \{ 0 \})$ such that $\sigma_2^k(\cdot \mid x_2) \geq 1/((|M_2|+1)k)$. 
	Each $\sigma_i^k$ represents a perturbed reporting strategy in $G$.
	We denote $\sigma^k \equiv (\sigma_1^k, \sigma_2^k)$ 
	and $\sigma^k(m \mid x) \equiv \sigma_1^k(m_1 \mid x_1) \sigma_2^k(m_2 \mid x_2)$
	for each $m$ and $x$. 
	Given an action profile $(\gamma^k, \sigma^k)$,
	the two agents of $x_1$ obtain the same payoff
	\begin{align*}
		&\gamma^k(x_1) E_{x_2} \left[\sum_{m}  
		\sigma^k(m \mid x) u_1(g(m), x) \right]
		+ (1-\gamma^k(x_1)) U_1^{f^*}(x_1),
	\end{align*}
	and the agent of $x_2$ obtains the payoff
	\begin{align*}
		&E_{x_1}^{\rho^k} \left[ \gamma^k(x_1) \sum_{m} 
		\sigma^k(m \mid x) u_2(g(m),x) + (1-\gamma^k(x_1)) u_2^{f^*}(x) \right],
	\end{align*}
	where $(\rho^k)_{k = 1}^\infty$ is an arbitrary sequence of interior beliefs in $\Delta(X_1)$ 
	that converges to the given belief $\pi$ 
	as $k \rightarrow \infty$ 
	and satisfies $\rho^k(\cdot) \geq 1/(|X_1|k)$ for each $k$.\footnote{
		The purpose of perturbing $\pi$ is to prove that limit posteriors are reasonable.
		This perturbation is unnecessary 
		if we prove only that the RSW allocation is an equilibrium. 
	}

	Each perturbed game $k$ has at least one Nash equilibrium. 
	For each $k$, fix any Nash equilibrium $(\gamma^k, \sigma^k)$ 
	and define the buyer's 
	interior 
	belief $\pi^k \in \Delta(X_1)$ as 
	\begin{align}
		\pi^k(x_1) &\equiv \frac{\rho^k(x_1)\gamma^k(x_1)}{\sum_{x'_1} \rho^k(x'_1)\gamma^k(x'_1)}.
		\label{Thm_EofIA_1}
	\end{align}
	With some abuse of notation, let $(\gamma^k, \sigma^k, \pi^k)_{k = 1}^\infty$
	denote a subsequence of the mother sequence that converges in the Euclidean space. 
	Let $(\gamma^G, \sigma^G, \pi^G) \equiv \lim_{k \rightarrow \infty} (\gamma^{k}, \sigma^{k}, \pi^{k})$.
	Then, $(\gamma^G, \sigma^G)$ 
	is a perfect equilibrium of the agent strategic form of the auxiliary game.
	Hence, $\sigma^G$ is a BNE of the continuation game $(G, \pi^G)$.
	Define
	the continuation BNE allocation $f$ as $f(x) \equiv \sum_{m}\sigma^G(m \mid x) g(m)$.

	Now, we prove that the limit posterior $\pi^G$ is reasonable, that is, 
	$\pi^G \in \Delta(X_1 \setminus X_1^{f^*}(G))$ whenever 
	$X_1^{f^*}(G) \subsetneq X_1$. 
	If the set of losing types is empty, there is nothing to prove.
	Fix any losing type $x_1 \in X_1^{f^*}(G)$.
	Then, $U_1^f(x_1) < U_1^{f^*}(x_1)$.
	The strict inequality 
	implies that, 
	if $k$ is sufficiently large, then 
	the losing type $x_1$ selects $G$ with the lowest probability
	$\gamma^{k}(x_1) = \delta^k(x_1) = 1/(2k^3)$, 
	and hence, 
	\begin{align*}
		\pi^{k}(x_1) 
		= \frac{\rho^k(x_1) \gamma^{k}(x_1)}{\sum_{x'_1} \rho^k(x'_1) \gamma^{k}(x'_1)}
		\leq \frac{\frac{1}{2 k^3}}{\sum_{x'_1 \not\in X_1^{f^*}(G)} \rho^k(x'_1) \gamma^{k}(x'_1)}
		\leq \frac{ \frac{1}{2 k^3} }{ |X_1 \setminus X_1^{f^*}(G)| \ \frac{1}{2|X_1| k^2} },
	\end{align*}
	where the second inequality follows from the fact that $\rho^k(\cdot) \geq 1/(|X_1|k)$ 
	and $\gamma^{k}(x'_1) \geq 1/(2k)$ for each $x'_1 \not\in X_1^{f^*}(G)$.
	If $|X_1 \setminus X_1^{f^*}(G)| > 0$, 
	then we obtain $\pi^G(x_1) = \lim_{k \rightarrow \infty} \pi^{k}(x_1) = 0$,
	and hence, $\pi^G \in \Delta(X_1 \setminus X_1^{f^*}(G))$.

	\noindent
	{\bf Step 2.}	
	Given the two allocations $f$ and $f^*$, 
	we show that 
	$U_1^{f}(x_1) \leq U_1^{f^*}(x_1)$ for each $x_1$. 
	We define the perfect equilibrium allocation $f'$ 
	as
	\begin{align*}
		f'(x)
		&\equiv \gamma^G(x_1) f(x)
		+ (1-\gamma^G(x_1)) f^*(x).
	\end{align*}
	Because $\gamma^G$ is the seller's best response to $\sigma^G$,
	$\gamma^G(x_1) = 0$ if $U_1^f(x_1) < U_1^{f^*}(x_1)$, and 
	$\gamma^G(x_1) = 1$ if $U_1^f(x_1) > U_1^{f^*}(x_1)$. 
	This 
	implies that $f'$ weakly dominates $f^*$.
	We claim that $f'$ is $\pi$-feasible. 
	Because $\sigma_2^G$ is the buyer's best response to $(\gamma^G, \sigma_1^G)$,
	we obtain
	\begin{align}
		E_{x_1}^{\pi} \left[
		\gamma^G(x_1) u_2^f(x)
		\right]
		\geq 
		E_{x_1}^{\pi} \left[
		\gamma^G(x_1) u_2^f(\hat{x}_2 \mid x)
		\right]
		\label{Thm_EofIA_2}
	\end{align}
	for each $x_2 \in X_2$ and $\hat{x}_2 \in X_2 \cup \{ 0 \}$.
	The RSW satisfies both \eqref{B-EPIC} and \eqref{B-EPIR}.
	Multiplying both sides of each constraint by $1-\gamma^G(x_1)$
	and taking the expectation $E_{x_1}^\pi$, 
	we obtain 
	\begin{align}
		E_{x_1}^{\pi} \left[
		(1-\gamma^G(x_1)) u_2^{f^*}(x)
		\right]
		\geq 
		E_{x_1}^{\pi} \left[
		(1-\gamma^G(x_1)) u_2^{f^*}(\hat{x}_2 \mid x)
		\right]
		\label{Thm_EofIA_3}
	\end{align}
	for each $x_2 \in X_2$ and $\hat{x}_2 \in X_2 \cup \{ 0 \}$. 
	Summing inequalities \eqref{Thm_EofIA_2} and \eqref{Thm_EofIA_3}, 
	we observe that $f'$ satisfies both \eqref{B-IC} and \eqref{B-IR}.
	Because $(\gamma^G, \sigma_1^G)$ is the seller's best response to $\sigma_2^G$,
	we obtain
	\begin{align*}
		\gamma^G(x_1) U_1^f(x_1) + (1-\gamma^G(x_1)) U_1^{f^*}(x_1)
		\geq 
		\gamma^G(\hat{x}_1) U_1^f(\hat{x}_1 \mid x_1) + (1-\gamma^G(\hat{x}_1)) U_1^{f^*}(x_1)
	\end{align*}
	for each $x_1$ and $\hat{x}_1$.
	Because $f^*$ satisfies \eqref{S-IC}, 
	$U_1^{f^*}(x_1) \geq U_1^{f^*}(\hat{x}_1 \mid x_1)$
	for each $x_1$ and $\hat{x}_1$.
	Thus, $f'$ satisfies \eqref{S-IC}.
	As $f^*$ 
	satisfies \eqref{S-IR}, so does $f'$. 
	These arguments establish that $f'$ is $\pi$-feasible. 
	The $\pi$-feasible $f'$ cannot dominate $f^*$,
	because $f^*$ is undominated for the given belief $\pi$.
	Hence, $U_1^{f^*} = U_1^{f'}$,  
	and
	the RSW $f^*$ weakly dominates $f$.

	From Steps 1 and 2,
	$f^*$ is an equilibrium of the mechanism-selection game
	that is  
	supported by the reasonable posterior $\pi^{G}$ with the continuation BNE allocation $f \in BN(G, \pi^G)$
	for each off-path 
	$G$. 
	This implies that $f^*$ is intuitive.
\end{proof}

\begin{proof}[Proof of Proposition \ref{Prop_PayoffEq}]
	Fix any allocation $f'=(q',t')$ that satisfies \eqref{S-IC} and (\hyperref[B-IC]{B-$p_1$-IC}).
	Given $q'$, let $q$ be a solution to problem \eqref{3.1_PMin} in Lemma \ref{Lem_QMon} for $\pi = p_1$.
	Denote $Q_2 \equiv Q_2^{p_1}$ and $Q_2^\prime \equiv Q_2^{\prime p_1}$. 
	The new 
	rule $q$ 
	satisfies 
	$Q_i = Q'_i$ for each $i$.
	Further, $q$ is increasing in $x_2$.
	Let $T'(\cdot) \equiv E_{x_1}[t'(x_1,\cdot)]$ denote the buyer's interim payments. 
	We define an ``adjusted'' private-value component 
	$\alpha \in \mathbb{R}^{X_2}$ such that 
	$\alpha(1) \equiv v_2^2(1)$, and for each $x_2 > 1$,
	\begin{align*}
		\alpha(x_2) \equiv 
		\displaystyle
		\frac{T'(x_2) - T'(x_2^-) - 
		E_{x_1} \left[ v_2^1(x_1) \left( q'(x) - q'(x_1,x_2^-) \right) \right] }
		{Q'_2(x_2) - Q'_2(x_2^-)}
	\end{align*}
	if $Q'_2(x_2) > Q'_2(x_2^-)$, and
	$\alpha(x_2) \equiv v_2^2(x_2)$
	if $Q'_2(x_2) = Q'_2(x_2^-)$.
	We denote $d \alpha(x_2) \equiv \alpha(x_2^+)-\alpha(x_2)$ for each $x_2 < \overline{x}_2$, 
	and 
	let $d \alpha(\overline{x}_2)$ be an arbitrary number.
	We then inductively define a new payment rule $t$ as follows:
	\begin{align}
		t(x_1, 1) &\equiv v_2(x_1, 1)q(x_1,1) + U_1^{f'}(x_1)
		\label{Prop_PayoffEq_0.1}
		\\
		&\ \ \ \ - E_{x_2} \left[ \left(v_2^1(x_1)+ \alpha(x_2) - \frac{1-P_2(x_2)}{p_2(x_2)}d \alpha(x_2) -v_1(x)
		\right) q(x) \right],
		\nonumber
		\\
		t(x_1, x_2) &\equiv t(x_1,x_2^-) + \left( v_2^1(x_1) + \alpha(x_2) \right)\left( q(x_1,x_2) - q(x_1,x_2^-) \right) 
		\label{Prop_PayoffEq_0.2}
	\end{align}
	for each $x_1$ and $x_2 > 1$.

	Because $f'$ satisfies (\hyperref[B-IC]{B-$p_1$-IC}), 
	it follows from the definition 
	of the function $\alpha$ that 
	$v_2^2(x_2^-) \leq \alpha(x_2) \leq v_2^2(x_2)$ for each $x_2 > 1$.
	These inequalities, 
	together with \eqref{Prop_PayoffEq_0.2} and the monotonicity of $q$ in $x_2$, 
	imply 
	that $f$ satisfies \eqref{B-EPIC}.

	Each type $x_1$ of seller obtains the following interim payoff from $f$: 
	\begin{align}		 
		U_1^f(x_1)
		&= E_{x_2} \left[ \left(v_2^1(x_1) + \alpha(x_2) - \frac{1-P_2(x_2)}{p_2(x_2)}d \alpha(x_2) -v_1(x)
		\right) q(x) \right] 
		- u_2^f(x_1,1)
		\nonumber
		\\
		&= U_1^{f'}(x_1),
		\label{Prop_PayoffEq_1}
	\end{align}
	where the second equality follows from \eqref{Prop_PayoffEq_0.1}. 
	The seller's interim-payoff equivalence \eqref{Prop_PayoffEq_1} implies that
	$f$ satisfies \eqref{S-IC} because for each $x_1, \hat{x}_1 \in X_1$,
	\begin{align*}
		U_1^f(x_1) = U_1^{f'}(x_1) 
		\geq U_1^{f'}(\hat{x}_1 \mid x_1)
		&= U_1^{f'}(\hat{x}_1) + \left(v_1^1(\hat{x}_1) - v_1^1(x_1) \right) Q'_1(\hat{x}_1)
		\\
		&= U_1^f(\hat{x}_1) + \left(v_1^1(\hat{x}_1) - v_1^1(x_1) \right) Q_1(\hat{x}_1)
		\\
		&= U_1^f(\hat{x}_1 \mid x_1),
	\end{align*}
	where the inequality follows from the hypothesis that $f'$ satisfies \eqref{S-IC}.

	We obtain the equivalence of the ex-ante expected social surpluses as follows:
	\begin{align*}
		&E_{x} \left[ \left(v_2(x)-v_1(x)\right) q(x) \right] 
		\\
		&= E_{x_1} \left[ \left(v_2^1(x_1)-v_1^1(x_1)\right) Q_1(x_1) \right] + E_{x_2} \left[ \left(v_2^2(x_2)-v_1^2(x_2)\right) Q_2(x_2) \right]
		\\
		&= E_{x_1} \left[ \left(v_2^1(x_1)-v_1^1(x_1)\right) Q'_1(x_1) \right] + E_{x_2} \left[ \left(v_2^2(x_2)-v_1^2(x_2)\right) Q'_2(x_2) \right]
		\\
		&= E_{x} \left[ \left(v_2(x)-v_1(x)\right) q'(x) \right].
	\end{align*}
	The equivalence implies $E_{x_1}[U_1^f(x_1)] + E_{x_2}[U_2^f(x_2)] = E_{x_1}[U_1^{f'}(x_1)] + E_{x_2}[U_2^{f'}(x_2)]$.
	From the seller's interim-payoff equivalence \eqref{Prop_PayoffEq_1}, 
	$E_{x_2}[U_2^f(x_2)] = E_{x_2}[U_2^{f'}(x_2)]$.

	Finally, we prove the buyer's interim-payoff equivalence. 
	It follows from the definitions of $t$ and $\alpha$ that,
	for each $x_2 > 1$,
	\begin{align*}
		U_2^f(x_2) - U_2^{f}(x_2^-)
		&= v_2^2(x_2)Q_2(x_2) - v_2^2(x_2^-)Q_2(x_2^-) - \alpha(x_2) \left( Q_2(x_2) - Q_2(x_2^-) \right)
		\\
		&= v_2^2(x_2)Q'_2(x_2) - v_2^2(x_2^-)Q'_2(x_2^-) - \alpha(x_2) \left( Q'_2(x_2) - Q'_2(x_2^-) \right)
		\\
		&= U_2^{f'}(x_2) - U_2^{f'}(x_2^-).
	\end{align*} 
	This, together with $E_{x_2}[U_2^f(x_2)] = E_{x_2}[U_2^{f'}(x_2)]$, implies $U_2^f = U_2^{f'}$.
\end{proof}

As explained in Section \ref{Section5.3}, 
we reconstruct the interim-payoff-equivalent allocation $f$ 
in Figure \ref{Fig3}. 
Given any candidate deviator $\tilde{x}_1$ from the equilibrium, 
the following lemma modifies 
the equivalent allocation
by using a coalition $\{ \tilde{x}_1,...,\tilde{x}'_1 \}$ of types 
such that the constraints \eqref{3_B-EPIR-B} are slack for these types
and 
the constraint \eqref{S-IC-D} is slack for $\tilde{x}_1^{\prime +}$. 
\begin{lem}\label{Lem_higher}
	Let $\hat{f} = (\hat{q}, \hat{t})$ be an allocation that satisfies \eqref{S-IC}, \eqref{S-IR}, and \eqref{B-EPIC}.
	Suppose that $u_2^{\hat{f}}(\tilde{x}_1, 1) > 0$ for some $\tilde{x}_1 \in X_1$. 
	Then, there exists an allocation $f$ and a type $\tilde{x}'_1 \geq \tilde{x}_1$ 
	with the following three properties:

	\noindent
	(i)
	$f(x_1, \cdot) = \hat{f}(x_1, \cdot)$ if $x_1 \leq \tilde{x}_1$.

	\noindent
	(ii)
	$u_2^{f}(x_1, 1) > 0$ if $\tilde{x}_1 \leq x_1 \leq \tilde{x}'_1$, 
	and $U_1^{f}(\tilde{x}_1^{\prime +}) = 0 > U_1^{f}(\tilde{x}'_1 \mid \tilde{x}_1^{\prime +})$ if $\tilde{x}'_1 < \bar{x}_1$. 

	\noindent
	(iii)
	$f$ satisfies \eqref{S-IC}, \eqref{S-IR}, and \eqref{B-EPIC}.
\end{lem}

\begin{proof}[Proof of Lemma \ref{Lem_higher}]
	Assume, without loss of generality, that $\tilde{x}_1 < \bar{x}_1$.\footnote{
		If the candidate 
		is the highest type, 
		we can complete the proof by setting $f \equiv \hat{f}$ and $\tilde{x}'_1 \equiv \bar{x}_1$.
	}
	For notational simplicity, we denote $\tilde{q}(\cdot) \equiv \hat{q}(\tilde{x}_1, \cdot)$. 
	As $U_1^{\hat{f}}(\tilde{x}_1) \geq 0$, $u_2^{\hat{f}}(\tilde{x}_1, 1) > 0$, and $\hat{f}$ satisfies \eqref{B-EPIC}, 
	the candidate $\tilde{x}_1$ delivers the good with positive probability (i.e., $\tilde{q}(\cdot) \not= 0$).
	Define an allocation $f' \equiv (q',t')$ as follows:
	for each $x_1 \leq \tilde{x}_1$, $f'(x_1, \cdot) \equiv \hat{f}(x_1, \cdot)$; 
	for each $x_1 > \tilde{x}_1$, $q'(x_1, \cdot) \equiv \tilde{q}(\cdot)$ and 
	\begin{align*}
		t'(x) \equiv v_2(x)\tilde{q}(x_2) - \sum_{\hat{x}_2 < x_2} dv_2(\hat{x}_2) \tilde{q}(\hat{x}_2) - u_2^{\hat{f}}(\tilde{x}_1, 1)
	\end{align*}
	for each $x_2$.
	Because $\tilde{q}$ is increasing in $x_2$ from the ex-post monotonicity and 
	$u_2^{f'}(x) = u_2^{f'}(x_2^- \mid x) $ for each $x$ with $x_1 > \tilde{x}_1$ and $x_2 > 1$, 
	$f'$ satisfies \eqref{B-EPIC}. 
	However, the allocation $f'$ may 
	violate 
	\eqref{S-IC-U}. 
	This is because 
	$v_2$ is increasing in $x_1$, 
	and hence, 
	for each $x'_1$, 
	$U_1^{f'}(\cdot \mid x'_1)$ is increasing in $x_1$ on $\{\tilde{x}_1,...,\bar{x}_1\}$.

	Then, we ``shrink'' $f'$ 
	to prevent each seller type from lying. 
	Define $\tilde{x}'_1 \equiv \min \{ x_1 > \tilde{x}_1 \mid U_1^{f'}(x_1^- \mid x_1) < 0 \} - 1$;
	we denote $\tilde{x}'_1 \equiv \bar{x}_1$ 
	if $U_1^{f'}(x_1^- \mid x_1) \geq 0$ for each $x_1 > \tilde{x}_1$. 
	Note that $\tilde{x}_1 < x_1 \leq \tilde{x}'_1$ implies 
	$U_1^{f'}(x_1^-) > U_1^{f'}(x_1^- \mid x_1) \geq 0$. 
	Here, the strict inequality follows from $q'(x_1^-, \cdot) =  \tilde{q}(\cdot) \not= 0$. 
	We define a decreasing function $\alpha \in [0,1]^{X_1}$ as follows:
	$\alpha(x_1) \equiv 1$ if $x_1 \leq \tilde{x}_1$, 
	\begin{align*}
		\alpha(x_1) 
		\equiv \alpha(x_1^-) \frac{U_1^{f'}(x_1^-)}{U_1^{f'}(x_1 \mid x_1^-)}
	\end{align*}
	if $\tilde{x}_1 < x_1 \leq \tilde{x}'_1$, 
	and $\alpha(x_1) \equiv 0$ if $x_1 > \tilde{x}'_1$.
	The function $\alpha$ is well-defined because $U_1^{f'}(x_1 \mid x_1^-) \geq U_1^{f'}(x_1^-) > 0$
	if $\tilde{x}_1 < x_1 \leq \tilde{x}'_1$.

	Using this function, 
	we define 
	an allocation $f=(q,t)$ 
	as $f(x) \equiv \alpha(x_1) f'(x)$ for each $x$. 
	It satisfies property (i).
	We also obtain two facts: 
	(a) $\tilde{x}_1 \leq x_1 \leq \tilde{x}'_1$ implies that $\alpha(x_1) > 0$ and $u_2^{f}(x_1, 1) = 
	\alpha(x_1) u_2^{f'}(x_1, 1) = \alpha(x_1) u_2^{\hat{f}}(\tilde{x}_1, 1) > 0$;
	and (b) $\tilde{x}'_1 < \bar{x}_1$ implies
	$U_1^{f}(\tilde{x}_1^{\prime +}) = 0 > \alpha(\tilde{x}'_1) U_1^{f'}(\tilde{x}'_1 \mid \tilde{x}_1^{\prime +})$. 
	Thus, $f$ has property (ii).
	Further, $f$ satisfies \eqref{S-IC}. 
	This is because $Q_1 = \alpha Q'_1$ is decreasing, 
	$\tilde{x}_1 < x_1 \leq \tilde{x}'_1$ implies $U_1^{f}(x_1 \mid x_1^-) = U_1^{f}(x_1^-)$,
	and $x_1 > \tilde{x}'_1$ implies both $f(x_1, \cdot) \equiv (0,0)$ and $0 > U_1^{f}(\tilde{x}'_1 \mid \tilde{x}_1^{\prime +})$. 
	It is clear from the definition of $\alpha$ that 
	$f$ satisfies \eqref{S-IR}.
	As $f'$ satisfies \eqref{B-EPIC}, so does $f = \alpha f'$.
	Thus, $f$ has property (iii).
\end{proof}

As explained in Section \ref{Section5.3}, 
we construct a less-trading allocation $\tilde{f}$ from the modified allocation $f$ 
with the coalition $\{ \tilde{x}_1,...,\tilde{x}'_1 \}$
in Lemma \ref{Lem_higher}. 
This allocation $\tilde{f}$ is defined as a convex combination of $f$  
and a fee, 
as in \eqref{4_convex}. 
However, 
no type 
$x_1 > \tilde{x}_1^\prime$ outside the coalition should charge this fee, 
because otherwise 
the buyer, believing that 
the seller's type is the non-coalition member $x_1$,
opts out of 
$\tilde{f}$.
Nonetheless, 
$\tilde{f}$ can be IC for the seller
because \eqref{S-IC-D} is slack for $\tilde{x}_1^{\prime +}$ in the original $f$. 
The next lemma shows how this allocation $\tilde{f}$ is constructed. 
\begin{lem}\label{Lem_perturb}
	Let $f$ be an allocation that 
	satisfies \eqref{S-IC}, \eqref{S-IR}, and \eqref{B-EPIC}.
	Suppose that there exists a pair of types
	$ \tilde{x}_1,\tilde{x}'_1 \in X_1$ 
	such that 
	$u_2^{f}(x_1, 1) > 0$ if $\tilde{x}_1 \leq x_1 \leq \tilde{x}'_1$,
	and 
	$U_1^{f}(\tilde{x}_1^{\prime +}) > U_1^{f}(\tilde{x}'_1 \mid \tilde{x}_1^{\prime +})$ if $\tilde{x}'_1 < \bar{x}_1$.
	Then, 
	there exists an allocation $\tilde{f}$ with the following three properties:

	\noindent
	(i) $\tilde{f}$ 
	satisfies \eqref{S-IC}, \eqref{S-IR}, and \eqref{B-EPIC}.

	\noindent
	(ii) $u_2^{\tilde{f}}(x_1, 1) > 0$ if $\tilde{x}_1 \leq x_1 \leq \tilde{x}'_1$.

	\noindent
	(iii)
	$U_1^{\tilde{f}}(x_1) > U_1^{f}(x_1)$ if and only if $\tilde{x}_1 \leq x_1 \leq \tilde{x}'_1$. 
\end{lem}
\begin{proof}[Proof of Lemma \ref{Lem_perturb}]
	Denote by $f = (q, t)$ the original allocation.
	Fix any $\delta \in (0,1)$.
	Define $\tau \equiv ( U_1^{f}(\tilde{x}_1) + U_1^{f}(\tilde{x}_1^-) ) / 2$ with
	$U_1^{f}(0) \equiv U_1^{f}(1) + 1$. 
	We also define an allocation $\tilde{f}$ as follows:
	$\tilde{f}(x_1, \cdot) \equiv (1 - \delta)f(x_1, \cdot) + \delta(0, \tau)$ if $x_1 \leq \tilde{x}'_1$, 
	and $\tilde{f}(x_1, \cdot) \equiv (1 - \delta)f(x_1, \cdot)$ if $x_1 > \tilde{x}'_1$.
	Note that 
	as $\delta \rightarrow 0$, $\tilde{f} \rightarrow f$.

	{\bf Properties (i) and (ii).}
	Using the hypotheses that 
	$\tilde{x}_1 \leq x_1 \leq \tilde{x}'_1$ implies
	$u_2^{f}(x_1, 1) > 0$ 
	and that $\tilde{x}'_1 < \bar{x}_1$ implies 
	$U_1^{f}(\tilde{x}_1^{\prime +}) > U_1^{f}(\tilde{x}'_1 \mid \tilde{x}_1^{\prime +})$, 
	we 
	make $\delta > 0$ sufficiently small such that 
	the nearby allocation $\tilde{f}$ satisfies property (ii) 
	and the seller's local IC constraints
	(which are equivalent to \eqref{S-IC} from Lemma \ref{Lem_local} (i)). 
	As $f$ satisfies both \eqref{S-IR} and \eqref{B-EPIC}, so does $\tilde{f}$.
	Thus, $\tilde{f}$ 
	satisfies property (i).

	{\bf Property (iii).}
	Because $f$ satisfies \eqref{S-IR},
	$U_1^{f}(\tilde{x}'_1) \geq 0$. 
	Then, the hypothesis that $u_2^{f}(\tilde{x}'_1, 1) > 0$ and $f$ satisfies \eqref{B-EPIC}
	implies $q(\tilde{x}'_1, \cdot) \not= 0$. 
	Because $q$ is interim monotone for the seller, $q(x_1, \cdot) \not= 0$ if $x_1 < \tilde{x}'_1$.
	This, together with the hypothesis that $f$ satisfies \eqref{S-IC-U}, 
	implies that $U_1^{f}$ is strictly decreasing in $x_1$ on $\{1,...,\tilde{x}'_1\}$. 
	Hence, for each $x_1 \leq \tilde{x}'_1$, 
	the interim 
	payoff $U_1^{\tilde{f}}(x_1) = U_1^{f}(x_1) + \delta (\tau - U_1^{f}(x_1))$ 
	is larger than $U_1^{f}(x_1)$ if $x_1 \geq \tilde{x}_1$ and 
	smaller than $U_1^{f}(x_1)$ otherwise.
	For each $x_1 > \tilde{x}'_1$, $U_1^{\tilde{f}}(x_1) = (1-\delta)U_1^{f}(x_1) \leq U_1^{f}(x_1)$.
	Thus, $\tilde{f}$ has property (iii).
\end{proof}

The following lemma constructs an AM mechanism $G = (M, g)$ 
described in Section \ref{Section5.3}.
Note that, for a pure strategy profile denoted by $s = (s_1, s_2) \in M_1^{X_1} \times (M_2 \cup \{0\})^{X_2}$, 
the composition $g \circ s$ is an allocation. 
See Appendix D in the Supplementary material for the proof.
\begin{lem}\label{Lem_AM}
	Suppose that 
	$(f, \tilde{x}_1, \tilde{x}'_1)$ satisfies 
	the hypotheses of Lemma \ref{Lem_perturb}.
	Then, 
	there exists a mechanism 
	$G = (M,g) \in \mathcal{G}$ with the following 
	three 
	properties:

	\noindent
	(i) 
	$X_1^{f}(G) = \{x_1 \in X_1 \mid x_1 < \tilde{x}_1 \textnormal{ or } x_1 > \tilde{x}'_1 \}$.

	\noindent
	(ii)
	There exists 
	a 
	profile 
	$s$
	such that,
	for each $\pi \in \Delta(X_1 \setminus X_1^{f}(G))$, 
	$s$ is the unique profile 
	that survives
	the iterative deletion of strictly dominated strategies
	in $(G, \pi)$.	

	\noindent
	(iii)
	The strategy profile $s$ satisfies
	$U_1^{g \circ s}(x_1) > U_1^{f}(x_1)$ for each $x_1 \in X_1 \setminus X_1^{f}(G)$.
\end{lem}

\begin{proof}[Proof of Proposition \ref{Prop_EqmAllocations}]
	From Theorem \ref{Thm_EofIA}, 
	every feasible allocation that weakly dominates the RSW allocations is an equilibrium.
	Fix any equilibrium $f'$ and RSW allocation $f^*$.
	To derive a contradiction, suppose that 
	$U_1^{f'}(x_1) < U_1^{f^*}(x_1)$
	for some $x_1$. 
	We show that this type $x_1$ can  
	profitably deviate from $f'$ to another mechanism.
	Take any $\varepsilon \in (0, U_1^{f^*}(x_1) - U_1^{f'}(x_1)]$. 
	Define an allocation $f \in A^X$ by $f \equiv f^* - (0, \varepsilon)$. 
	The triplet
	$(f, \tilde{x}_1, \tilde{x}'_1)$ with $\tilde{x}_1 \equiv 1$ and $\tilde{x}'_1 \equiv \bar{x}_1$ 
	satisfies the hypotheses of Lemma \ref{Lem_perturb}. 
	Lemma \ref{Lem_AM} 
	then provides a mechanism $G = (M, g)$ with properties (i)--(iii).
	Property (i) implies $X_1^{f}(G) = \varnothing$.
	From properties (ii) and (iii), we obtain a pure strategy profile $s$ 
	such that 
	$\cup_{\pi \in \Delta(X_1)} BN(G, \pi) = \{ g \circ s \}$ 
	and $U_1^{g \circ s}(x_1) > U_1^{f}(x_1) \geq U_1^{f'}(x_1)$.
	This contradicts the hypothesis that 
	$f'$ is an equilibrium.
\end{proof}

\begin{proof}[Proof of Theorem \ref{Thm_IC1}]
	Fix any intuitive equilibrium $f'$.
	Let $f^*$ be an RSW allocation.
	It follows from the equilibrium characterization (Proposition \ref{Prop_EqmAllocations})
	that $f'$ weakly dominates $f^*$ (i.e., $U_1^{f'} \geq U_1^{f^*} $).
	To derive a contradiction, suppose that $f'$ dominates $f^*$ (i.e., $U_1^{f'} \not= U_1^{f^*} $).
	Using the interim-payoff equivalence (Proposition \ref{Prop_PayoffEq}), 
	we obtain a feasible allocation $\hat{f}$ that satisfies \eqref{B-EPIC}
	and $U_i^{\hat{f}} = U_i^{f'}$ for each $i$.
	Because the new $\hat{f}$ dominates $f^*$, 
	it 
	violates \eqref{B-EPIR}. 
	It follows that
	$u_2^{\hat{f}}(x_1, 1) < 0$ for some $x_1$. 
	However, $\hat{f}$ satisfies \textnormal{(\hyperref[B-IR]{B-$p_1$-IR})}.
	Hence, 
	we can find some $\tilde{x}_1$ with
	$u_2^{\hat{f}}(\tilde{x}_1, 1) > 0$.

	We show that this type $\tilde{x}_1$ can convincingly deviate from $f'$.
	Using Lemma \ref{Lem_higher}, we obtain $f \in A^X$ and $\tilde{x}'_1 \geq \tilde{x}_1$ with the following properties:
	first, $f(x_1, \cdot) = \hat{f}(x_1, \cdot)$ if $x_1 \leq \tilde{x}_1$;
	second, $u_2^{f}(x_1, 1) > 0$ if $\tilde{x}_1 \leq x_1 \leq \tilde{x}'_1$, and 
	$U_1^{f}(\tilde{x}_1^{\prime +}) = 0 > U_1^{f}(\tilde{x}'_1 \mid \tilde{x}_1^{\prime +})$ if $\tilde{x}'_1 < \bar{x}_1$; 
	and third, $f$ satisfies \eqref{S-IC}, \eqref{S-IR}, and \eqref{B-EPIC}.
	The triplet $(f, \tilde{x}_1, \tilde{x}'_1)$ satisfies the hypotheses of Lemma \ref{Lem_perturb}.
	Lemma \ref{Lem_AM} then provides a mechanism $G = (M, g)$ with properties (i)--(iii).
	Because $U_1^{f}(x_1) = U_1^{\hat{f}}(x_1)$ for each $x_1 < \tilde{x}_1$ 
	and $U_1^{f}(x_1) \leq U_1^{f}(\tilde{x}_1^{\prime +}) =  0 \leq U_1^{\hat{f}}(x_1)$ for each $x_1 > \tilde{x}'_1$, 
	property (i) implies $X_1^{f}(G) 
	\subseteq X_1^{\hat{f}}(G) = X_1^{f'}(G)$. 
	The last equality follows from the interim-payoff equivalence $U_1^{\hat{f}} = U_1^{f'}$. 
	Further, from properties (ii) and (iii), we obtain a pure strategy profile $s$ 
	such that, 
	if $\pi \in \Delta(X_1 \setminus X_1^{f}(G))$, 
	$BN(G, \pi) = \{ g \circ s \}$ 
	and $U_1^{g \circ s}(\tilde{x}_1) > U_1^{f}(\tilde{x}_1) = U_1^{\hat{f}}(\tilde{x}_1)$.
	Hence, if the off-path belief $\pi^G$ given $G$ 
	is reasonable (i.e., $\pi^G \in \Delta(X_1 \setminus X_1^{f'}(G)) \subseteq \Delta(X_1 \setminus X_1^{f}(G))$),
	the type $\tilde{x}_1$ is 
	better off deviating from $f'$ to $G$. 
	This contradicts the hypothesis that 
	$f'$ is intuitive.
\end{proof}

\begin{proof}[Proof of Theorem \ref{Thm_IC2}]
	Fix any intuitive equilibrium $f'$.
	Theorem \ref{Thm_IC1} implies that $U_1^{f'}$ is the seller's RSW payoff vector. 
	Using the interim-payoff equivalence (Proposition \ref{Prop_PayoffEq}), 
	we obtain a feasible allocation $f$ that satisfies \eqref{B-EPIC}
	and $U_i^{f} = U_i^{f'}$ for each $i$. 
	We complete the proof by showing that $u_2^{f}(x_1,1) = 0$ for each $x_1$, 
	and hence, 
	$f$ is an RSW allocation.
	To derive a contradiction, suppose that $u_2^{f}(x_1,1) \not= 0$ for some $x_1$. 
	Because $f$ satisfies \textnormal{(\hyperref[B-IR]{B-$p_1$-IR})}, 
	we obtain a type $x'_1$ with $u_2^{f}(x'_1,1) > 0$.
	Let $f^{*}$ be an RSW allocation. 
	We define an allocation $f^{**}$ as follows: 
	$f^{**}(x_1, \cdot) \equiv f(x_1, \cdot)$ if $x_1 = x'_1$, 
	and $f^{**}(x_1, \cdot) \equiv f^{*}(x_1, \cdot)$ otherwise.
	Because $U_1^{f^{**}}$ is the RSW payoff vector and $f^{**}$ is safe, 
	the new allocation $f^{**}$ is an RSW allocation.
	This, together with $u_2^{f^{**}}(x'_1,1) > 0$, contradicts the characterization of RSW (Theorem \ref{Thm_RSW}).
\end{proof}

\section*{Acknowledgments}
I am extremely grateful to an editor and three anonymous referees whose insightful comments significantly improved the paper.
I want to express my gratitude to Akira Okada, 
Reiko Aoki, Hideshi Itoh, Takashi Kunimoto, and Takashi Ui for their comments and suggestions. 
Further, I want to thank Yu Awaya, Gary Biglaiser, Wonki Cho, Makoto Hanazono, Benjamin E. Hermalin, Akifumi Ishihara, Daniel Kr\"{a}hmer, Joosung Lee, 
Shintaro Miura, Toshiji Miyakawa, Kota Murayama, Nozomu Muto, Tymofiy Mylovanov, Patrick Rey, David Salant, Ryuji Sano, Takashi Shimizu, Yasuhiro Shirata, Yiman Sun, Takuro Yamashita, 
and the participants at the Kansai Game Theory Seminar, GAMES 2012, 
7th International Workshop on Evolution of Standards and Technology, Workshop on Economics of Procurement, 
2018 AMES, and EEA-ESEM Cologne 2018, 
for their thoughtful comments and advice. 
This work was supported by JSPS KAKENHI Grant Numbers 26885081, 15H03346, and 18K12747.



\newpage
\begin{center}
{\Large Supplementary Material for \\ ``Informed Principal Problems in Bilateral Trading''}
\end{center}

\renewcommand{\thesection}{B}
\makeatletter
\@addtoreset{equation}{section}
\def\theequation{\thesection.\arabic{equation}}
\@addtoreset{figure}{section}
\def\thefigure{\thesection.\arabic{figure}}
\def\thetable{\thesection.\arabic{table}}
\makeatother

Appendix \ref{SectionS1} provides the equilibrium characterization presented in Section 4 of the paper.
Appendix \ref{SectionS2} gives an example of an equilibrium in which 
the seller cannot make a convincing deviation to any ``menu." 
Appendix \ref{SectionS3} 
constructs 
an Abreu--Matsushima (AM) mechanism that is described in Section 5.3 of the paper.

\section{Equilibrium characterization in Section 4}\label{SectionS1}

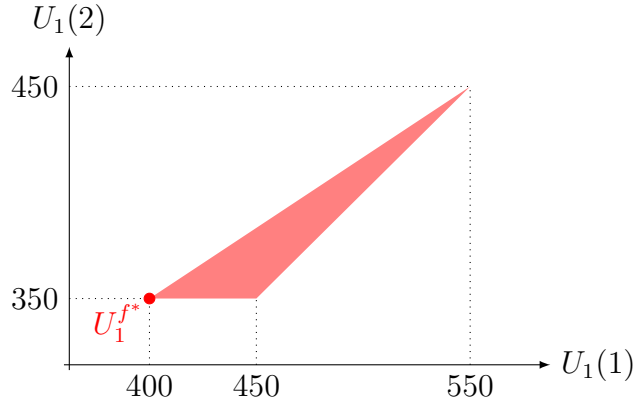
\begin{figure}[h]
\begin{center}
\begin{tikzpicture}[x=1,y=1]
	\path[draw,->,>=latex] (0,-2) -- (0,120);
	\path[draw,->,>=latex] (-2,0) -- (180,0);
	\path (180,0) node[right] {$U_1(1)$};
	\path (0,120) node[above] {$U_1(2)$};

		\coordinate (A) at (30,25);
	\coordinate (B) at (70,25);
	\coordinate (C) at (150,105);
	\fill[red, opacity=0.5] (A)--(B)--(C)--cycle;

		\path[draw, dotted] (0,25) -- (30,25);
	\path (0,25) node[left] {$350$};
	\path[draw, dotted] (30,0) -- (30,25);
	\path (30,0) node[below] {$400$};
	\path[draw, dotted] (0,105) -- (150,105);
	\path (0,105) node[left] {$450$};
	\path[draw, dotted] (70,0) -- (70,25);
	\path (70,0) node[below] {$450$};
	\path[draw, dotted] (150,0) -- (150,105);
	\path (150,0) node[below] {$550$};

		\fill[draw, red](30,25) circle[radius=2];
	\path (32,27) node[below left] {\color{red}$U_1^{f^*}$};
\end{tikzpicture}
\caption{Seller's equilibrium payoff vectors}\label{FigS1}
\end{center}
\end{figure}

In this appendix, we characterize the seller's equilibrium payoff vectors 
in Section 4 of the paper.
We assume that
$X_1 = X_2 = \{1, 2 \}$, $p_1 = p_2 = 1/2$, $v_1(x)=100(x_1 - 1)$, and $v_2(x)=300 x_1 + 100 x_2$.
Let $f=(q,t)$ denote an 
equilibrium.
We show that the set of equilibrium payoff vectors is characterized by the following conditions
(i.e., the red triangle in Figure \ref{FigS1}):
\begin{align}
	U_1^f(2) &\geq U_1^f(1) - 100,
	\label{S.1.1}
	\\
	U_1^f(2) &\leq \frac{2}{3}U_1^f(1) + \frac{250}{3},
	\label{S.1.2}
	\\
	U_1^f(2) &\geq 350.
	\label{S.1.3}
\end{align}
Denote by $f^*$ the RSW allocation in Table 1 of Section 4.
The seller's RSW payoff vector is given by $U_1^{f^*} = (400, 350)$.

(i) We show that $U_1^f$ must satisfy \eqref{S.1.1}--\eqref{S.1.3}.
The equilibrium characterization in the paper (Proposition 1 with Corollary 1) implies that $U_1^f$ satisfies
\begin{align}
	U_1^f(2) \geq U_1^{f^*}(2) = 350.
	\label{S.1.5}
\end{align}
From \eqref{S.1.5}, \eqref{S.1.3} follows immediately.
The equilibrium $f$ is feasible, 
and hence, 
\begin{align}
	U_1^f(1) - U_1^f(2) &\leq 50(q(1,1)+q(1,2)),
	\label{S.1.6}
	\\
	U_1^f(1) - U_1^f(2) &\geq 50(q(2,1)+q(2,2)),
	\label{S.1.7}
	\\
	U_1^f(1) + U_1^f(2) &\leq 150 q(1,1) + 250 q(1,2) + 250 q(2,1) + 350 q(2,2),
	\label{S.1.8}
\end{align}
where 
the last inequality follows from 
\begin{align*}
	E_{x_1}[U_1^f(x_1)] &= E_{x}[(v_2(x)-v_1(x))q(x)] - E_{x_2}[U_2^f(x_2)],
	\\
	U_2^f(2) &\geq U_2^f(1) + 50 (q(1,1)+q(2,1)),
	\\
	U_2^f(1) &\geq 0.
\end{align*}
Because $q \leq 1$, \eqref{S.1.6} implies \eqref{S.1.1}. 
By substituting \eqref{S.1.7} into \eqref{S.1.8}, we obtain 
\begin{align*}
	U_1^f(1) + U_1^f(2) &\leq 5(U_1^f(1) - U_1^f(2)) + 150 q(1,1) + 250 q(1,2) + 100 q(2,2).
\end{align*}
This inequality with $q\leq 1$ implies \eqref{S.1.2}.

(ii)
We complete the proof by finding 
two feasible allocations $f^{**}$ and $f$ with $U_1^{f^{**}} = (550, 450)$ and $U_1^{f} = (450, 350)$.
Then, the constant allocations defined by $f^{**} \equiv (1, 550)$ and $f \equiv (1, 450)$ are what we need.

\renewcommand{\thesection}{C}
\section{Deviation to menus}\label{SectionS2}

Using an example with three seller types, 
we 
construct an equilibrium in which
no type 
can make a convincing deviation to any menu. 

We assume that
$X_1 = \{1, 2, 3\}$, $X_2 = \{1, 2 \}$, $p_1 = 1/3$, $p_2 = 1/2$, 
$v_1(x)=100(x_1 - 1)$, and $v_2(x)=300 x_1 + 200 x_2$.
From Theorem 1 in the paper, 
the unique RSW allocation $f^*$ is given by Table \ref{TblS1}. 
We also consider the allocation $f$ in Table \ref{TblS2}. 
The seller's interim payoff vectors in these allocations are given by 
\begin{align}
	U_1^{f^*}=(500, 450, 825/2), \ \ \ \ 
	U_1^{f}=(525, 460, 425).
	\label{S.2.1}
\end{align}
Thus, $f$ dominates $f^*$. 
Simple computations show that $f$ is feasible. 
With these facts, the equilibrium characterization in the paper implies that $f$ is an equilibrium. 
Note that the low-type buyer benefits from $f$ only if the seller has the middle type.

\setcounter{table}{0}
\begin{table}[htb]
\begin{minipage}{0.5\hsize}
\begin{center}
	\begin{tikzpicture}[x=1,y=1]			
		\path[draw, black, thick] (0,54) -- (140,54);
		\path[draw, black, thick] (0,36) -- (140,36);
		\path[draw, black, thick] (0,18) -- (140,18);
		\path[draw, black, thick] (0,0) -- (140,0) -- (140,54) -- (0,54) -- (0,0);
		\path[draw, black, thick] (70,0) -- (70,36);
		\path[draw, black, thick] (140,0) -- (140,36);
		\path[draw, black, thick] (140,0) -- (140,54);
	
		\path (0,44) node[left] {$x_1 = 1$};
		\path (0,26) node[left] {$x_1 = 2$};
		\path (0,8) node[left] {$x_1 = 3$};		
		\path (35,54) node[above] {$x_2 = 1$};
		\path (105,54) node[above] {$x_2 = 2$};
	
		\path (70,44) node {$1, \ \ \ 500$};
		\path (35,26) node {$0, \ \ \ 0$};
		\path (105,26) node {$1, \ \ \ 1000$};
		\path (35,8) node {$0, \ \ \ 0$};
		\path (105,8) node {$3/4, \ \ \ 975$};
	\end{tikzpicture} 
\end{center}
	\caption{RSW allocation $f^*$}
	\label{TblS1}
\end{minipage}
\begin{minipage}{0.5\hsize}
\begin{center}
	\begin{tikzpicture}[x=1,y=1]			
		\path[draw, black, thick] (0,54) -- (140,54);
		\path[draw, black, thick] (0,36) -- (140,36);
		\path[draw, black, thick] (0,18) -- (140,18);
		\path[draw, black, thick] (0,0) -- (140,0) -- (140,54) -- (0,54) -- (0,0);
		\path[draw, black, thick] (70,0) -- (70,36);
		\path[draw, black, thick] (140,0) -- (140,36);
		\path[draw, black, thick] (140,0) -- (140,54);
	
		\path (0,44) node[left] {$x_1 = 1$};
		\path (0,26) node[left] {$x_1 = 2$};
		\path (0,8) node[left] {$x_1 = 3$};		
		\path (35,54) node[above] {$x_2 = 1$};
		\path (105,54) node[above] {$x_2 = 2$};
	
		\path (70,44) node {$1, \ \ \ 525$};
		\path (35,26) node {$3/10, \ \ \ 175$};
		\path (105,26) node {$1, \ \ \ 875$};
		\path (35,8) node {$0, \ \ \ 40$};
		\path (105,8) node {$7/10, \ \ \ 950$};
	\end{tikzpicture} 
\end{center}
	\caption{Equilibrium allocation $f$}
	\label{TblS2}
\end{minipage}
\end{table}

We call a mechanism $G \in \mathcal{G}$ a {\it menu} (of outcomes) if the seller has no reporting opportunity in $G$
(i.e., $M_1 = \varnothing$). 
For each posterior $\pi \in \Delta(X_1)$, 
we denote by $(q,t) \in BN(G, \pi)$ an allocation generated by a ``BNE" (i.e., possibly mixed optimal choices for the buyer)
in the game $(G, \pi)$. 
Because this allocation does not depend on the seller's type, 
we denote $(q,t) = (q(x_2), t(x_2))_{x_2 \in X_2}$.

Now, fix any menu $G$. 
In the following, 
we show that no type $x_1 \in \{1,2,3\}$ of seller can convincingly deviate from the equilibrium $f$ to the menu $G$. 

\noindent
{\bf Low type's deviation.}
The low-type seller's interim payoff $U_1^f(1) = 525$ in the equilibrium $f$ is higher than 
her full-information payoff $U_1^{f^*}(1) = 500$. 
From this fact, it is clear that the low type cannot convincingly deviate from $f$ to $G$. 

\noindent
{\bf Middle type's deviation.}
To derive a contradiction, suppose that the middle type can convincingly deviate from the equilibrium $f$ to the menu $G$. 
Denote by $\pi$ the posterior with $\pi(2) = 1$. 
Fix any BNE allocation $(q,t) \in BN(G, \pi)$. 
Because the middle type benefits from the revealing deviation, 
we obtain 
\begin{align}
	\frac{1}{2} \left( t(1)+t(2)-100(q(1)+q(2)) \right) > 460 = U_1^f(2).
	\label{S.2.2}
\end{align}
The BNE allocation is IC and IR for the buyer with the posterior $\pi$, 
and hence, 
\begin{align}
	1000 (q(2)-q(1)) &\geq t(2) - t(1),
	\label{S.2.3}
	\\
	800 q(1) &\geq t(1).
	\label{S.2.4}
\end{align}
By substituting these two inequalities with $t(2) \geq t(1)$ into \eqref{S.2.2}, 
we obtain
\begin{align}
	t(2) \geq \frac{3680}{7} > 525 = U_1^f(1). 
	\label{S.2.5}
\end{align}
Let $\pi'$ denote the most optimistic posterior (i.e., $\pi'(3) = 1$). 
From inequalities \eqref{S.2.3} and \eqref{S.2.4}, 
it is optimal for each type of buyer with $\pi'$ to 
choose an outcome
with a payment 
that is equal to (or higher than) $t(2)$.
With this fact, the two inequalities in \eqref{S.2.5}
imply that
the deviation from 
$f$ to $G$ can be profitable for the low type
(i.e., $1 \not\in X_1^f(G)$). 
Thus, the middle type must benefit from the deviation given the most pessimistic posterior (i.e., $\pi''(1)=1$). 
This is impossible. 

\noindent
{\bf High type's deviation.}
To derive a contradiction, suppose that the high type can convincingly deviate from $f$ to $G$. 
First, we claim that the middle type is a non-lowing type (i.e., $2 \not\in X_1^f(G)$). 
Denote by $\pi$ the most optimistic posterior (i.e., $\pi(3) = 1$). 
Fix any BNE $(q,t) \in BN(G, \pi)$. 
Denote $Q \equiv E_{x_2}[q(x_2)]$ and $T \equiv E_{x_2}[t(x_2)]$. 
The high type benefits from the revealing deviation, 
and hence, 
\begin{align}
	T-200Q &> 425 = U_1^f(3).
	\label{S.2.6}
\end{align}
The BNE allocation $(q,t)$ is IR for the buyer with $\pi$, 
and hence, 
\begin{align}
	Q \geq \frac{T}{1300} > \frac{425 + 200Q}{1300}
	\ \ 
	\Rightarrow
	\ \ 
	Q > \frac{17}{44}.
	\label{S.2.7}
\end{align}
From \eqref{S.2.6} and \eqref{S.2.7},
the middle type 
is a non-losing type 
because
\begin{align}
	T-100Q = (T-200Q) +100Q > 460 = U_1^f(2).
	\label{S.2.8}
\end{align}

Second, we claim that the low type is a non-lowing type (i.e., $1 \not\in X_1^f(G)$). 
Denote by $\pi$ the posterior with $\pi(2) = 1$. 
Note that 
$\pi$ is reasonable.
Fix any BNE $(q,t) \in BN(G, \pi)$.
Denote $Q \equiv E_{x_2}[q(x_2)]$ and $T \equiv E_{x_2}[t(x_2)]$. 
Because the high type benefits from the deviation given the 
reasonable 
$\pi$, 
we obtain the same inequality as \eqref{S.2.6}.
The BNE $(q,t)$ is IR for the buyer with $\pi$, 
and hence, 
\begin{align}
	Q \geq \frac{T}{1000} > \frac{425 + 200Q}{1000}
	\ \ 
	\Rightarrow
	\ \ 
	Q > \frac{17}{32}.
	\label{S.2.9}
\end{align}
Hence, 
the low type is a non-losing type 
because
\begin{align}
	T = (T-200Q) +200Q > 525 = U_1^f(1).
	\label{S.2.10}
\end{align}

The two claims imply $X_1^f(G) = \varnothing$.
Thus, the high type must benefit from the deviation given the most pessimistic posterior. 
This is impossible.

%
%

\renewcommand{\thesection}{D}
\section{AM mechanism}\label{SectionS3}

In this appendix, 
we 
construct an AM 
mechanism that is described in Section 
5.3 of the paper.
Because the formal proof is quite complicated, we begin with a simpler example 
to provide some intuition behind the proof.

\begin{ex}\label{Ex1}
	Consider the example in Appendix \ref{SectionS2}.
	Let $f$ be the equilibrium in Table \ref{TblS2}.
	The middle-type seller is the {\it candidate deviator} from the equilibrium.
	As in Lemma 4 of the paper, 
	we define a nearby allocation $\tilde{f}$ as follows: 
	\begin{align}
		\tilde{f}(x_1, \cdot) \equiv (1-\delta)f(x_1, \cdot) + \delta (0, \tau)
	\end{align}
	if $x_1 \not= 3$, 
	and $\tilde{f}(3, \cdot) \equiv (1-\delta)f(3, \cdot)$, 
	where $\tau \in (460, 525) = (U_1^{f}(2), U_1^{f}(1))$ is a fixed fee
	and $\delta \in (0,1)$ is a small weight with $u_2^{\tilde{f}}(2,1) > 0$ and $U_1^{\tilde{f}}(3) > U_1^{\tilde{f}}(2 \mid 3)$.

	The middle-type seller prefers the allocation $\tilde{f}$ to the equilibrium $f$, 
	while the other types have the opposite preference
	(i.e., $U_1^{\tilde{f}}(2) > U_1^{f}(2)$ and $U_1^{\tilde{f}}(x_1) < U_1^{f}(x_1)$ for each $x_1 \not= 2$).
	The direct mechanism $\tilde{f}$ is IC for the seller and EPIC for the buyer. 
	However, 
	the middle type {\it cannot} convincingly deviate from the equilibrium $f$ to the mechanism $\tilde{f}$.
	This is because the game $(\tilde{f}, \pi)$ given the posterior $\pi(2) = 1$
	has BNE allocations for which the middle type suffers from the deviation.\footnote{
		For example, the game $(\tilde{f}, \pi)$ has a BNE in which 
		the seller always reports that she has the highest type
		and the buyer always opts out of $\tilde{f}$. 
		As discussed by \cite{JET1991Gresik2}, 
		this multiplicity of equilibria 
		is 
		typical of
		interdependent-values environments. 
	}

	Then, we construct an AM mechanism to eliminate $f$ as unintuitive.
	For simplicity, we assume here that the set of posterior beliefs for the buyer
	is finite and does not depend on off-path mechanisms. 
	Let $\mathcal{P} \subset \Delta(X_1)$ denote a 
	nonempty finite set of posterior beliefs $\pi$ with which
	the buyer benefits from the allocation $\tilde{f}$
	(i.e., $U_2^{\tilde{f}, \pi}(1)  > 0$).

	This AM mechanism is defined as a finite strategic game form 
	$G = (M, g) \in \mathcal{G}$. 
	In this static mechanism, each party simultaneously makes $(K+1)$ reports given a large integer $K \geq 1$.
	Each party reports both its type and the buyer's belief about the seller's type.	
	A typical message profile is denoted by $m = (m_1, m_2) \in M_1 \times M_2$
	with  
	$m_i = (m_i^0,...,m_i^K)$, 
	$m_1^0 = x_1^0$, 
	$m_2^0 = (x_2^0, \pi_2^0)$, 
	and $m_i^k = (x_i^k, \pi_i^k)$ for each $i$ and $k \geq 1$.
	The outcome function $g$ (with $g(\cdot, 0) \equiv (0,0)$)
	is a convex combination of some functions defined as follows:
	For some small $\varepsilon > 0$,
	\begin{align*}
		g(m) &\equiv \varepsilon f_1(x_1^0) 
		+ \varepsilon^2 \left( f_2(x_1^0, x_2^0) + b(x_1^0, \pi_2^0) \right)
		+ \varepsilon^3 d(m) 
		+ \frac{1- \varepsilon - \varepsilon^2 -\varepsilon^3 }{K} 
		\sum_{k=1}^{K} \varphi(m^k)
	\end{align*}
	and $m^k \equiv (m_1^k, m_2^k)$.
	First, $f_1$ is a ``dictatorial'' allocation 
	that is strictly IC for the seller. 
	Second, $f_2$ is an allocation that is strictly EPIC 
	and EPIR 
	for the buyer. 
	We can construct each $f_i$ using the strict monotonicity of $v_i$ in $x_i$
	(see Lemmas \ref{Lem_Pfunction} and \ref{Lem_Afunction} below).
	Third, $b$ is a 
	{\it strictly proper scoring rule}
	in which the buyer reports his belief $\pi_2^0$ about the seller's type report $x_1^0$.\footnote{
		See, for example, \cite{EE1998Selten} for the 
		scoring rule.
	}
	Fourth, $d$ is an {\it AM penalty rule} in which ``the first to lie'' pays a small amount (say, \$1) to the other party. 
	Here, $i$'s lie at $k$ means $(x_i^k, \pi_i^k) \not= (x_i^0, \pi_2^0)$.
	Finally, $\varphi$ is defined by 
	\begin{align*}
		\varphi(x_1^k, \pi_1^k, x_2^k, \pi_2^k) \equiv
		\begin{cases}
		\tilde{f}(x_1^k, x_2^k) & \textnormal{if } \pi_1^k \in \mathcal{P}, \ \pi_2^k \in \mathcal{P}, \\
		(0, 0) & \textnormal{otherwise}.
		\end{cases}	
	\end{align*}
	That is, the direct mechanism $\tilde{f}$ is played 
	if each party reports that the buyer's posterior is in the set $\mathcal{P}$,  
	and no trade takes place otherwise.

	Now, let $\pi$ be a posterior after this off-path mechanism $G$ is offered. 
	We show that 
	the static Bayesian game $(G, \pi)$ has a unique strategy profile 
	that survives the iterative 
	deletion of strictly dominated strategies. 
	This is shown by using the following inductive argument.
	Note that the off-path belief $\pi$ is common knowledge. 
	First, lying in $f_1$ (i.e., $x_1^0 \not= x_1$) is dominated for the seller because $\varepsilon \approx 0$.\footnote{
		We construct this allocation $f_1$ in a manner such that the seller pays sufficient amounts to the buyer to induce his participation in $G$.
	}
	Second, lying in $f_2$ (i.e., $x_2^0 \not= x_2$) is dominated for the buyer because $\varepsilon \approx 0$. 
	Third, lying in $b$ (i.e., $\pi_2^0 \not= \pi$) is dominated for the buyer because the scoring rule is strictly IC for him.
	Finally, mathematical induction implies that
	lying in $\varphi$ at each $k$ (i.e., $(x_i^k, \pi_i^k) \not= (x_i, \pi)$) is dominated for each $i$.
	This is because 
	$\varphi(\cdot, \pi, \cdot, \pi)$ 
	is IC and IR for both parties given $\pi$,
	and 
	$1/K$ 
	can be small enough that the small penalty imposed by $d$ outweighs 
	all benefits
	from lying at each $k$.\footnote{
		This penalty rule $d$ also requires a liar at each $k$ to make a small additional payment to the other party. 
		Hence, lying is strictly dominated by truth-telling.
	}

	Thus, each party tells the truth at each $k$. 
	For each $x$, the ex-post outcome is 
	\begin{align*}
		g(m) &= \varepsilon f_1(x_1) 
		+ \varepsilon^2 \left( f_2(x_1, x_2) + b(x_1, \pi) \right)
		+ \left( 1- \varepsilon - \varepsilon^2 -\varepsilon^3 \right)
		\varphi(x_1, \pi, x_2, \pi),
	\end{align*}
	where $\varphi(x_1, \pi, x_2, \pi) = \tilde{f}(x_1, x_2)$ if $\pi \in \mathcal{P}$, 
	and $\varphi(x_1, \pi, x_2, \pi) = (0, 0)$ otherwise.
	Because $\varepsilon \approx 0$, the low type and the high type are {\it always} worse off deviating from $f$ to $G$, 
	while the middle type is better off deviating from $f$ to $G$ 
	{\it if the deviation convinces the buyer that the seller has the middle type.}
	Thus, we can conclude that the equilibrium $f$ is unintuitive.
\end{ex}

In Example \ref{Ex1}, we make the ad hoc assumption that the set of posterior beliefs is finite
and does not depend on off-path mechanisms.
Dropping the assumption, we redesign an AM mechanism more formally.
As a preliminary step, we construct two allocations incorporated into this AM mechanism.

\setcounter{lem}{5}

The first auxiliary result provides the seller's ``dictatorial'' allocation that is strictly IC for her. 
\begin{lem}\label{Lem_Pfunction}
	There exists a function $\tilde{f}_1 \in A^{X_1}$ such that 
	$ x_1 \not= \hat{x}_1$ implies
	\begin{align}
		U_1^{\tilde{f}_1}(x_1) - U_1^{\tilde{f}_1}(\hat{x}_1 \mid x_1) 
		> 0.
		\label{Lem_Pfunction_1}
	\end{align}
\end{lem}
\begin{proof}[Proof of Lemma \ref{Lem_Pfunction}]
	We define the function $\tilde{f}_1 = (q,t) \in A^{X_1}$ as follows: 
	\begin{align*}
		&q(x_1) \equiv \frac{\overline{x}_1 - x_1 + 1}{\overline{x}_1}, \ \ \ 
		t(x_1) \equiv  
		\sum_{\hat{x}_1 = x_1}^{\overline{x}_1} 
		\frac{ v_1^1(\hat{x}_1^+) + v_1^1(\hat{x}_1) + 2 E_{x_2}[v_1^2(x_2)] }{2 \overline{x}_1},
	\end{align*}
	where $v_1^1(\overline{x}_1^+) \equiv v_1^1(\overline{x}_1) + 1$.
	For each $x_1, \hat{x}_1 \in X_1$ with $\hat{x}_1 > 1$,
	\begin{align*}
		E_{x_2} \left[u_1(\tilde{f}_1(\hat{x}_1), x) - u_1(\tilde{f}_1(\hat{x}_1^-), x) \right]
		&= \frac{1}{\overline{x}_1} \left( v_1^1(x_1) 
		- \frac{v_1^1(\hat{x}_1) + v_1^1(\hat{x}_1^-)}{2} \right).
	\end{align*}
	This payoff difference is positive if $x_1 \geq \hat{x}_1$ 
	and negative if $x_1 < \hat{x}_1$. 
	Hence, $\tilde{f}_1$ is strictly IC for the seller.
\end{proof}

The second auxiliary result provides a direct mechanism that is strictly EPIC 
and strictly EPIR 
for the buyer. 
\begin{lem}\label{Lem_Afunction}
	There exists a function $f_2 \in A^{X_1 \times (X_2 \cup \{ 0 \})}$ such that 
	\begin{align}
		u_2^{f_2}(x) - u_2^{f_2}(\hat{x}_2 \mid x) > 0
		\label{Lem_Afunction_1}
	\end{align}
	for each $x = (x_1, x_2) \in X$ and $\hat{x}_2 \in X_2 \cup \{ 0 \}$ with $x_2 \not= \hat{x}_2$.
\end{lem}
\begin{proof}[Proof of Lemma \ref{Lem_Afunction}]
	We define the function $f_2 = (q,t) \in A^{X_1 \times (X_2 \cup \{ 0 \})}$ as follows: 
	\begin{align*}
		&q(x_1, x_2) \equiv \frac{x_2}{\overline{x}_2}, \ \ \ 
		t(x_1, x_2) \equiv 
		\sum_{\hat{x}_2 = 1}^{x_2} 
		\frac{v_2^2(\hat{x}_2) + v_2^2(\hat{x}_2^-) + 2 v_2^1(x_1) }{2 \overline{x}_2},
	\end{align*}
	where $v_2^2(0) \equiv v_2^2(1) - 1$. 
	For each $x \in X$ and $\hat{x}_2 \in X_2$,
	\begin{align*}
		u_2(f_2(x_1, \hat{x}_2), x) - u_2(f_2(x_1, \hat{x}_2^-), x) 
		&= \frac{1}{\overline{x}_2} \left( v_2^2(x_2) 
		- \frac{v_2^2(\hat{x}_2) + v_2^2(\hat{x}_2^-)}{2} \right).
	\end{align*}
	This payoff difference is positive if $x_2 \geq \hat{x}_2$ 
	and negative if $x_2 < \hat{x}_2$. 
	Hence, $f_2$ is strictly EPIC and strictly EPIR for the buyer.
\end{proof}

\setcounter{lem}{4}

We now prove Lemma \ref{Lem_AM_S} in the paper to construct an AM mechanism.
\begin{lem}\label{Lem_AM_S}
	Suppose that 
	$(f, \tilde{x}_1, \tilde{x}'_1)$ satisfies 
	the hypotheses of Lemma 4 in the paper.
	Then, 
	there exists a mechanism 
	$G = (M,g) \in \mathcal{G}$ with the following 
	three 
	properties:

	\noindent
	(i) 
	$X_1^{f}(G) = \{x_1 \in X_1 \mid x_1 < \tilde{x}_1 \textnormal{ or } x_1 > \tilde{x}'_1 \}$.

	\noindent
	(ii)
	There exists 
	a 
	profile 
	$s$
	such that,
	for each $\pi \in \Delta(X_1 \setminus X_1^{f}(G))$, 
	$s$ is the unique profile 
	that survives 
	the iterative deletion of strictly dominated strategies
	in $(G, \pi)$.	

	\noindent
	(iii)
	The strategy profile $s$ satisfies
	$U_1^{g \circ s}(x_1) > U_1^{f}(x_1)$ for each $x_1 \in X_1 \setminus X_1^{f}(G)$.
\end{lem}

\begin{proof}[Proof of Lemma \ref{Lem_AM_S}]
We define an AM mechanism $G = (M, g) \in \mathcal{G}$ as follows:

\noindent
{\bf Message spaces.}
Let $K, H \in \mathbb{N}$ be positive integers.
The message spaces are defined by 
\begin{align*}
	&M \equiv M_1 \times (M_2 \cup \{ 0 \}), 
	\ \ \ \ \ \ 
	M_i \equiv M_i^0 \times \cdots \times M_i^K, 
	\ \ \ \ \ \ M^k \equiv M_1^k \times M_2^k,
	\\
	&M_i^k \equiv \underline{M}_i^k \times \overline{M}_i^k, \ \ \ \ \
	\underline{M}_1^k \equiv X_1, \ \ \ \ \
	\underline{M}_2^k \equiv X_2 \cup \{ 0 \}, \ \ \ \ \
	\overline{M}_i^k \equiv \{0,...,H\}
\end{align*}
for each $i \in \{1,2\}$ and $k \in \{0,...,K\}$.
Denote typical elements of these sets by
$m = (m_1,m_2) \in M$, $m_i = (m_i^0,...,m_i^K) \in M_i$, 
$m^k = (m_1^k,m_2^k) \in M^k$, and $m_i^k = (x_i^k, h_i^k) \in M_i^k$
for each $i \in \{1,2\}$ and $k \in \{0,...,K\}$.

\noindent
{\bf Outcome function.}
First, we define some allocations. 
For the allocation $f$ in the statement, let $\tilde{f}$ denote the nearby allocation given by Lemma 4 
of 
the paper. 
Let $\tilde{f}_1$ and $f_2$ be the functions given by Lemmas \ref{Lem_Pfunction} and \ref{Lem_Afunction}, respectively.
Denote by $\underline{\eta} > 0$ the smallest number among the left sides of inequalities \eqref{Lem_Pfunction_1} and \eqref{Lem_Afunction_1}.
We define $\overline{\eta} > 0$ as 
\begin{align*}
	\overline{\eta} \equiv \max \{ |u_i(a,x)| \in \mathbb{R}_+ 
	\mid i \in \{1,2\}, a \in \tilde{f}_1(X_1) \cup f_2(X) \cup \tilde{f}(X) 
	\cup \{(1,0)\}, x \in X \}.
\end{align*}
Let $f_1 \equiv \tilde{f}_1 - (0, 2\overline{\eta})$.

Second, we construct a scoring rule to elicit the buyer's posterior.
Denote $\tilde{X}_1 \equiv \{\tilde{x}_1,...,\tilde{x}'_1\}$.
Let $\pi^H$ be the center of $\Delta(\tilde{X}_1)$.\footnote{
	That is, $\pi^H(x_1) \equiv 1/|\tilde{X}_1|$ if $x_1 \in \tilde{X}_1$, 
	and $\pi^H(x_1) \equiv 0$ otherwise.
}
Let $\rho \in [0,1]^{X_1}$ be the vector such that $\rho(x_1) \equiv 0$ if $x_1 \in \tilde{X}_1$, 
and $\rho(x_1) \equiv 1/|X_1 \setminus \tilde{X}_1|$ otherwise.
Pick a small $\delta \in (0,1)$ such that, if $\pi$ is in the $\sqrt{2}\delta$-neighborhood of $\Delta(\tilde{X}_1)$, 
then $U_2^{\tilde{f},\pi}(1) > 0$.
For each $h \in \{0,...,H \}$, we set $\pi^h \equiv \delta(1 - h/H) (\rho - \pi^H) + \pi^H$ and define
\begin{align}
	\overline{\mathcal{P}}^h \equiv \{ \pi \in \Delta(X_1) \mid 
	\| \pi - \pi^h \| \leq \| \pi - \pi^{h'} \| \ \ \forall h' \not= h \}
\end{align}
with $\mathcal{P}^h \equiv \overline{\mathcal{P}}^{h} \setminus \overline{\mathcal{P}}^{h - 1}$ 
and $\overline{\mathcal{P}}^{- 1} \equiv \varnothing$.
Here, $\| \cdot \|$ denotes the Euclidean norm.\footnote{
	\samepage
	For each $\pi \in \Delta(\tilde{X}_1)$, $\pi - \pi^H$ is orthogonal to $\rho - \pi^H$. 
	This implies $\Delta(\tilde{X}_1) \subseteq \mathcal{P}^H$.
	The inclusion holds true even if $\tilde{X}_1 = X_1$ 
	because 
	$h < H$ then implies both $\pi^h \not\in \Delta(X_1)$ and $\overline{\mathcal{P}}^h = \varnothing$.}
Figure \ref{FigS2} illustrates how the choice set $\{ \pi^h \}$ is constructed 
in the case that $|X_1| = 3$, $\tilde{X}_1 = \{2,3\}$, and $H=2$.
We define a quadratic scoring rule $\tau \in [-1, 1]^{\underline{M}_1^0 \times \overline{M}_2^0}$ as 
\begin{align*}
	\tau(x_1, h) \equiv - 2 \pi^h(x_1) + \left\| \pi^{h} \right\|^2.
\end{align*}
For each $\pi$ and $h$, the buyer's expected payment is 
\begin{align*}
	E_{x_1}^\pi [\tau(x_1, h)]
	= - \sum_{x_1} 2 \pi^h(x_1)\pi(x_1) + \left\| \pi^{h} \right\|^2 
	= - \left\| \pi \right\|^2 + \left\| \pi - \pi^{h} \right\|^2.
\end{align*}
Denote $b \equiv (0, \tau)$.
Given the scoring rule $b$, if the seller always reports her true type $x_1$,
then it is optimal for the buyer to report some $h$ with $\pi \in \overline{\mathcal{P}}^h$.
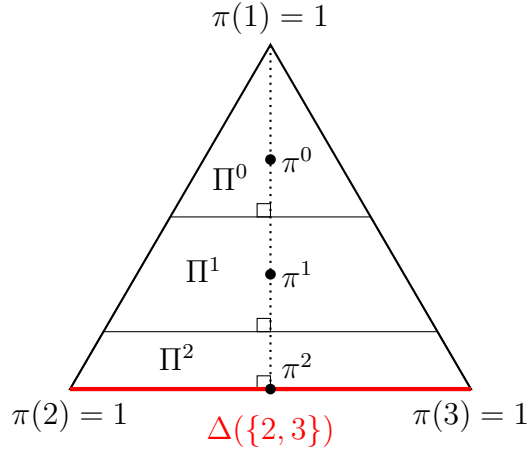
\begin{figure}[htbp]
\begin{center}
\begin{tikzpicture}[x=1,y=1]
	\path[draw, dotted, thick, name path=P0] (75,0) -- (75,130);
    \path[draw, name path=P2] (0,0) -- (0:150) -- ++(120:75) -- ++(180:75) -- cycle;
    \path[draw, name path=P4] (0,0) -- (0:150) -- ++(120:25) -- ++(180:125) -- cycle;
    \path[name path=P1] (0,0) -- (0:150) -- ++(120:100) -- ++(180:50) -- cycle;
    \path[name path=P3] (0,0) -- (0:150) -- ++(120:50) -- ++(180:100) -- cycle;
	\path[name intersections={of= P0 and P1}];
	\fill (intersection-2) circle (2) node[right]{$\pi^0$};
	\path[name intersections={of= P0 and P3}];
	\fill (intersection-2) circle (2) node[right]{$\pi^1$};

	\path[name intersections={of= P0 and P2}];
    \path[draw] (intersection-2) -- ++(90:5) -- ++(180:5) -- ++(270:5) -- cycle;
	\path[name intersections={of= P0 and P4}];
    \path[draw] (intersection-2) -- ++(90:5) -- ++(180:5) -- ++(270:5) -- cycle;
	\path[name intersections={of= P0 and P4}];
    \path[draw] (intersection-1) -- ++(90:5) -- ++(180:5) -- ++(270:5) -- cycle;

	\path (60,80) node {$\mathcal{P}^0$};
	\path (50,45) node {$\mathcal{P}^1$};
	\path (40,12) node {$\mathcal{P}^2$};

	\path (75,130) node[above] {$\pi(1)=1$};
	\path (0,0) node[below] {$\pi(2)=1$};
	\path (150,0) node[below] {$\pi(3)=1$};

    \path[draw, thick, name path=P5] (0,0) -- (0:150) -- ++(120:150) -- cycle;
	\path[name intersections={of= P0 and P5}];
	\fill (intersection-2) circle (2) node[right]{$\rho$};

	\path[draw, red, ultra thick] (0,0) -- (150,0);
	\fill (intersection-1) circle (2) node[above right]{$\pi^2$};
	\fill (intersection-1) node[below=2mm]{\color{red}$\Delta(\{2,3\})$};
\end{tikzpicture}
\caption{Choice set $\{\pi^h\}$ in $\Delta(X_1)$}
\label{FigS2}
\end{center}
\end{figure}

Third, we introduce some penalty (or reward) rules.
Let $\overline{X} \equiv \{ x \in X \mid v_2(x) > v_1(x) \}$ 
and $\underline{X} \equiv \{ x \in X \mid v_2(x) < v_1(x) \} $.
The assumption of non-zero social surplus 
in Section 2 of the paper 
implies $X = \overline{X} \cup \underline{X}$.
Let $y$ and $z$ be the allocations such that, 
if $x \in \overline{X}$, then $y(x) \equiv (1, v_2(x))$ and $z(x) \equiv (0, 0)$, 
and if $x \in \underline{X}$, then $y(x) \equiv (0, 0)$ and $z(x) \equiv (1, v_2(x))$.
We define a function $c \in A^{M^0}$ as
\begin{align*}
	c(x_1, h_1, x_2, h_2) \equiv 
	\begin{cases}
	y(x_1, x_2) & \textnormal{if } h_1 = h_2, \\
	z(x_1, x_2) & \textnormal{if } h_1 \not= h_2, \\
	\end{cases}	
\end{align*}
and $c(\cdot, \cdot, 0, \cdot) \equiv (0,0)$.
Note that, for each $h, h'$ with $h \not= h'$ and $x \in X$, 
it holds
\begin{align}
	u_1(c(x_1, h, x_2, h), x) - u_1(c(x_1, h', x_2, h), x) &= |v_2(x) - v_1(x)| > 0,
	\\
	u_2(c(x_1, h, x_2, h), x) = u_2(c(x_1, h, x_2, h'), x) &= 0.
	\label{Lem_AM_-1}
\end{align}
Define an AM penalty rule $t_0 \in \{0,1\}^M$ as 
\begin{align*}
	t_0(m) \equiv 
	\begin{cases}
	1 & \textnormal{if } \exists k \in \{1,...,K\} \textnormal{ s.t. } 
	m_1^{k} = m_1^0, \ 
	m_2^{k} \not\in \{ x_2^0 \} \times \{h_1^0 - 1, h_1^0, h_1^0 + 1 \}, \\
	& \ \ \ \forall l \in \{1,...,k-1\}, \ \ m_1^{l} = m_1^0, 
	\ m_2^{l} \in \{ x_2^0 \} \times \{h_1^0 - 1, h_1^0, h_1^0 + 1 \}, \\
	0 & \textnormal{otherwise}.
	\end{cases}	
\end{align*}
We also define $t_1 \in [0,1]^{M_1}$ as $t_1(m_1) \equiv \sum_{k=1}^K \mathbb{I}(m_1^{k} \not= m_1^{0}) / K$
and $t_2 \in [0,1]^{\overline{M}_1^0 \times M_2}$ as $t_2(h_1^0, m_2) \equiv \sum_{k=1}^K (\mathbb{I}(x_2^k \not= x_2^{0}) + |h_2^k - h_1^{0}|/H) / (2K)$.\footnote{
	Here, $\mathbb{I}$ is the indicator function (i.e., $\mathbb{I}(\cdot) = 1$ if $\cdot$ is true, 
and $\mathbb{I}(\cdot) = 0$ otherwise).} 
Denote $t \equiv t_0 - t_1 + t_2$ and $d \equiv (0,t)$.
Note that this penalty rule $d$ is, by definition, independent of $h_2^0$.

Finally, we define an outcome function $g \in A^M$ as follows: $g(\cdot, 0) \equiv (0,0)$ and
\begin{align*}
	g(m) &\equiv \varepsilon f_1(x_1^0) 
	+ \varepsilon^2 \left( f_2(x_1^0,x_2^0) + b(x_1^0, h_2^0) \right)
	+ \varepsilon^3 c(m^0) 
	+ \varepsilon^4 d(m)
	\\ 
	&\ \ \ \ \ \ \ \ \ \ \ 
	+ \frac{1- \varepsilon - \varepsilon^2 -\varepsilon^3 - \varepsilon^4 }{K} 
	\sum_{k=1}^{K} \varphi(m^k)
\end{align*}
for each $m \in M_1 \times M_2$,
where $\varepsilon \in (0, 1/2)$ 
and $\varphi \in A^{M^k }$ is defined by
\begin{align}
	\varphi(x_1^k, h_1^k, x_2^k, h_2^k) &\equiv \frac{\min \{ h_1^k, h_2^k \}}{H} \tilde{f}(x_1^k, x_2^k)
	\label{Lem_AM_0}
\end{align}
given the mechanism $\tilde{f}$ with $\tilde{f}(\cdot, 0) \equiv (0,0)$.
We assume that $\varepsilon$, $1/K$, and $1/H$ are small enough that
\begin{align}
	\overline{\eta} &> \varepsilon + 2\varepsilon^3,
	\label{Lem_AM_1}
	\\
	\underline{\eta} &> (2\overline{\eta} + 2) \varepsilon + 4\overline{\eta} \varepsilon^2 + 2 \varepsilon^3,
	\label{Lem_AM_2}
	\\
	\min\{ |v_2(x)-v_1(x)| \mid x \in X \} &> 5 \varepsilon,
	\label{Lem_AM_3}
	\\
	\varepsilon^4 \min\{ p_2(x_2) \mid x_2 \in X_2 \} &> 2 \overline{\eta} / K,
	\label{Lem_AM_4}
	\\
	\varepsilon^4 &> 2 \overline{\eta} / H.
	\label{Lem_AM_5}
\end{align}

\noindent
{\bf Properties of $G$.}
In the rest of the proof, fix a posterior $\pi \in \Delta(X_1)$. 
Let $h$ be the integer for which $\pi \in \mathcal{P}^h$.
Denote by $s = (s_1, s_2)$ 
a pure strategy profile in the Bayesian game $(G, \pi)$. 
For each $k \in \{0,...,K\}$, we denote by $R(k)$ the statement: 
``If
$s$ is iteratively undominated in $(G, \pi)$, 
then for each $i$, $x_i$, and $l \in \{0,...,k\}$, $\underline{s}_i^{l}(x_i) = x_i$, 
$\overline{s}_1^{0}(x_1) = \overline{s}_1^{l}(x_1) \in \{ h, h+1 \}$, and 
$\overline{s}_2^{l}(x_2) \in \{ h, h+1 \}$.''
Here, for each $i$ and $k$, we denote $s_i = (s_i^0,...,s_i^K)$ and $s_i^k = (\underline{s}_i^k,\overline{s}_i^k) \in (M_i^k)^{X_i}$.

Suppose that $s$ is iteratively undominated in $(G, \pi)$.
First, we prove $R(0)$.

\noindent
{\bf Step 1.}
Fix any $x_2 \in X_2$.
We show that $s_2(x_2) \not= 0$ (i.e., the opt-out option is strictly dominated).
To derive a contradiction, suppose the opposite. 
Define $m_2$ by $m_2^k \equiv (0, h)$ for each $k$.
Then, $m_2$ strictly dominates $s_2(x_2) = 0$ because 
it holds that, for each $x_1$,
\begin{align*}
	&u_2(g(s_1(x_1), m_2), x) - u_2(g(s(x)), x)
	\nonumber
	\\
	&= \varepsilon \left( u_2(\tilde{f}_1(\underline{s}_1^0(x_1)), x) + 2 \overline{\eta} \right)
	- \varepsilon^2 \tau(\underline{s}_1^0(x_1), h) 
	- \varepsilon^4 t(s_1(x_1),m_2)
	\nonumber
	\\
	&\geq \varepsilon \left( \overline{\eta} - \varepsilon - 2 \varepsilon^3 \right)
	> 0,
\end{align*}
where the last inequality follows from \eqref{Lem_AM_1}.
Hence, $s_2(x_2) \not= 0$.

\noindent
{\bf Step 2.}
Fix any $x_1 \in X_1$.
We show that $\underline{s}_1^0(x_1) = x_1$.
To derive a contradiction, suppose the opposite. 
Define $m_1$ by $m_1^0 \equiv (x_1, \overline{s}_1^0(x_1))$ and $m_1^k \equiv s_1^k(x_1)$ for each $k > 0$.
Then, $m_1$ strictly dominates $s_1(x_1)$ because
\begin{align*}
	&E_{x_2} \left[ u_1(g(m_1, s_2(x_2)), x) - u_1(g(s(x)), x) \right]
	\nonumber
	\\
	&= \varepsilon \left( U_1^{f_1}(x_1) - U_1^{f_1}(\underline{s}_1^0(x_1) \mid x_1) \right)
	\nonumber
	\\
	&\ \ \ 
	+ \varepsilon^2 E_{x_2} \left[ u_1(f_2(x_1, \underline{s}_2^0(x_2)), x) - u_1(f_2(\underline{s}^0(x)), x) 
	+ \tau(x_1, \overline{s}_2^0(x_2)) - \tau(\underline{s}_1^0(x_1), \overline{s}_2^0(x_2)) \right]
	\nonumber
	\\
	&\ \ \ 
	+ \varepsilon^3 E_{x_2} \left[ u_1(c(m_1^0, s_2^0(x_2)), x) - 
	u_1(c(s^0(x)), x) \right] 
	+ \varepsilon^4 E_{x_2} \left[ t(m_1, s_2(x_2)) - t(s(x)) \right] 
	\nonumber
	\\
	&\geq \varepsilon \left( \underline{\eta} - (2\overline{\eta} + 2) \varepsilon - 4\overline{\eta} \varepsilon^2  
	- 2 \varepsilon^3 \right)
	> 0,
\end{align*}
where the last inequality follows from \eqref{Lem_AM_2}.
Hence, $\underline{s}_1^0(x_1) = x_1$.

\noindent
{\bf Step 3.}
Fix any $x_2 \in X_2$.
We show that $\underline{s}_2^0(x_2) = x_2$.
To derive a contradiction, suppose the opposite. 
Define $m_2$ by $m_2^0 \equiv (x_2, \overline{s}_2^0(x_2))$ 
and $m_2^k \equiv s_2^k(x_2)$ for each $k > 0$.
Then, $m_2$ strictly dominates $s_2(x_2)$ because 
it holds that, for each $x_1$, 
\begin{align*}
	& u_2(g(s_1(x_1), m_2), x) - u_2(g(s(x)), x) 
	\nonumber
	\\
	&= \varepsilon^2 \left( u_2^{f_2}(x) - u_2^{f_2}(\underline{s}_2^0(x_2) \mid x) \right)
	- \varepsilon^3 u_2(c(s^0(x)), x) 
	- \varepsilon^4 \left( t(s_1(x_1), m_2) - t(s(x)) \right)
	\nonumber
	\\
	&\geq \varepsilon^2 \left( \underline{\eta} - 2\overline{\eta} \varepsilon - (3/2)\varepsilon^2 \right)
	> 0,
\end{align*}
where the last inequality follows from \eqref{Lem_AM_2}.
Hence, $\underline{s}_2^0(x_2) = x_2$.

\noindent
{\bf Step 4.}
Fix any $x_2 \in X_2$.
We show that $\overline{s}_2^0(x_2) \in \{h, h+1\}$.
To derive a contradiction, suppose the opposite. 
Define $m_2$ by $m_2^0 \equiv (x_2, h)$ and $m_2^k \equiv s_2^k(x_2)$ for each $k > 0$.
The $c$-scheme is irrelevant to the buyer's ex-post payoffs, as shown by \eqref{Lem_AM_-1}. 
The $d$-punishment is, by definition, independent of his report $h_2^0$.
These facts imply 
\begin{align*}
	&E_{x_1}^\pi \left[ u_2(g(s_1(x_1), m_2), x) - u_2(g(s(x)), x) \right]
	= \varepsilon^2 \left( \left\| \pi - \pi^{\overline{s}_2^0(x_2)} \right\|^2 - \left\| \pi - \pi^h \right\|^2
	\right).
\end{align*}
This payoff difference is positive because
$\pi^h$ is closer to $\pi$ than $\pi^{\overline{s}_2^0(x_2)}$ is.
Thus, $m_2$ strictly dominates $s_2(x_2)$.
Hence, $\overline{s}_2^0(x_2) \in \{h, h+1\}$.\footnote{
	Note that, if $\pi \in \mathcal{P}^h \setminus \overline{\mathcal{P}}^{h+1}$, then 
	$\| \pi - \pi^h \| < \| \pi - \pi^{h+1} \|$, and hence, $\overline{s}_2^0(x_2) = h$.
}

\noindent
{\bf Step 5.}
Fix any $x_1 \in X_1$.
We show that $\overline{s}_1^0(x_1) \in \{h, h+1\}$.
To derive a contradiction, suppose the opposite. 
Step 4 then implies $\overline{s}_1^0(x_1) \not= \overline{s}_2^0(x_2)$ for each $x_2$. 
Let $m_1, n_1$ be the two messages such that 
$m_1^0 = (x_1, h)$, 
$n_1^0 = (x_1, \min \{h+1, H \})$, 
and $m_1^k = n_1^k = s_1^k(x_1)$ for each $k > 0$.
Let $\sigma_1(\cdot \mid x_1)$ be the mixed action randomizing $m_1$ and $n_1$ with equal probabilities
(i.e., $\sigma_1(m_1 \mid x_1) = \sigma_1(n_1 \mid x_1) = 1/2$).
Then, $\sigma_1(\cdot \mid x_1)$ strictly dominates $s_1(x_1)$ because 
it holds that, for each $x_2$, 
\begin{align*}
	&\left( u_1(g(m_1, s_2(x_2)), x) + u_1(g(n_1, s_2(x_2)), x) \right)/2
	- u_1(g(s(x)), x) 
	\nonumber
	\\
	&= \varepsilon^3 \left( u_1(y(x),x) - u_1(z(x),x) \right) / 2 
	+ \varepsilon^4 \left( 
	\left( t(m_1, s_2(x_2)) + t(n_1, s_2(x_2)) \right) / 2 - t(s(x)) \right) 
	\nonumber
	\\
	&\geq \varepsilon^3 \left( |v_2(x)-v_1(x)| / 2 - (5/2) \varepsilon \right)
	> 0,
\end{align*}
where the last inequality follows from \eqref{Lem_AM_3}. 
Hence, $\overline{s}_1^0(x_1) \in \{h, h+1\}$.
\newline

Second, we prove $R(k)$ for each $k$.
Fix any $k \in \{ 1,...,K\}$ and suppose $R(k-1)$.

\noindent
{\bf Step 6.}
Fix any $x_1 \in X_1$.
We show that $s_1^k(x_1) = (x_1, \overline{s}_1^0(x_1))$.
To derive a contradiction, suppose the opposite. 
Define $m_1$ by $m_1^k \equiv (x_1, h_1)$ with $h_1 \equiv \overline{s}_1^0(x_1)$ 
and $m_1^{l} \equiv s_1^{l}(x_1)$ for each $l \not= k$.

First, assume that $s_2^k(x_2) \not\in \{ x_2 \} \times \{ h_1 - 1, h_1, h_1 + 1 \}$ for some $x_2$.
By reporting $m_1$, the seller can make the buyer be ``the first to lie'' with a probability of at least 
$\min \{ p_2(x_2) \mid x_2 \in X_2 \}$.
This implies 
\begin{align}
	&E_{x_2} \left[ u_1(g(m_1, s_2(x_2)), x)
	- u_1(g(s(x)), x) \right]
	\nonumber
	\\
	&= \varepsilon^4 E_{x_2} \left[ t_0(m_1, s_2(x_2)) \right] 
	+ \frac{\varepsilon^4}{K} \mathbb{I}(s_1^{k}(x_1) \not= s_1^0(x_1))
	\nonumber
	\\
	&\ \ \ + \frac{1-\varepsilon-\varepsilon^2-\varepsilon^3-\varepsilon^4 }{K}
	E_{x_2} \left[ u_1(\varphi(m_1^k, s_2^k(x_2)), x) - u_1(\varphi(s_1^k(x_1), s_2^k(x_2)), x) \right]	
	\nonumber
	\\
	&\geq \varepsilon^4 \min \{ p_2(x_2) \mid x_2 \in X_2 \} 
	+ \frac{\varepsilon^4}{K}
	- \frac{1-\varepsilon-\varepsilon^2-\varepsilon^3-\varepsilon^4 }{K} 2 \overline{\eta}
	> 0,
	\label{Lem_AM_11}
\end{align}
where the last inequality follows from \eqref{Lem_AM_4}.

Second, assume that $s_2^k(x_2) \in \{ x_2 \} \times \{ h_1 -1, h_1, h_1 + 1 \}$ for each $x_2$.
For notational simplicity, we denote $h_2(\cdot) \equiv \min \{ h_1, \overline{s}_2^k(\cdot) \}$, 
$h'_2(\cdot) \equiv \min \{ \overline{s}_1^k(x_1), \overline{s}_2^k(\cdot) \}$, 
and $\hat{x}_1 \equiv \underline{s}_1^k(x_1)$. 
We then obtain
\begin{align}
	&E_{x_2} \left[ u_1(g(m_1, s_2(x_2)), x)
	- u_1(g(s(x)), x) \right]
	\label{Lem_AM_12}
	\\
	&\geq \frac{\varepsilon^4}{K}
	+ \frac{1-\varepsilon-\varepsilon^2-\varepsilon^3-\varepsilon^4 }{K}
	E_{x_2} \left[ \frac{h_2(x_2)}{H} u_1^{\tilde{f}} \left(x \right) - \frac{h'_2(x_2)}{H} 
	u_1^{\tilde{f}} \left(\hat{x}_1 \mid x \right) \right],
	\nonumber
\end{align}
where $u_1^{\tilde{f}}(\hat{x}_1 \mid x) \equiv u_1(\tilde{f}(\hat{x}_1, x_2), x)$
and $u_1^{\tilde{f}}(x) \equiv u_1(\tilde{f}(x), x)$.
We claim that the right side of \eqref{Lem_AM_12} is positive. 
Consider two cases. 
First, suppose $\overline{s}_1^k(x_1) \geq h_1 -1$. 
Then, the payoff difference in \eqref{Lem_AM_12} is bounded from below as follows:
\begin{align}
	&E_{x_2} \left[ \frac{h_2(x_2)}{H}u_1^{\tilde{f}} \left(x \right) - \frac{h'_2(x_2)}{H}u_1^{\tilde{f}} \left(\hat{x}_1 \mid x \right) \right]
	\nonumber
	\\
	&= \frac{h_1}{H} \left( U_1^{\tilde{f}} (x_1) - U_1^{\tilde{f}} (\hat{x}_1 \mid x_1) \right) 
	+ E_{x_2} \left[ \frac{h_2(x_2) - h_1}{H} u_1^{\tilde{f}} \left(x \right) + \frac{h_1 - h'_2(x_2)}{H} 
	u_1^{\tilde{f}} \left(\hat{x}_1 \mid x \right) \right]	
	\nonumber
	\\
	&\geq -\frac{2 \overline{\eta}}{H},
	\label{Lem_AM_13}
\end{align}
where the inequality follows from the fact that $| h_2(x_2) - h_1 | \leq 1$, 
$| h_1 - h'_2(x_2)| \leq 1$, and
$\tilde{f}$ is IC for the seller from Lemma 4 in the paper.
Inequalities \eqref{Lem_AM_5} and \eqref{Lem_AM_13} verify our claim. 
Next, suppose $\overline{s}_1^k(x_1) < h_1 -1$.
Then, $h'_2(\cdot) = \min \{ \overline{s}_1^k(x_1), \overline{s}_2^k(\cdot) \} = \overline{s}_1^k(x_1)$.
Because $\tilde{f}$ is both IC and IR for the seller from Lemma 4, 
we obtain
\begin{align*}
	&E_{x_2} \left[ \frac{h_1 - 1}{H}u_1^{\tilde{f}} \left(x \right) - \frac{h'_2(x_2)}{H}u_1^{\tilde{f}} \left(\hat{x}_1 \mid x \right) \right]
	\\
	&= \frac{\overline{s}_1^k(x_1)}{H} \left( U_1^{\tilde{f}} (x_1) - U_1^{\tilde{f}} (\hat{x}_1 \mid x_1) \right) 
	+ \frac{h_1-1-\overline{s}_1^k(x_1)}{H} U_1^{\tilde{f}} (x_1)
	\geq 0.
\end{align*}
This inequality, together with the first case,
verifies our claim.

Thus, $m_1$ strictly dominates $s_1(x_1)$. 
Hence, $s_1^k(x_1) = s_1^0(x_1)$.

\noindent
{\bf Step 7.}
Fix any $x_2 \in X_2$.
We show that $\underline{s}_2^k(x_2) = x_2$ and $\overline{s}_2^k(x_2) \in \{ h, h+1 \}$.
To derive a contradiction, suppose the opposite. 
First, suppose $\underline{s}_2^k(x_2) \in X_2 \setminus\{ x_2 \}$.
This yields a contradiction because $\underline{s}_1^k(x_1) \equiv x_1$ from Step 6, $\tilde{f}$ is EPIC for the buyer 
from Lemma 4 in the paper, 
and $t_2$ imposes a penalty (i.e., $\mathbb{I}(\underline{s}_2^k(x_2) \not= x_2) = 1$) on the buyer.
Second, suppose $\underline{s}_2^k(x_2) = 0$.
This yields a contradiction because 
$\pi \not\in \mathcal{P}^0 \setminus \overline{\mathcal{P}}^1$ implies $U_2^{\tilde{f},\pi}(x_2) > 0$, 
and $\pi \in \mathcal{P}^0 \setminus \overline{\mathcal{P}}^1$ implies $\overline{s}_1^k = \overline{s}_1^0 = 0$
(and thus, $\varphi(\cdot, \overline{s}_1^k(\cdot), \cdot, \cdot)$ always specifies the no-trade outcome)
from Steps 4 and 5.
Finally, suppose $\overline{s}_2^k(x_2) \not\in \{ h, h+1 \}$.
We suppose, without loss of generality, that $\overline{s}_2^k(x_2) \geq h+2$
because $\overline{s}_1^k(\cdot) \in \{ h, h+1\}$, and $h \geq 1$ implies $U_2^{\tilde{f},\pi}(x_2) > 0$.
The message $s_2(x_2)$ is strictly dominated by $m_2$ such that 
$m_2^k \equiv (x_2, h+1)$ 
and $m_2^{l} \equiv s_2^{l}(x_2)$ for each $l \not= k$ 
because for each $x_1$, $t_2(\overline{s}_1^0(x_1), m_2) - t_2(\overline{s}_1^0(x_1), s_2(x_2)) \geq 1/(2HK) > 0$.
This is a contradiction.
\newline

Third, we characterize the ex-post outcomes 
given 
the iteratively undominated strategy profile $s = (s_1, s_2)$ in the game $(G, \pi)$.

\noindent
{\bf Step 8.}
Fix any $x \in X$.
Steps 1--7 imply that, for each $i$ and $k$, 
$\underline{s}_i^k(x_i) = x_i$, 
$\overline{s}_1^0(x_1) = \overline{s}_1^k(x_1) \in \{h,h+1\}$, and 
$\overline{s}_2^k(x_i) \in \{h, h+1\}$.
The ex-post outcome is 
\begin{align*}
	g(s(x)) &= \varepsilon f_1(x_1) 
	+ \varepsilon^2 \left( f_2(x) + b(x_1, \overline{s}_2^0(x_2)) \right)  
	+ \varepsilon^3 c(x_1, \overline{s}_1^0(x_1), x_2, \overline{s}_2^0(x_2)) 
	\nonumber
	\\ 
	&\ 
	+ \varepsilon^4 (0, t_2(\overline{s}_1^0(x_1), s_2(x_2)))
	+ \frac{1- \varepsilon - \varepsilon^2 -\varepsilon^3 - \varepsilon^4 }{K}
	\sum_{k=1}^K \frac{\min\{\overline{s}_1^{k}(x_1), \overline{s}_2^{k}(x_2)\}}{H} \tilde{f}(x). 
\end{align*}
From this equation, we obtain the following inequality:\footnote{
		The term $\overline{\eta}/H$ is needed if for some $k \geq 1$, 
$\overline{s}_1^{k}(x_1) = h + 1$ and $\overline{s}_2^{k}(X_2) = \{ h, h+1 \}$ 
(i.e., the seller is uncertain about the probability of implementing $\tilde{f}$ at $k$).} 
\begin{align}
	U_1^{g \circ s}(x_1) &\leq -\overline{\eta}\varepsilon  + (\overline{\eta} + 1)\varepsilon^2  
	+ 2 \overline{\eta} \varepsilon^3
	+ \frac{\varepsilon^4}{2}
	+ \frac{\overline{\eta}}{H} + U_1^{\tilde{f}}(x_1).
	\label{Lem_AM_18}
\end{align}
If $\pi \in \mathcal{P}^H$ (i.e., $h = H$), then $\overline{s}_i^k(x_i) = H$ for each $i$ and $k$, 
and thus, $s$ is the unique iteratively undominated strategy profile in $(G, \pi)$ and 
\begin{align}
	g(s(x)) &= \varepsilon f_1(x_1) 
	+ \varepsilon^2 \left( f_2(x) + b(x_1, H) \right) 
	+ \varepsilon^3 c(x_1, H, x_2, H) 
	\nonumber
	\\ 
	&\ \ \ \ \ \ \ \ \ \ \ 
	+ \left( 1- \varepsilon - \varepsilon^2 -\varepsilon^3 - \varepsilon^4 \right) \tilde{f}(x). 
	\label{Lem_AM_17}
\end{align}
Equation \eqref{Lem_AM_17} implies that, if $\pi \in \mathcal{P}^H$, then
\begin{align}
	U_1^{g \circ s}(x_1) &\geq - 3\overline{\eta}\varepsilon - (\overline{\eta}+1)\varepsilon^2  
	+ \left( 1- \varepsilon - \varepsilon^2 -\varepsilon^3 - \varepsilon^4 \right) U_1^{\tilde{f}}(x_1).
	\label{Lem_AM_19}
\end{align}

Finally, we show that the mechanism $G$ has properties (i)--(iii) in Lemma \ref{Lem_AM_S}.

\noindent
{\bf Step 9.}
Lemma 4 in the paper implies that $U_1^{\tilde{f}}(x_1) > U_1^{f}(x_1)$ if and only if $x_1 \in \tilde{X}_1$. 
Using this fact with \eqref{Lem_AM_18} and \eqref{Lem_AM_19}, 
we assume, without loss of generality, that $\varepsilon$ and $1/H$ are small enough that, 
for each $\pi' \in \Delta(X_1)$ and
$f' \in BN(G,\pi')$, 
\begin{align}
	U_1^{f'}(x_1) &< U_1^{f}(x_1) \textnormal{ if } x_1 \not\in \tilde{X}_1,
	\label{Lem_AM_20}
	\\
	U_1^{f'}(x_1) &> U_1^{f}(x_1) \textnormal{ if } x_1 \in \tilde{X}_1 \textnormal{ and } 
	f' = g \circ s, 
	\label{Lem_AM_21}
\end{align}
where the allocation $g \circ s$ is given by \eqref{Lem_AM_17}.
It follows that $X_1^{f}(G) = X_1 \setminus \tilde{X}_1$, that is, 
$G$ has property (i) in Lemma \ref{Lem_AM_S}.
Because $\Delta(X_1 \setminus X_1^{f}(G)) = \Delta(\tilde{X}_1) \subseteq \mathcal{P}^H$, 
Step 8 with \eqref{Lem_AM_21} implies that $G$ has 
properties (ii) and (iii)
in Lemma \ref{Lem_AM_S}.
\end{proof}

\bibliographystyle{econ-jet}
\bibliography{1}

\end{document}